\documentclass[twocolumn,amsmath,amssymb,prd]{revtex4-1}

\usepackage{graphicx}
\usepackage{pstricks}
\usepackage{textcomp}
\usepackage{multirow}

\def\al{\alpha}
\def\be{\beta}
\def\ga{\gamma}
\def\de{\delta}
\def\ep{\epsilon}

\def\ze{\zeta}
\def\et{\eta}
\def\th{\theta}

\def\ka{\kappa}
\def\la{\lambda}

\def\rh{\rho}

\def\si{\sigma}

\def\ta{\tau}
\def\up{\upsilon}
\def\ph{\phi}

\def\ch{\chi}
\def\ps{\psi}
\def\om{\omega}
\def\Ga{\Gamma}

\def\La{\Lambda}

\def\da{^\dagger}
\def\tr{\textrm{tr}}
\def\pr{^\prime}

\def\D#1{D_{(#1)}}
\def\Da#1{D_{\al_{#1}}}

\def\Db#1{D_{\be_{#1}}}

\def\Dn#1{D_{(n_{#1})}}

\def\hc{\textrm{h.c.}}
\def\simn{\si_{\mu\nu}}
\def\scdots{\!\cdots\!}

\newcounter{tc1}\newcounter{tc2}
\newcounter{tr1}\newcounter{tr2}
\newlength{\h}
\def\newtableau#1#2{\psset{unit=12pt,linewidth=0.5pt}%
  \setlength{\h}{#2\psunit}\setlength{\h}{0.5\h}\addtolength{\h}{-0.3\psunit}
  \begin{pspicture}[shift=-\h](#1,#2)\small%
    \setcounter{tc1}{0}\setcounter{tc2}{1}%
    \setcounter{tr1}{#2}\setcounter{tr2}{#2}\addtocounter{tr1}{-1}%
    \psline(0,0)(0,#2)(#1,#2)}
\def\endtableau{\end{pspicture}}
\def\newbox#1{%
  \psline(\value{tc1},\value{tr1})(\value{tc2},\value{tr1})(\value{tc2},\value{tr2})%
  \rput(\value{tc1},\value{tr1}){\rput(0.5,0.5){#1}}
  \addtocounter{tc1}{1}\addtocounter{tc2}{1}}

\def\newrow{%
  \addtocounter{tr1}{-1}\addtocounter{tr2}{-1}%
  \setcounter{tc1}{0}\setcounter{tc2}{1}}

\def\co{{\cal O}}

\def\cL{{\cal L}}

\def\Ftil{\widetilde{F}}

\def\half{{\textstyle{\frac 1 2}}}
\def\quar{{\textstyle{\frac 1 4}}}
\def\eigh{{\textstyle{\frac 1 8}}}

\def\lsim{\mathrel{\rlap{\lower4pt\hbox{\hskip1pt$\sim$}}
    \raise1pt\hbox{$<$}}}
\def\gsim{\mathrel{\rlap{\lower4pt\hbox{\hskip1pt$\sim$}}
    \raise1pt\hbox{$>$}}}

\def\prt{\partial}

\newcommand{\beq}{\begin{equation}}
\newcommand{\eeq}{\end{equation}}
\newcommand{\bea}{\begin{eqnarray}}
\newcommand{\eea}{\end{eqnarray}}
\newcommand{\rf}[1]{(\ref{#1})}
\newcommand{\nn}{\nonumber}

\def\etal{{\it et al.}}

\def\psb{\overline{\ps}{}}

\def\ov{\overline}

\def\lR#1#2#3#4{R_{#1#2#3#4}}
\def\uR#1#2#3#4{R^{#1}_{{\pt{#1}}#2#3#4}}
\def\LC#1#2#3{\Ga^{#1}_{{\pt{#1}}#2#3}}

\def\cO{\mathcal O}

\def\pt#1{\phantom{#1}}

\def\nsc#1#2#3{\om_{#1}^{{\pt{#1}}#2#3}}

\def\lulsc#1#2#3{\om_{#1\pt{#2}#3}^{{\pt{#1}}#2}}
\def\llusc#1#2#3{\om_{#1#2}^{{\pt{#1#2}}#3}}

\def\vb#1#2{e_{#1}^{{\pt{#1}}#2}}
\def\ivb#1#2{e^{#1}_{{\pt{#1}}#2}}
\def\uvb#1#2{e^{#1#2}}
\def\lvb#1#2{e_{#1#2}}

\def\abcd#1{\al_{#1}\be_{#1}\ga_{#1}\de_{#1}}
\def\uk{\breve{k}}
\def\bk{\overline{k}}

\def\L{\textrm{L}}

\def\ol#1{\overline{#1}{}}
\def\vev#1{\langle {#1}\rangle}

\def\ab{\ol{a}}
\def\kb{\ol{k}}
\def\eb{\ol{e}}
\def\gb{\ol{g}}

\def\av{\vev{a}}
\def\kv{\vev{k}}
\def\ev{\vev{e}}
\def\gv{\vev{g}}

\def\xx{g,\de}
\def\xxa{A,\de}

\def\ua{\breve{a}}
\def\ub{\breve{b}}
\def\uc{\breve{c}}
\def\ud{\breve{d}}
\def\ue{\breve{e}}
\def\uf{\breve{f}}
\def\ug{\breve{g}}
\def\uG{\breve{G}}
\def\uH{\breve{H}}
\def\um{\breve{m}}

\def\V#1{V^{(#1)}}

\def\Lb{\ol{L}}
\def\Rb{\ol{R}}
\def\Qb{\ol{Q}}
\def\Ub{\ol{U}}
\def\Db{\ol{D}}

\begin{document}

\title{Backgrounds in gravitational effective field theory}

\author{
V.\ Alan Kosteleck\'y and Zonghao Li}

\affiliation{
Physics Department, Indiana University, 
Bloomington, IN 47405, USA}

\date{August 2020} 

\begin{abstract}

Effective field theories describing gravity coupled to matter are investigated,
allowing for operators of arbitrary mass dimension.
Terms violating local Lorentz and diffeomorphism invariance
while preserving internal gauge symmetries are included.
The theoretical framework for violations 
of local Lorentz and diffeomorphism invariance 
and associated conceptual issues are discussed,
including transformations in curved and approximately flat spacetimes,
the treatment of various types of backgrounds,
the implications of symmetry breaking,
and the no-go constraints for explicit violation in Riemann geometry.
Techniques are presented for the construction of effective operators,
and the possible terms in the gravity, gauge, fermion, and scalar sectors
are classified and enumerated.
Explicit expressions are obtained for terms containing operators 
of mass dimension six or less in the effective action
for General Relativity coupled to the Standard Model of particle physics.
Special cases considered include Einstein-Maxwell effective field theories 
and the limit with only scalar coupling constants.

\end{abstract}

\maketitle

\section{Introduction}
\label{Introduction}

The theory of General Relativity (GR)
coupled to the Standard Model (SM) of particle physics
provides an impressive description of many features of the Universe
over a wide range of distance scales.
Obtaining a fully satisfactory combination of gravity with quantum physics 
remains an open challenge,
however,
and a deeper underlying unified theory 
is expected to emerge at the Planck scale.
This theory could be founded 
on Riemann geometry or one of its extensions,
on a non-Riemann geometry,
or on nongeometric mathematics. 
In any scenario,
observable effects from the Planck scale at low energies
can be expected to include small deviations from the known physics
described by GR coupled to the SM,
and their detection would offer guidance in the construction
of the underlying theory. 
It is thus of central interest to ask how key properties
of the underlying theory could in principle manifest themselves 
in experiments and observations
performed using current and near-future technology.

A powerful tool for investigating prospective signals
from the Planck scale in a model-independent way
is provided by effective field theory
\cite{sw09}.
Tiny remnant couplings from the underlying theory
can be expected to emerge at the level of effective field theory 
as terms correcting the action of GR coupled to the SM,
with the size of the physical effects governed by small coefficients 
reflecting features of the underlying theory.
An individual term in the effective Lagrange density
is the product of an operator constructed from fields observed in nature
with a controlling coupling coefficient.
In typical applications of effective field theory,
the coupling coefficients are assumed to be constant scalars
and are called coupling constants.
However,
the solutions of the underlying theory
describing our Universe may include nontrivial backgrounds,
which would then be reflected
at the level of the effective field theory
as coupling coefficients that 
could vary with spacetime position 
and that could be tensorial rather than scalar.
Tensorial couplings could arise directly 
from features of the underlying theory,
but even if the underlying theory generates 
only a nonconstant scalar background
then the ensuing effective field theory
can contain vector and tensor coupling coefficients
determined by the derivatives of the scalar.

For the above reasons,
a comprehensive investigation
of the effective description of the underlying theory
requires the inclusion of nonconstant tensor backgrounds
as couplings in the effective field theory.
Backgrounds of this type have anomalous symmetry properties
under the spacetime transformations of GR and the SM.
In particular,
their existence implies that the effective field theory
can contain apparent violations of local Lorentz invariance
and of diffeomorphism symmetry.
It is therefore of interest
to construct the general effective field theory
describing gravity coupled to matter
while allowing for terms violating these symmetries.

The framework for the general effective field theory
based on GR coupled to the SM is presented in Ref.\ \cite{ak04}.
This framework allows for violations
of local Lorentz and diffeomorphism invariances.
Terms in the corresponding effective field theory
can be organized according to their mass dimension $d$ in natural units,
with terms of larger $d$ expected to have smaller effects at low energies.
All terms with $d\leq 4$,
called minimal terms,
are presented in Ref.\ \cite{ak04},
and they have been the subject of numerous investigations 
\cite{tables,reviews}. 
In the pure-gravity sector,
phenomenological studies 
\cite{bk06,se09,abk10,bt11,macf14,yb15,afm18,18ms,blps18,%
nd18,18ap,rx19,19cl,clcc20,20ms,20bonder,20obn}
and experiments 
\cite{2007Battat,2007MullerInterf,2009Chung,2010Bennett,2012Iorio,%
2013Bailey,2014Shao,he15,le16,bo16,2017Abbott,19xu,20liu}
have constrained most minimal terms to impressive sensitivities.
The minimal matter-gravity sector has also been widely explored 
both phenomenologically
\cite{kt09,kt11,jt12,13bonder,jt15,19escobar,19xiao,19lshao}
and experimentally
\cite{2010Panjwani,2010Altschul,2011Hohensee,ms16,%
2016Flowers,bourgoin17,19microscope},
with the focus to date being primarily on spin-independent effects.
A subset of terms in the pure-gravity sector
with operators of larger mass dimensions $d\geq5$ has been constructed
and some experimental constraints obtained 
\cite{bkx15,lk,hust15,hustiu,hust16,2017Chen,%
bh17,kt15,jt16,km16,qb16,yu16,km17,2017Shao,km18,18shaobailey,%
bl18,amnp18,19ba,19mewes,19shaobailey,2019Shao,2017Wang,20lshao}.

One goal of this work is to extend the analysis
of terms in the effective field theory
to include nonminimal terms involving both gravity and matter
as well as nonminimal terms in the pure-gravity sector.
We present a systematic methodology for constructing all these terms,
and we obtain explicit results for $d\leq6$
for a generic theory,
for GR coupled to the SM,
and for its limits including Einstein-Maxwell theories
and the restriction to scalar coupling constants. 
To achieve this,
we develop further the framework presented in Ref.\ \cite{ak04},
combining it with techniques recently developed
for the construction of terms of arbitrary $d$
in nonabelian effective field theories in Minkowski spacetime 
\cite{kl19}.
We discuss the relevant spacetime transformations,
establishing the properties of various types of backgrounds
and their symmetry violations,
and we characterize the relationships 
between different types of effective terms 
and their linearizations to approximately flat spacetimes.
The analysis in Ref.\ \cite{ak04} revealed unexpected constraints 
on the form of the effective field theory,
arising from compatibility requirements 
between the variational procedure and the Bianchi identities,
which have recently been the subject of extensive study 
\cite{rb15,15rb2,bs16,bbw19}.
Here,
we revisit these no-go constraints to clarify their impact 
in the context of perturbative corrections to known physics,
showing that they can determine 
whether the underlying theory is based on Riemann geometry
or instead emerges from an alternative geometry
or nongeometric mathematics. 

The outline of this work is as follows.
In Sec.\ \ref{Framework},
we study the framework for the gravitational effective field theory. 
Essential definitions and conventions are presented in 
Sec.\ \ref{Metric, vierbein, and covariant derivative},
while Sec.\ \ref{Spacetime transformations}
discusses key concepts about spacetime transformations 
in both curved and approximately flat spacetimes.
The treatment of backgrounds is initiated in Sec.\ \ref{Backgrounds},
and the implications for violations of symmetries in curved spacetimes
are described in Sec.\ \ref{Symmetry violations}.
The relationships between broken symmetries in curved and flat spacetimes
are elucidated in Sec.\ \ref{Linearization}.
Our discussion of the framework concludes
with a treatment of the no-go constraints 
in Sec.\ \ref{No-go constraints}.

The methodology for the construction 
of terms in an effective Lagrange density
built within this framework
is presented in Sec.\ \ref{Effective gravity},
along with a compact notation for various types of backgrounds.
The procedure to obtain gauge-covariant spacetime operators
and related results is described in Sec.\ \ref{Dynamical operators}.
Using these results,
in Sec.\ \ref{Pure-gravity sector}
we enumerate and classify operators 
involving pure-gravity fields and backgrounds.
We turn attention to matter fields in Sec.\ \ref{Matter sector},
which presents the explicit form of operators with $d\leq 6$
for gauge fields,
Dirac fermions,
and scalars.
The application of all these results 
to realistic effective field theories
is considered in Sec.\ \ref{Applications}.
Terms with $d\leq 6$ in the Lagrange density 
for various Einstein-Maxwell theories
are tabulated in Sec.\ \ref{sec:abelian}.
Section \ref{sec:SME} enumerates the explicit form of terms 
in effective field theories based on GR coupled to the SM.
The limit with backgrounds acting only as scalar coupling constants
is discussed in Sec.\ \ref{sec:LI}.
We conclude with a summary in Sec.\ \ref{Summary}.

Throughout this work,
we adopt the conventions of Ref.\ \cite{ak04}.
We assume vanishing torsion and nonmetricity
except where otherwise indicated,
so the definitions and results in Appendix A of Ref.\ \cite{ak04}
apply with the torsion and contortion set to zero.
In particular,
Greek indices are used for tensorial components on the spacetime manifold,
and Latin ones for ones on the tangent space. 
The Minkowski metric $\et_{ab}$ has positive signature +2,
and the Levi-Civita tensor $\ep_{abcd}$ 
is fixed by $\ep_{0123}=+1$.
The Dirac matrices satisfy
$\{\ga^a,\ga^b\}=-2\et^{ab}$,
with $\si^{ab}$ defined as $\si^{ab} = i [\ga^a,\ga^b]/2$.

\section{Framework}
\label{Framework}

In this section,
we describe the concepts and framework
appropriate for the construction
of the general effective field theory
based on GR coupled to the SM.
Individual subsections treat
the basic setup,
spacetime transformations,
backgrounds,
symmetry violations,
linearization,
and the no-go constraints.

\subsection{Metric, vierbein, and covariant derivative}
\label{Metric, vierbein, and covariant derivative}

The geometric underpinning for GR coupled to the SM 
is a four-dimensional smooth manifold called spacetime
that contains a dynamical four-dimensional metric $g_{\mu\nu}$.
While the underlying unified theory may be nongeometric,
its low-energy approximation must reduce to known physics.
The corresponding effective field theory extending GR coupled to the SM
can therefore reasonably be taken as based on a manifold with a metric.

The SM incorporates fermions as spinor fields,
which are conveniently described using the vierbein formalism
\cite{uk56}.
In this approach,
spinor fields at each spacetime point
and related objects including the Dirac gamma matrices 
are all defined in a coordinate frame in the tangent space,
called a local frame.
In contrast,
a coordinate frame on the manifold is called a spacetime frame.
The vierbein $\vb\mu a(x)$ connects local frames with spacetime frames.

Since tangent spaces are flat,
the vierbein is related to the metric by 
\beq
g_{\mu\nu}=\vb\mu a\vb\nu b\et_{ab}.
\label{gee}
\eeq
For simplicity,
we assume the connection is both metric and torsion free,
so the metric $g_{\mu\nu}$, 
the spacetime connection $\LC\la\mu\nu$,
and the spin connection $\nsc\mu ab$
are all fixed by the vierbein $\vb\mu a$.
The vierbein can therefore be taken as the sole field
describing the gravitational dynamics.
We adopt a covariant derivative satisfying
\beq
D_\mu \vb \nu a
\equiv \prt_\mu \vb \nu a- \LC \la\mu\nu \vb\la a+\lulsc \mu ab \vb \nu b
=0,
\eeq
which then also implies $D_\la g_{\mu\nu} =0$.
Note,
however,
that the theoretical framework used in this work
can readily be extended to include nonzero nonmetricity and torsion
\cite{ak04}.
In particular,
key implications such as the no-go constraints 
also hold when the requirements of zero torsion and nonmetricity
are relaxed.

The SM incorporates bosons in gauge field theory. 
The corresponding covariant derivative in curved spacetime
then incorporates a gauge connection $A_\mu$ 
along with the spacetime connection $\LC \la\mu\nu$
and the spin connection $\nsc \mu ab$.
The gauge connection 
is the connection for the internal gauge degrees of freedom
and acts on objects with gauge indices,
the spacetime connection is the connection on the spacetime manifold
and acts on spacetime indices,
and the spin connection is the connection in local frames
and acts on local indices.

Following standard convention,
the explicit use of gauge and spinor indices is avoided in this work.
For example,
the action of the covariant derivative on a fermion field $\ps$
with suppressed spinor and gauge indices
can be written in this convention as
\beq
D_\mu \ps=\prt_\mu \ps +\quar i \nsc \mu a b \si_{ab}\ps -ig A_\mu \ps,
\eeq
where $g$ is the gauge coupling constant.
As another example,
the gauge field strength $F_{\mu\nu}$
contains both spacetime and gauge indices,
and its covariant derivative can be written as 
\beq
D_\la F_{\mu\nu}=
\prt_\la F_{\mu\nu} -\LC \rh\la\mu F_{\rh\nu}-\LC \rh\la\nu F_{\mu\rh}
-ig[A_\la, F_{\mu\nu}].
\eeq
Note that the 16 Dirac gamma matrices 
$\Ga^A\in \{I,i\ga_5,\ga^a,\ga_5 \ga^a, \si^{ab}/2\}$
forming a complete basis for spinor matrices 
typically appear in fermion bilinears containing local-frame indices.
We define $D_\mu \Ga^A$ by imposing the product rule
\beq
D_\mu(\psb \Ga^A \ps)=
(\ov{D_\mu \ps})\Ga^A \ps+\psb (D_\mu \Ga^A)\ps+\psb\Ga^A(D_\mu\ps),
\label{dga}
\eeq
which can be shown to imply that $D_\mu \Ga^A=\prt_\mu \Ga^A=0$.

\subsection{Spacetime transformations}
\label{Spacetime transformations}

Spacetime transformations play a key role in GR
\cite{texts} 
and hence also in the construction of the general effective field theory
based on GR coupled to the SM.
In this subsection,
we discuss some essential concepts for transformations 
in curved spacetime and in approximately flat spacetime,
focusing on requisite aspects for the present work.

\subsubsection{Particle versus observer}
\label{Particle versus observer}

It is convenient and useful to distinguish
two notions of transformations,
called particle and observer transformations
\cite{ck97,ak04}.
Particle transformations change dynamical particles and fields,
while observer transformations change the observer frame. 
In the absence of backgrounds,
the component forms of the two transformations are inverses of each other
and in that context are sometimes called active and passive.
However,
this equivalence fails in the presence of backgrounds.
A particle transformation affects dynamical particles and fields
but leaves any backgrounds invariant,
which can modify the physics associated with couplings
between the dynamical variables and the background.
In contrast,
an observer transformation amounts to a coordinate transformation,
which changes the components of fields and backgrounds
but is assumed to leave invariant the physics.
A physical symmetry associated with a given particle transformation
can therefore be violated in the presence of backgrounds,
even though the physics remains invariant 
under the corresponding observer transformation.

Mathematically,
particle transformations involve mappings 
of the spacetime manifold and its tangent and cotangent bundles,
whereas observer transformations
are implemented on the atlas of the manifold.
Since physics is independent of the coordinate frames used for the atlas
but can depend on the manifold mappings,
discussions of symmetry violations
are best conducted in the language of particle transformations
without invoking frame changes.
For this reason,
we focus here primarily on particle transformations,
often omitting the word particle for simplicity.
Unless indicated otherwise,
every transformation in this work 
should be understood as a particle transformation.

Note that distinguishing between the two types of transformations 
can be subtle in practice.
For instance,
special relativity is typically introduced in textbooks 
from the perspective of observer Lorentz transformations.
This approach works well for Lorentz-invariant theories,
where the component forms of particle and observer Lorentz transformations
are essentially equivalent.
However,
analyzing generic violations of Lorentz invariance
in the context of observer Lorentz transformations is challenging at best.
Physical Lorentz-violating effects are features 
of experimental configurations of particles and fields
rather than features of the observer,
so the general treatment of Lorentz violation 
cannot readily be described using modified observer transformations
\cite{ak07}.

\subsubsection{Transformations in curved spacetime}
\label{Transformations in curved spacetime}

In Minkowski spacetime,
the central spacetime transformations are global transformations
that include spatial rotations, Lorentz boosts, and translations.
The rotations and Lorentz boosts form the group of Lorentz transformations,
which is enlarged by translations to the Poincar\'e group.
All these Minkowski-spacetime transformations are isometries
of the Minkowski metric $\et_{\mu\nu}$,
and they move spacetime points. 
For example,
a global rotation about a point $P$ in the spacetime
maps all points other than $P$ into different points. 

In contrast,
the metric $g_{\mu\nu}$ in a generic curved spacetime 
typically has no isometries,
and so the usual notions 
of global Lorentz transformations and translations
play no particular role.
Instead,
it is useful to study local Lorentz transformations 
and diffeomorphisms.

Local Lorentz transformations are Lorentz transformations
in the tangent space at each spacetime point,
leaving the spacetime point unmoved.
Under a local Lorentz transformation,
the vierbein and metric transform as
\beq
e_\mu^{\pt{\mu}a}(x) \to \La^a_{\pt{a}b}(x)e_\mu^{\pt{\mu}b}(x),
\quad
g_{\mu\nu}(x) \to g_{\mu\nu}(x),
\label{eq:LLT}
\eeq
where $\La^a_{\pt{a}b}(x)$ 
are the components of the matrix $\La(x)$ 
for the local Lorentz transformation at the point $x$. 
Other dynamical boson fields transform similarly,
with spacetime indices unchanged
and local indices acted on by the components of $\La(x)$.
Fermion fields are transformed
by the corresponding matrices $S(\La(x))$
in the appropriate spinor representation of the local Lorentz group.

Note that
local Lorentz transformations at different spacetime points
are typically different.
However,
an associated global transformation can be defined 
in any curved spacetime
by requiring that the same local Lorentz transformation
is performed simultaneously at every spacetime point.
This can be termed a global local Lorentz transformation,
and it is the analogue of a global gauge transformation
constructed from a local gauge transformation in a gauge field theory.
Global local Lorentz transformations leave spacetime points fixed,
so they cannot be the analogues of global Lorentz transformations
in Minkowski spacetime.
Instead,
the analogues can be taken to be 
certain types of Lorentz transformations 
defined in approximately flat spacetimes,
as described in Sec.\ \ref{approximatelyflat} below.

Diffeomorphisms in a curved spacetime 
capture the idea of moving spacetime points.
Under a diffeomorphism,
a spacetime point at position $x$ is mapped to another point at $x\pr$
according to
\beq
x^\mu \to x^{\prime\mu}=x^\mu+\xi^\mu (x),
\label{eq:movepoints}
\eeq
where $\xi^\mu(x)$ is smooth and the mapping is assumed invertible.
Here,
$x^{\prime\mu}$ denotes the components of the new position 
in the original coordinates,
which remain unchanged by the transformation.
Dynamical fields on the manifold transform 
according to the pushforward or pullback
induced by the diffeomorphism.
For example,
the vierbein and metric transform as
\bea
\vb \mu a (x) &\to& 
e\pr_\mu{}^a (x\pr)
=\frac{\prt x^\rh}{\prt x^{\prime\mu}} \vb \rh a(x),
\nn\\
g_{\mu\nu}(x) &\to& g\pr_{\mu\nu}(x\pr)
=\frac{\prt x^\rh}{\prt x^{\prime\mu}} 
\frac{\prt x^\si}{\prt x^{\prime\nu}} g_{\rh\si}(x),
\label{eq:DT}
\eea
where $e\pr_\mu{}^a (x\pr)$ and $g\pr_{\mu\nu}(x\pr)$ 
are the new vierbein and metric at the point $x\pr$ 
after the diffeomorphism.
In contrast,
dynamical fields valued in local frames,
including spinor fields,
transform like scalar fields under a diffeomorphism.

Although the expressions \rf{eq:DT} appear similar 
to those for a general coordinate transformation,
the physical interpretation is different.
Only the coordinates change
under general coordinate transformations,
leaving physical particles and fields invariant.
General coordinate transformations 
can thus be identified as observer diffeomorphisms.
In contrast,
the particle diffeomorphisms of interest here
change physical particles and fields while leaving 
the coordinate system unchanged.

Fields can be valued at any position on the manifold.
When valued at the same position,
dynamical fields undergoing a diffeomorphism \rf{eq:movepoints}
with infinitesimal $\xi^\mu(x)$
change by the corresponding Lie derivative.
For example,
under an infinitesimal diffeomorphism the vierbein and metric transform as
\bea
e\pr_\mu{}^a (x)
&=&
\vb \mu a(x) - \cL_\xi \vb \mu a (x)
\nn\\
&=&
\vb \mu a - \vb \rh a \prt_\mu \xi^\rh -\xi^\la \prt_\la \vb \mu a,
\nn\\
g\pr_{\mu\nu}(x)&=&
g_{\mu\nu}(x)-\cL_\xi g_{\mu\nu}(x)
\nn\\ 
&=&
g_{\mu\nu}-g_{\rh\nu} \prt_\mu \xi^\rh -g_{\mu\si} \prt_\nu \xi^\si
-\xi^\la \prt_\la g_{\mu\nu}.
\label{eq:Lie}
\eea
These results can be derived directly 
from Eqs.\ \rf{eq:movepoints} and \rf{eq:DT}.
Equations where dynamical fields are valued at the same position
are often used in calculations involving diffeomorphisms,
and they are distinct from ones like Eq.\ \rf{eq:DT}
in which the fields are valued at different positions.

\subsubsection{Transformations in approximately flat spacetime}
\label{approximatelyflat}

As discussed above,
a local Lorentz transformation in a curved spacetime
leaves spacetime points fixed while acting on local frames,
and a diffeomorphism moves spacetime points
while leaving local frames unchanged.
Next,
we show that 
suitable combinations of local Lorentz transformations and diffeomorphisms 
in approximately flat spacetimes
can mimic the roles of global Lorentz transformations 
and translations in Minkowski spacetime.

Most experiments are performed in weak gravitational fields 
such as those found in the solar system.
The corresponding spacetimes are therefore approximately flat.
The vierbein and metric can then be decomposed as
\bea
e_{\mu a} (x) &=& \et_{\mu a} + \ep_{\mu a}(x) 
\approx \et_{\mu a} + \half h_{\mu a}(x) + \ch_{\mu a}(x),
\nn\\
g_{\mu\nu}(x) &=& \et_{\mu\nu}+h_{\mu\nu}(x),
\label{eq:linear}
\eea
where $\ep_{\mu a}\ll 1$, $h_{\mu\nu}\ll 1$, and $\ch_{\mu a}\ll 1$
with $h_{\mu\nu}$ symmetric and $\ch_{\mu a}$ antisymmetric.  
Under the diffeomorphism \rf{eq:movepoints},
the vierbein and metric transform according to Eq.\ \rf{eq:DT}.
We can assign the resulting changes in the vierbein and metric
to the fluctuations $h_{\mu\nu}(x)$ and $\ch_{\mu a}(x)$,
with the Minkowski metric defined as invariant.
For infinitesimal diffeomorphisms
and at lowest order in the fluctuations,
we then find
\bea
h_{\mu\nu}(x) &\to& 
h\pr_{\mu\nu}(x\pr)
= h_{\mu\nu}(x)-\et_{\rh\nu}\prt_\mu \xi^\rh(x)
-\et_{\mu\si}\prt_\nu \xi^\si(x),
\nn\\
\ch_{\mu a}(x) &\to&
\ch\pr_{\mu a}(x\pr)
= \ch_{\mu a}(x)
-\half\et_{\rh a}\prt_\mu \xi^\rh(x)
\nn\\
&&
\hskip 80pt
+\half\et^\rh_{\pt{\rh}a}\et_{\mu\si}\prt_\rh \xi^\si(x).
\label{eq:lindiffeo}
\eea
These transformations are called linearized diffeomorphisms
in approximately Minkowski spacetime.

Since the spacetime is assumed to be approximately flat,
one might anticipate the existence of notions similar 
to the global Lorentz transformations and translations
in Minkowski spacetime.
Indeed,
when the displacements $\xi^\mu(x)$ are independent of spacetime position,
the linearized diffeomorphisms \rf{eq:lindiffeo} 
take the same form as Minkowski-spacetime translations.
It is therefore natural to define translations
in an approximately Minkowski spacetime
as linearized diffeomorphisms with constant $\xi^\mu$.
This definition applies to all dynamical quantities,
including matter fields.

The linearized diffeomorphisms \rf{eq:lindiffeo}
can also be expressed as field transformations valued at the same position,
\bea
h_{\mu\nu}(x) &\to& 
h\pr_{\mu\nu}(x)
\approx h_{\mu\nu}-\et_{\rh\nu}\prt_\mu \xi^\rh
-\et_{\mu\si}\prt_\nu \xi^\si
\nn\\
&&
\hskip 40pt
-\xi^\la\prt_\la h_{\mu\nu},
\nn\\
\ch_{\mu a}(x) &\to&
\ch\pr_{\mu a}(x)
\approx \ch_{\mu a}
-\half\et_{\rh a}\prt_\mu \xi^\rh
+\half\et^\rh_{\pt{\rh}a}\et_{\mu\si}\prt_\rh \xi^\si
\nn\\
&&
\hskip 40pt
-\xi^\la\prt_\la \ch_{\mu a},
\label{eq:lindiffeosame}
\eea
in parallel with the result \rf{eq:Lie}.
If we make the further approximation 
of keeping only terms at leading order in small quantities,
the linearized diffeomorphisms \rf{eq:lindiffeosame} reduce to
\bea
h_{\mu\nu}(x) &\to& 
h_{\mu\nu}-\et_{\rh\nu}\prt_\mu \xi^\rh
-\et_{\mu\si}\prt_\nu \xi^\si,
\nn\\
\ch_{\mu a}(x) &\to&
\ch_{\mu a}
-\half\et_{\rh a}\prt_\mu \xi^\rh
+\half\et^\rh_{\pt{\rh}a}\et_{\mu\si}\prt_\rh \xi^\si.
\nn\\
\label{eq:gaugetrans}
\eea
These transformations in approximately Minkowski spacetime
are called gravitational gauge transformations,
or simply gauge transformations if there is no risk of confusion
with internal gauge transformations in the matter sector.
For $h_{\mu\nu}$ and $\ch_{\mu a}$,
gauge transformations and 
linearized diffeomorphisms valued at the same position 
thus differ by contributions involving the operator $\xi^\la\prt_\la$
that originates from the Lie derivative \rf{eq:Lie}.
However,
nongravitational fields are unaffected by the linearization procedure,
and so for consistency the corresponding contributions
must be kept when expanding nongravitational expressions
at leading order in small quantities.
A gauge transformation of a nongravitational operator $\cO(x)$ 
in the Lagrange density therefore can be defined as 
\beq
\cO(x)\to \cO(x) -  \cL_\xi \cO(x),
\label{eq:mattergaugetr}
\eeq
which retains the contribution $-\xi^\la \prt_\la \cO(x)$.

\renewcommand\arraystretch{1.2}
\begin{table*}
\caption{
\label{tab:transformations}
Some transformations in curved and approximately flat spacetimes.}
\setlength{\tabcolsep}{6pt}
\begin{tabular}{lll}
\hline
\hline
Manifold	&	Transformation	&	Definition	\\	
\hline						
General spacetime	&	Local Lorentz transformation	&	Eq.\ \rf{eq:LLT}	\\	
	&	Diffeomorphism	&	Eqs.\ \rf{eq:movepoints} and \rf{eq:DT}	\\	
	&	Infinitesimal diffeomorphism	&	Eq.\ \rf{eq:Lie}	\\	
	&	Global local Lorentz transformation	&	Local Lorentz transformation with $\La^a{}_b$ constant	\\	
	&	Manifold Lorentz transformation	&	Diffeomorphism $x^\mu \to \La^\mu{}_\nu x^\nu$, $\La^\mu{}_\nu$ constant	\\	
	&		&	\qquad and global local Lorentz transformation	\\	
	&	Translation	&	Diffeomorphism with $\xi^\mu$ constant	\\	[8pt]
Approximately flat spacetime	&	Lorentz transformation	&	Manifold Lorentz transformation	\\	
	&	Linearized diffeomorphism	&	Eqs.\ \rf{eq:lindiffeo} and \rf{eq:lindiffeosame}	\\	
	&	Gauge transformation	&	Eqs.\ \rf{eq:gaugetrans} and \rf{eq:mattergaugetr}	\\	
	&	Translation	&	Linearized diffeomorphism with $\xi^\mu$ constant	\\	
\hline
\hline
\end{tabular}
\end{table*}

To identify the analogues of Minkowski-spacetime Lorentz transformations
in an approximately flat spacetime,
it is useful to introduce a special set of transformations
on the curved manifold
called manifold Lorentz transformations.
These transformations are distinct 
both from the usual local Lorentz transformations
and also from the global local Lorentz transformations
described in the previous subsection. 
By definition,
manifold Lorentz transformations 
act both on spacetime points and on local frames. 
Under a transformation of this type 
specified by a fixed element $\La$ of the Lorentz group,
every spacetime point at position $x$ 
is mapped to another point according to the special diffeomorphism
\beq
x^\mu \to x^{\prime\mu}=\La^\mu_{\pt{\mu}\nu} x^\mu,
\label{eq:LT}
\eeq
where $x\pr$ is the new position 
expressed in the original coordinate system
and $\La^\mu_{\pt{\mu}\nu}$ are the components of $\La$.
In addition,
the vierbein and the metric on the manifold 
are defined to transform as
\bea
\vb \mu a (x) &\to& e^{\prime}_\mu{}^a(x\pr)
=(\La^{-1})^\rh_{\pt{\rh}\mu} \La^a_{\pt{a}b} \vb \rh b (x),
\nn\\
g_{\mu\nu}(x) &\to& g\pr_{\mu\nu}(x\pr)
=(\La^{-1})^\rh_{\pt{\rh}\mu} (\La^{-1})^\si_{\pt{\si}\nu} g_{\rh\si}(x),
\label{eq:GLT}
\eea
where 
the new vierbein $e_\mu^{\prime \hskip3pt a}(x\pr)$
and the new metric $g\pr_{\mu\nu}(x\pr)$ 
are at the new position $x\pr$ after the diffeomorphism,
and where $(\La^{-1})^\rh_{\pt{\rh}\mu}=\La_\mu^{\pt{\mu}\rh}$
are the components of the inverse of the matrix $\La$.
Other dynamical boson fields are defined to transform similarly
under manifold Lorentz transformations,
with both spacetime and local indices transforming according to $\La$
and its inverse,
while fermion fields transform according to the
corresponding spinor transformation $S(\La)$.

The manifold Lorentz transformations 
can be defined on any single coordinate chart 
in the atlas of a curved spacetime,
but the dependence on the coordinate chart
limits the value of the definition in the generic case.
However,
approximately flat spacetimes 
permit natural choices of coordinate systems
in which the vierbein and metric are given approximately by
$\vb\mu a\approx\delta_\mu^{\pt{\mu}a}$
and $g_{\mu\nu}\approx\eta_{\mu\nu}$.
These natural coordinate systems are not strictly unique,
but all manifold Lorentz transformations 
defined on them are closely related.
So for practical applications
and in particular in the context of experimental analyses,
we can select any one of them without loss of physical generality.
Moreover,
the existence of these natural coordinates 
in approximately flat spacetimes
insures that manifold Lorentz transformations are the natural analogues 
of Minkowski-spacetime Lorentz transformations.
This follows because Minkowski spacetime is a special manifold
on which a coordinate system can be chosen 
such that the vierbein and the metric take the form
$\vb\mu a=\delta_\mu^{\pt{\mu}a}$ and $g_{\mu\nu}=\eta_{\mu\nu}$.
A Minkowski-spacetime Lorentz transformation 
can then be viewed as 
a manifold Lorentz transformation 
defined in this chosen coordinate system for Minkowski spacetime.

Within this setup,
we can verify that manifold Lorentz transformations 
are combinations of local Lorentz transformations and diffeomorphisms.
Assume first that the local Lorentz transformations 
are global local Lorentz transformations,
so that the components $\La^a_{\pt{a}b}(x)$ in Eq.\ \rf{eq:LLT} 
are independent of spacetime position.
Assume further that the diffeomorphisms 
are the special transformations 
$x^{\prime\mu}=\La^\mu_{\pt{\mu}\nu}x^\nu$,
which preserve the Minkowski metric.
The transformations in Eq.\ \rf{eq:GLT} 
can then be identified as combinations 
of those in Eqs.\ \rf{eq:LLT} and \rf{eq:DT} in this limit.
Other dynamical boson fields transform appropriately,
as do the spinor fields.
It is therefore natural to identify 
global Lorentz transformations in an approximately Minkowski spacetime
as manifold Lorentz transformations.
The global Lorentz transformations can thus be understood as
suitable combinations of global local Lorentz transformations
and special diffeomorphisms that preserve the Minkowski metric.

Table \ref{tab:transformations} summarizes the various transformations
introduced above.
The first column identifies the type of manifold.
An entry in the second column specifies the transformation of interest,
while one in the third column provides its definition, 
either via a brief descriptive statement
or as a reference to defining equations in the text.
All these transformations play a key role in the present work.

\subsection{Backgrounds}
\label{Backgrounds}

A given term in the Lagrange density $\cL$ 
of the general effective field theory extending GR coupled to the SM
is the product of a field operator $\cO(x)$ 
with a coupling coefficient $k(x)$
or its derivatives.
Since it plays the role of a coupling,
$k$ can be viewed as a background in the theory or,
equivalently,
as a nonzero vacuum value of a field
\cite{ak04}.
This perspective holds irrespective 
of the detailed origin or nature of the coefficient 
in the context of the underlying theory.

Since the field operator $\cO$ may behave nontrivially
under spacetime transformations
and since the Lagrange density is a scalar density
under general coordinate transformations,
the background $k$ can carry spacetime and local indices.
For example, 
backgrounds with tensorial indices may arise in string theory 
\cite{ksp}.
In effective field theory,
the operator $\cO$ is bosonic
and so contains spinor fields only as combinations of fermion bilinears.
The background $k$ therefore carries no spinor indices.
For definiteness and simplicity,
we assume here that $k$ carries 
no indices associated with any internal gauge degrees of freedom.
Backgrounds carrying gauge indices are possible in principle 
and would appear in an effective field theory
that violates internal gauge invariance,
but investigating this possibility lies outside our scope.
For present purposes,
we can therefore treat $k$ 
as a tensor under general coordinate transformations 
and under observer local Lorentz transformations.
Note that $k$ must remain invariant under all particle transformations,
including both diffeomorphisms and local Lorentz transformations,
because it is nondynamical by construction.

The distinction between any upper and lower local indices carried by $k$
is physically irrelevant because the two types of indices
can be interconverted using the Minkowski metric,
which by definition is a nondynamical quantity.
However,
upper spacetime indices on $k$ 
can represent physically different effects from lower spacetime indices
because the two are connected by the metric $g_{\mu\nu}$,
which is a dynamical field.
As an example,
when $k$ is a nondynamical background
then a term in $\cL$ of the form $k_\mu \cO^\mu$ 
generates different contributions to the equations of motion 
than does the term $k^\mu g_{\mu\nu} \cO^\nu$.
These considerations imply that
we can limit attention to three types of indices on $k$
without loss of generality:
upper spacetime indices,
lower spacetime indices,
and local indices in any position.
An arbitrary background can thus be denoted as
$k^{\mu\ldots}{}_{\nu\ldots}{}^{a\ldots}(x)$.
Note that adopting a nonstandard definition of the vierbein,
which could lead to local frames with a nondynamical local metric
and nontrivial local curvature,
cannot introduce new physical effects
because the nonstandard and conventional vierbeins
are related via nondynamical algebraic equations.
Modulo possible derivatives acting on the background,
the general structure of a term in the Lagrange density $\cL$
can therefore be written in the form 
\beq
\cL \supset 
k^{\mu\ldots}{}_{\nu\ldots}{}^{a\ldots}(x) 
\cO_{\mu\ldots}{}^{\nu\ldots}{}_{a\ldots}(x), 
\label{genlagterm}
\eeq
where $\cO$ contains all dynamical fields
including any factors 
involving the vierbein $\vb \mu a$ and metric $g_{\mu\nu}$.
If derivatives acting on the background are present,
their indices must also be contracted
to insure that $\cL$ remains a scalar density.

Two classes of backgrounds $k$ can conveniently be identified,
according to whether they are spontaneous or explicit.
As the two classes have different physical implications,
for clarity in much of what follows
we denote spontaneous backgrounds by $\kv$ and explicit ones by $\kb$.
Spontaneous backgrounds $\kv$ arise 
as solutions of the equations of motion in the underlying theory
and hence are vacuum expectation values of underlying fields.
They satisfy the equations of motion and are thus on-shell quantities.
Fluctuations of the underlying fields about $\kv$ then exist
and can represent additional modes in the effective theory
\cite{bk05,bfk08},
including Nambu-Goldstone 
\cite{ng60}
and massive modes.
In contrast,
explicit background fields $\kb$ are specified by fiat 
and so are nondynamical.
They are unconstrained by equations of motion and hence can be off shell.
Moreover,
no dynamical fluctuations about them exist.
Intuitively,
a spontaneous background $\kv$ can be viewed 
as a special nondynamical background $\kb$ that must be on shell
and that has accompanying dynamical fluctuations.
The on-shell restriction and the presence of dynamical fluctuations
imply that the backgrounds $\kv$ and $\kb$ 
are associated with distinct physics.

Among the set of possible general background fields $k$
is a subset consisting of background vierbeins and background metrics.
In realistic applications,
the usual vierbein and metric have nonzero values in the vacuum,
which insures nonzero distances between points.
We can view these quantities as a background vierbein 
and a background metric.
In many theories,
including GR,
they emerge spontaneously as solutions of the equations of motion 
in otherwise empty regions of spacetime, 
where the energy-momentum tensor and other relevant sources vanish.
Like other background fields,
they are invariant under particle transformations.
In the present context,
it is thus appropriate to denote
the usual background vierbein by $\ev$
and the usual background metric by $\gv$.
For example,
the Minkowski solution to the GR field equations
has $\ev_\mu{}^a =\et_\mu{}^a$ and $\gv_{\mu\nu}=\et_{\mu\nu}$,
and for the background the basic relation \rf{gee} reduces to the identity 
\cite{bk05,bfk08}
\beq
\gv_{\mu\nu}=\ev_\mu{}^a \ev_\nu{}^b \et_{ab}.
\label{geevev}
\eeq
It follows that the general effective field theory
based on GR coupled to the SM
must contain at least one 
spontaneous background vierbein and background metric.
Note that one or more explicit backgrounds $\kb$ in the theory
may also have the same index structure as the usual vierbein and metric
and may therefore be identified as 
one or more explicit background vierbeins and metrics.
The notion of an explicit background vierbein
that relates explicit backgrounds $\kb$ with local and spacetime indices
was introduced and investigated in Refs.\ \cite{bs16,bbw19}.
In what follows,
any explicit background vierbeins and background metrics
are denoted as $\eb$ and $\gb$.
They are nondynamical and by definition cannot arise
from a dynamical vierbein or metric on the manifold,
so they can be treated in the same way 
as other explicit backgrounds $\kb$.
An effective field theory based on GR 
containing an explicit background vierbein and background metric
must therefore have at least two metrics and two vierbeins.

To illustrate some implications of these various results,
consider a background $k$ carrying a single index.
The above discussion reveals that 
six versions of this $k$ can usefully be distinguished,
with the two classes of spontaneous and explicit $k$ 
being further subdivided according to the three possible index types,
\beq
k \in \{ \kv^\mu, \kv_\mu, \kv^a, \kb^\mu, \kb_\mu, \kb^a \}.
\label{koneindex}
\eeq
Consider first the spontaneous case.
Since any spontaneous background $\kv$ 
arises as the solution of dynamical equations of motion,
the three types of spontaneous $\kv$ are related
by the spontaneous background vierbein and background metric
\cite{bk05},
\beq
\kv_\mu = \ev_{\mu a}\kv^a = \gv_{\mu\nu}\kv^\nu .
\label{eq:spon3}
\eeq
All three spontaneous backgrounds $\kv$ thus represent the same physics.
In contrast,
the three types of explicit backgrounds $\kb$ 
correspond to different physics
because they are nondynamical and couple differently 
with the usual vierbein and metric.
For instance,
given an explicit background $\kb^\mu$ with a contravariant index,
we can take advantage of the existence of the usual metric $g_{\mu\nu}$ 
to form the product $g_{\mu\nu}\kb^\nu$,
which might naively seem to represent an explicit background
with a covariant index.
However,
this product involves the dynamical operator $g_{\mu\nu}$
and hence cannot be treated as an explicit background
in the variational procedure.
Attempting instead to take advantage of the existence 
of the usual metric background $\gv_{\mu\nu}$ 
to write the product $\gv_{\mu\nu}\kb^\nu$
also fails to generate a satisfactory explicit background 
with a covariant index
because the product mixes on-shell and off-shell quantities.
The situation for explicit backgrounds is further complicated
in scenarios with an explict background vierbein $\eb_\mu{}^a$
and background metric $\gb_{\mu\nu}$
in addition to the usual background vierbein $\ev_\mu{}^a$
and background metric $\gv_{\mu\nu}$
\cite{bs16,bbw19}.
Formal equations such as  
$\kb_\mu\equiv \eb_{\mu a}\kb^a$
or $\kb_\mu\equiv \gb_{\mu\nu}\kb^\nu$
can then be introduced,
but these must be understood as definitions 
instead of physical relations.
Since generic theories lack 
an explicit background vierbein and background metric,
the three explicit backgrounds $\kb$ typically cannot be related
even by definitions of this type.
All these examples generalize to backgrounds $k$
with more complicated index structures.

\subsection{Symmetry violations}
\label{Symmetry violations}

The presence of a background can violate spacetime symmetries
because backgrounds behave differently from dynamical fields
under particle spacetime transformations.
Both backgrounds and dynamical fields
behave covariantly under observer transformations,
which insures invariance of the physics 
under coordinate changes.
For instance,
physical invariance under general coordinate transformations,
which are observer diffeomorphisms,
is assumed to be a property of a realistic theory.
However,
backgrounds are invariant under particle transformations,
while dynamical fields transform covariantly.
This difference can lead to physical symmetry violations
in observables that involve dynamical fields coupled to a background.

\renewcommand\arraystretch{1.6}
\begin{table*}
\caption{
\label{tab:fulltypes}
Examples of terms in the Lagrange density
with different transformation properties.}
\setlength{\tabcolsep}{8pt}
\begin{tabular}{c|c|c}
\hline
\hline
Local Lorentz transformations	&	Diffeomorphisms	&	Examples	\\
\hline\multirow{3}{*}{invariance (LLI)}	&	invariance (DI)	&	GR	\\
\cline{2-3}	&	spontaneous violation (SDV)	&	$\kv^{ab} \cO_{ab}$, $~\kv^{ab}= \et^{ab}$	\\
\cline{2-3}	&	explicit violation (EDV)	&	$\kb^\mu \cO_\mu$	\\
\hline\multirow{3}{*}{spontaneous violation (SLLV)}	&	invariance (DI)	&	none	\\
\cline{2-3}	&	spontaneous violation (SDV)	&	$\kv^\mu \cO_\mu$	\\
\cline{2-3}	&	explicit violation (EDV)	&	$k^{\mu a} \cO_{\mu a}$, $~k^{\mu a} \equiv \kb^\mu \kv^a$	\\
\hline\multirow{3}{*}{explicit violation (ELLV)}	&	invariance (DI)	&	$\kb^a\cO_a$, $~\kb^a$ constant 	\\
\cline{2-3}	&	spontaneous violation (SDV)	&	$k^{\mu a} \cO_{\mu a}$, $~k^{\mu a} \equiv \kv^\mu \kb^a$	\\
\cline{2-3}	&	explicit violation (EDV)	&	$\kb^a \cO_a$, $~\kb^a$ nonconstant	\\
\hline
\hline
\end{tabular}
\end{table*}

Consider,
for example,
a generic background $k^{a\ldots}$ in a local frame.
This can be viewed as specifying an orientation in the frame,
sometimes called a preferred direction,
which is invariant under local Lorentz transformations.
Unless $k^{a\ldots}$ happens to have no indices
and is independent of position,
or unless it is proportional to combinations
of the Lorentz-group invariants $\et_{ab}$ and $\ep_{abcd}$,
the coupling of a dynamical field to $k^{a\ldots}$ 
can produce changes of physical observables
under local rotations or local Lorentz boosts.
These are violations of local Lorentz invariance,
which can thus be traced to
a direction-dependent background in a local frame
\cite{ak04}.
Note that even a scalar background $k(x)$ without indices
but varying with spacetime position
can introduce violations of local Lorentz invariance
because the derivatives of $k(x)$ 
specify an orientation in a local frame
\cite{klp03}.
Similarly,
a generic background $k^{\mu\ldots}{}_{\nu\ldots}$
on the spacetime manifold
can lead to violations of diffeomorphism invariance
unless it has no indices and is independent of spacetime position.
Only a background serving as a scalar coupling constant,
such as the expectation value of the Higgs field in the SM,
can preserve local Lorentz invariance and diffeomorphisms.

For explicit backgrounds,
the above results hold without further subtleties.
An explicit background $\kb^{\mu\ldots}{}_{\nu\ldots}{}^{a\ldots}(x)$
defined both on the manifold and in local frames
violates local Lorentz and diffeomorphism invariance
in ways determined directly by its index structure
and by its nonvanishing derivatives.
For spontaneous backgrounds,
however,
conditions like Eq.\ \rf{eq:spon3}
relate the different types of indices.
A spontaneous background $\kv^{\mu\ldots}{}_{\nu\ldots}{}^{a\ldots}(x)$
can therefore be viewed equivalently 
as defined entirely on the manifold, 
entirely in local frames,
or as a mixture of the two.
Consequently,
we recover the result obtained in Ref.\ \cite{bk05}:
a generic theory contains spontaneous local Lorentz violation (SLLV)
if and only if it contains spontaneous diffeomorphism violation (SDV),
\beq
{\rm SLLV} \iff {\rm SDV} \quad {\rm (generic~theories)} .
\label{lemma}
\eeq
Two exceptions to this result exist,
one due to an accidental symmetry and the other to convention.
The first exception arises when the spontaneous background
happens to be proportional to combinations of 
the Lorentz-group invariants $\et_{ab}$ and $\ep_{abcd}$,
in which case it has accidental local Lorentz invariance 
but can still violate diffeomorphism invariance.
The other is specific 
to the usual background vierbein $\ev_\mu{}^a =\et_\mu{}^a$ 
and metric $\gv_{\mu\nu}=\et_{\mu\nu}$
in an approximately Minkowski spacetime.
As described in the previous subsection,
these quantities are taken by convention to be invariant 
under both local Lorentz transformations and diffeomorphisms,
with special transformation rules \rf{eq:lindiffeo}
assigned to the fluctuations around them
to compensate for this defined invariance.

The action of the effective field theory
is defined as usual via integration over the spacetime manifold,
\beq
S = \int d^4x ~ e \cL ,
\label{action}
\eeq
and is assumed invariant by construction 
under general coordinate transformations,
which can be understood as observer diffeomorphisms
as described in Sec.\ \ref{Transformations in curved spacetime}.
A generic term in $\cL$ involving a background 
takes the form \rf{genlagterm}
or its generalization incorporating background derivatives.
The properties of the term 
under local Lorentz transformations and diffeomorphisms
are determined by the index structure and spacetime dependence 
of the background $k$. 

Table \ref{tab:fulltypes} shows some examples of terms in $\cL$ 
and their properties under local Lorentz transformations and diffeomorphisms:
invariance,
spontaneous violation, 
or explicit violation.
In principle,
this yields nine possible classes of terms
identified according to the transformation properties 
displayed in the first two columns of the table,
which we denote by the abbreviations 
LLI, SLLV, ELLV, DI, SDV, EDV
as shown in the parentheses.
However,
the generic result \rf{lemma} insures that the SLLV-DI class is empty,
while the LLI-SDV class contains only the exceptions to the result
mentioned above.
Note that terms in the SLLV-EDV and ELLV-SDV classes
must involve mixed backgrounds arising 
partly from spontaneous violation and partly from explicit violation,
so in this sense the corresponding examples are more complicated. 
More generally,
a given model may contain several terms lying in distinct classes 
and can therefore be expected to exhibit multiple features
associated with different types of symmetry violations.

The third column of Table \ref{tab:fulltypes} 
provides examples of individual terms in each of the eight nonempty classes.
Other than the exceptions in the LLI-SDV class,
for which the backgrounds must be formed from the Minkowski metric
or the Levi-Civita tensor,
the chosen examples all involve the comparatively simple backgrounds 
with a single index taken from the set \rf{koneindex}.
The operators $\cO$ are understood 
to have transformation properties determined by their index structure
and to enter the action \rf{action} as dynamical fields on the manifold,
which can include the usual vierbein $\vb \mu a$ and metric $g_{\mu\nu}$
along with matter fields. 
Any local indices on an operator $\cO$ therefore arise from the presence 
of vierbeins and fermion bilinears involving Dirac matrices
rather than from spacetime fields expressed in local coordinates.
For instance,
the choice $\cO_{ab} = R_{ab}$ is excluded
to avoid the spurious appearance of the usual GR combination 
$\et^{ab} R_{ab} = g^{\mu\nu} R_{\mu\nu}= R$
in the LLI-SDV class.
Some classes also contain additional simple examples
beyond those shown in the table.
As an illustration,
the term $\kb_\mu \cO^\mu$ lies in the LLI-EDV class
and is distinct from the term $\kb^\mu \cO_\mu$ listed in the table.
Also,
the three possible SLLV-SDV terms 
$\kv^\mu \cO_\mu$, $\kv_\mu \cO^\mu$, and $\kv^a \cO_a$
are related by virtue of the equivalence \rf{eq:spon3}.

The transformations listed in Table \ref{tab:fulltypes} 
refer to properties of terms in the action 
rather than to physical observables.
The relationship between properties of the action
and experimental measurements can be subtle.
Consider,
for example,
a term of the form $\kb_\mu \cO^\mu$,
which exhibits local Lorentz invariance at the level of the action.
Experiments searching for local Lorentz violation
may nonetheless be sensitive to this term
because the nonzero vierbein $\ev_\mu{}^a$ 
implies that the combination $\kb_\mu \ev^\mu{}_a$ 
provides a definite orientation in the local experimental frame,
which can yield observable local Lorentz violation.
An example is the fermion-sector term
\cite{ak04}
\beq
\cL \supset - {\ol b}_\mu \ivb \mu a \psb \ga_5 \ga^a \ps,
\label{btype}
\eeq 
which causes orientation-dependent splittings in
the fermion energy spectrum
and has been studied in numerous experiments
searching for Lorentz violation in quantum electrodynamics
\cite{tables}.
A kind of converse is also possible:
some terms of the form $\kb_\mu \cO^\mu$
explicitly violate diffeomorphism invariance at the level of the action
but are undetectable in experiments.
This can often be confirmed directly for a given case
by identifying a suitable field or coordinate redefinition 
that removes the term from the action
and thereby demonstrates its physical irrelevance
\cite{ck97,ak04,kt11}.
A well-known example is the term
\beq
\cL \supset - {\ol a}_\mu \ivb \mu a \psb \ga^a \ps,
\label{atype}
\eeq 
for which one component of the background ${\ol a}_\mu(x)$ 
can be removed using the field redefinition
$\ps = \exp[i f(x)]\ps^\prime$,
representing a position-dependent change of phase 
\cite{ak04}.

Another subtlety arises in spontaneous symmetry violation,
where the underlying theory is invariant under
local Lorentz transformations and diffeomorphisms.
In any spontaneous symmetry violation,
the symmetry of the full theory remains unbroken
but  becomes hidden when the Lagrange density of the full theory
is expressed in terms of field fluctuations about the background
\cite{coleman}.
In the context of spontaneous breaking of spacetime symmetries,
the field fluctuations transform in unconventional ways,
which insure that the full theory retains the complete spacetime symmetry 
\cite{bfk08}.
However,
experiments cannot change the background
by performing local Lorentz and diffeomorphism transformations, 
and they treat the fluctuations as conventional tensor fields.
As a result,
experiments can be sensitive to the existence of spontaneous backgrounds
despite the hidden invariance of the underlying theory.
Note, however,
that in approximately Minkowski spacetime
the usual spontaneous background vierbein $\ev_\mu{}^a =\et_\mu{}^a$ 
and spontaneous background metric $\gv_{\mu\nu}=\et_{\mu\nu}$
form an exception to this picture
because the fluctuations \rf{eq:linear}
are conventionally assumed to transform 
so that the full vierbein and metric behave as tensor fields.
In experimental analyses,
the backgrounds $\ev_\mu{}^a =\et_\mu{}^a$ and $\gv_{\mu\nu}=\et_{\mu\nu}$
can therefore be viewed as preserving  
local Lorentz invariance and diffeomorphism invariance.

The experimental situation can be further complicated
by dynamical fields from objects outside the control of the experimentalist
that can mimic the effects of a background 
and hence play the role of one or more nonzero coefficients $k$ 
in the effective field theory.
A simple example is the gravitational field of the Earth,
which acts as a nontrivial background 
and provides a preferred direction in the laboratory.
Some coefficients $k$
then depend on the local gravitational acceleration $\vec g$.
This introduces apparent signals 
for local Lorentz and diffeomorphism violation
even in a scenario with an invariant theory.
The invariance would be manifestly evident 
under transformations of the experimental conditions
only if the Earth could be transformed as well.
Similarly,
a background distribution of particles or a thermal bath
establishes a preferred inertial frame
and hence can also create apparent signals for Lorentz violation.
For instance,
a neutrino beam that travels through the body of the Earth
interacts with the electrons in the Earth's material
\cite{msw}.
This acts as a background 
described by an $a$-type coefficient 
similar to that in Eq.\ \rf{atype},
producing apparent Lorentz violation
in neutrino flavor oscillations
\cite{km04}.
Note that mimic backgrounds can occur at various scales,
including cosmological ones.
For example,
the cosmic microwave background 
fixes a rest frame throughout the Universe.
This leads to apparent violations of Lorentz invariance,
including subtle effects such as the observed dipole temperature anisotropy
due to the velocity of the Earth relative to this frame
\cite{planck14}.
Preferred spacetime directions
can also be expected from other mimic backgrounds at large scales,
including the cosmic neutrino background
and perhaps also dark matter and dark energy.

The effects of known mimic backgrounds
must be removed in any experimental analysis
searching for violations of spacetime symmetries
arising in an underlying theory.
Alternatively,
since mimic backgrounds can play the role of the coefficients $k$,
laboratory searches for violations of spacetime symmetries 
can be reinterpreted as providing constraints on unknown dynamical fields,
even in theories that are invariant under 
local Lorentz transformations and diffeomorphisms.
For instance,
extensions of Riemann geometry to include 
spacetime torsion or nonmetricity tensors
typically generate nontrivial backgrounds in nature,
and matching these to the above framework
permits sensitive experimental constraints on the components of
both torsion 
\cite{torsion}
and nonmetricity
\cite{nonmetricity}
to be achieved by reinterpreting experimental bounds 
obtained in laboratory searches for Lorentz violation.

The frame dependence of the backgrounds 
implies that meaningful comparisons of results 
obtained in different experiments
must be made in a specified frame.
For this purpose,
it is desirable to choose a standard frame
that is approximately inertial over the time scale of typical measurements 
and that is experimentally accessible.
No Earth-based frame is a suitable choice 
due to the rotation of the Earth about its axis
and its revolution around the Sun,
which imply consequent experimental effects 
such as sidereal variations of observables
\cite{ak98}.
Instead,
the canonical frame adopted in the literature is the Sun-centered frame
\cite{sunframe},
which uses a right-handed coordinate system
determined by the Earth's rotational axis
and the direction to the 2000 vernal equinox. 
This frame has been used to report results
of numerous experimental investigations 
performed in the last two decades
\cite{tables}.

\subsection{Linearization}
\label{Linearization}

Experimental and observational tests of spacetime symmetries 
mostly involve weak gravitational fields in approximately flat spacetime,
for which it is appropriate to adopt 
the linearized description \rf{eq:linear} of the vierbein and metric
introduced in Sec.\ \ref{approximatelyflat}.
From the viewpoint of the whole manifold,
these experiments probe local Lorentz and diffeomorphism invariance.
In the linearized description,
however,
this reduces to studying the analogues in approximately flat spacetime
of Minkowski-spacetime Lorentz transformations and translations,
which can mix local Lorentz transformations and diffeomorphisms
as shown in Sec.\ \ref{approximatelyflat}.
The spacetime symmetries of a given theory on the manifold
therefore can correspond nontrivially to spacetime symmetries 
of its linearized limit. 
For example,
a Lorentz transformation in experiments
searching for sidereal or annual variations 
involves changes both of the velocity in the local frame 
and of the spacetime position,
so even backgrounds $k$ having only spacetime indices 
can generate Lorentz violation in experiments.
In this subsection,
we consider some aspects of this correspondence.

In the limit of weak gravitational fields in approximately flat spacetime,
the action \rf{action} is linearized to $S^\L$ according to
\beq
S = \int d^4x ~ e \cL \quad \to \quad
S^\L = \int d^4x ~  \cL^\L ,
\label{linearizedaction}
\eeq
where the linearized Lagrange density $\cL^\L$
incorporates relevant contributions
from the linearization of the vierbein determinant $e$.
The pure-gravity sector of $\cL^\L$
is understood to contain terms up to second order 
in the fluctuations $h$ and $\ch$,
which permits exploration of effects on gravitational waves
and graviton propagation,
except that contributions from the cosmological-constant term
are kept only to first order in $h$.
For the matter-gravity sector, 
$\cL^\L$ is restricted to contain terms at first order in $h$ and $\ch$
but to include other fields at all orders.
These choices are the usual ones adopted for the linearization procedure
in GR coupled to matter.

In the linearized limit,
three kinds of spacetime transformations in approximately flat spacetime
are of interest: 
Lorentz transformations, gauge transformations, and translations.
Their definitions and basic properties are presented 
in Sec.\ \ref{approximatelyflat}.
The three symmetries can be broken spontaneously or explicitly.

\renewcommand\arraystretch{1.6}
\begin{table*}
\caption{
\label{tab:lineartypes}
Examples of terms in the linearized Lagrange density
with different transformation properties.}
\setlength{\tabcolsep}{8pt}
\begin{tabular}{c|c|c|c}
\hline
\hline
Lorentz transformations	&	Gauge transformations	&	Translations	&	Examples	\\
\hline\multirow{4}{*}{invariance (LI)}	&	\multirow{2}{*}{invariance (GI)}	&	invariance (TI)	&	GR	\\
\cline{3-4}	&		&	violation (TV)	&	none	\\
\cline{2-4}	&	\multirow{2}{*}{violation (GV)}	&	invariance (TI)	&	$k^{\mu\nu}\cO^\prime_{\mu\nu}$, $~k^{\mu\nu}= \et^{\mu\nu}$	\\
\cline{3-4}	&		&	violation (TV)	&	none	\\
\hline\multirow{4}{*}{violation (LV)}	&	\multirow{2}{*}{invariance (GI)}	&	invariance (TI)	&	$k^a \cO_a$, $~k^a$ constant	\\
\cline{3-4}	&		&	violation (TV)	&	$k^\mu \cO_\mu$, $k^\mu$ nonconstant	\\
\cline{2-4}	&	\multirow{2}{*}{violation (GV)}	&	invariance (TI)	&	$k^\mu \cO^\prime_\mu$, $~k^\mu$ constant	\\
\cline{3-4}	&		&	violation (TV)	&	$k^a \cO^\prime_a$, $~k^a$ nonconstant	\\
\hline
\hline
\end{tabular}
\end{table*}

Consider first Lorentz transformations in the linearized theory.
These are combinations of special local Lorentz transformations
and diffeomorphisms,
so theories with either local Lorentz violation or diffeomorphism violation 
typically have linearized limits violating Lorentz invariance.
In some special scenarios,
however,
the backgrounds on the manifold reduce in the linearized limit
to combinations of the Lorentz-group tensors
$\et_{\mu\nu}$ and $\ep_{\ka\la\mu\nu}$.
Lorentz invariance is then preserved in the linearized theory 
despite the presence of local Lorentz violation or diffeomorphism violation 
in the original theory.

Gauge transformations of linearized gravitational fields 
in approximately flat spacetime are given by Eq.\ \rf{eq:gaugetrans},
while for nongravitational fields 
they are given by the linearized diffeomorphisms \rf{eq:mattergaugetr}.
In the linearized limit,
the Riemann curvature $R_{\mu\nu\rh\si}$ 
and its contractions $R_{\mu\nu}$ and $R$
are gauge invariant at first order,
while the combination $eR$ is gauge invariant at second order.
These features insure gauge invariance 
of the linearized action for GR coupled to the SM.
Gauge invariance also holds for other theories
with unbroken local Lorentz and diffeomorphism invariance,
provided any dynamical backgrounds $k(x)$ without indices 
are treated like matter scalar fields in the linearization procedure. 
Note that
terms with nondynamical backgrounds $\kb(x)$
break diffeomorphism invariance
and may lead to gauge violations in the linearized limit,
irrespective of their spacetime- or local-index structures.

With the above understandings 
of the linearization procedure and gauge transformations,
calculation shows that when a theory is diffeomorphism invariant (DI)
then its linearized limit is gauge invariant (GI).
This holds irrespective of the properties of the theory
under local Lorentz transformations.
It implies a linearized theory with gauge violation (GV)
comes from a theory with diffeomorphism violation (DV), 
\beq
{\rm DI} \Rightarrow {\rm GI} ,
\quad
{\rm GV} \Rightarrow {\rm DV} .
\label{digi}
\eeq
Note that the converses are false.
For example,
a term in $\cL$ of the form 
$\cL\supset k^{\al\be\ga\de\ka\la\mu\nu}R_{\al\be\ga\de}R_{\ka\la\mu\nu}$ 
violates diffeomorphism invariance on the manifold 
but preserves gauge invariance in the linearized theory $\cL^\L$,
so DV$\not\Rightarrow$GV.

Translations in approximately flat spacetime
are special cases of linearized diffeomorphisms
for which the displacements $\xi^\mu(x)$ of spacetime points 
are independent of spacetime position.
Therefore,
if a theory is DI 
then its linearized limit is also translation invariant (TI),
and hence if a linearized theory has translation violation (TV)
then the full theory is DV, 
\beq
{\rm DI} \Rightarrow {\rm TI} ,
\quad
{\rm TV} \Rightarrow {\rm DV} .
\label{diti}
\eeq
As before,
the converses are false.

Translation violation also implies the existence of
at least one nonconstant background $k$.
Since nonzero derivatives of $k$ at a spacetime point
determine preferred directions at that point, 
translation violation in a linearized theory 
is accompanied by Lorentz violation (LV).
The contrapositive insures that Lorentz invariance (LI)
in the linearized theory implies translation invariance.
We thus have
\beq
{\rm LI} \Rightarrow {\rm TI} ,
\quad
{\rm TV} \Rightarrow {\rm LV} .
\label{tvlv}
\eeq
Again,
the converses are false,
as a constant background can violate Lorentz symmetry
while preserving translation invariance.

Given the three kinds of transformations at the linearized level,
each of which allows the two options of invariance or violation,
one might expect to classify any term in a linearized theory 
as one of eight types.
However,
the relation \rf{tvlv}
implies that two of these eight classes must be empty.
Table \ref{tab:lineartypes} displays the eight possibilities
and provides examples of terms in the linearized Lagrange density $\cL^\L$.
The first three columns list the possible properties
under Lorentz transformations, gauge transformations, and translations,
which we denote by the abbreviations shown in parentheses.
The final column shows representative terms in $\cL^\L$
for each of the five nonempty classes.
The backgrounds $k$ are assumed to be generic 
unless otherwise indicated.
Some entries involve the comparatively simple single-index backgrounds 
listed in the set \rf{koneindex},
and some contain two-index backgrounds determined 
by the Minkowski and Levi-Civita tensors.
In this table, 
gauge-invariant operators are denoted by $\cO$ 
while gauge-violating ones are denoted by $\cO^\prime$.
The operators are taken to have the same basic properties
as those adopted for Table \ref{tab:fulltypes},
except that any gravitational field they contain is linearized.
For example,
the gauge-invariant operators displayed could include 
linearizations of products of the scalar curvature and the Ricci tensor,
which are gauge invariant in $\cL^\L$ 
because each factor is separately gauge invariant
at first order in the fluctuations.
The entries in the final column are only representative,
and other simple examples exist. 
Note also that combinations of terms in a given model
can produce more complicated combinations of effects.

\renewcommand\arraystretch{1.6}
\begin{table*}
\caption{
\label{tab:relation}
Symmetry properties of sample terms and their linearizations.}
\setlength{\tabcolsep}{6pt}
\begin{tabular}{c|cccccc}
\hline
\hline
Theory	&	 \multicolumn{6}{c}{Linearization}											\\
\cline{2-7}	&	LI, GI, TI	&	LV, GV, TI	&	LI, GV, TI	&	LV, GV, TV	&	LV, GI, TV	&	LV, GI, TI	\\
\hline LLI, DI	&	GR	&	none	&	none	&	none	&	none	&	none	\\
\multirow{2}{*}{SLLV, SDV} 	&	\multirow{2}{*}{none}	&	 $\kv^\mu \cO_\mu$,	&	\multirow{2}{*}{none}	&	$\kv^\mu \cO_\mu$,	&	 $\kv^{\mu\nu} R_{\mu\nu}R$,	&	 $\kv^{\mu\nu} R_{\mu\nu}R$,	\\[-5pt]
	&		&	$\kv^\mu$ constant 	&		&	$\kv^\mu$ nonconstant	&	$\kv^{\mu\nu}$ nonconstant	&	$\kv^{\mu\nu}$ constant	\\
\multirow{2}{*}{LLI, SDV}	&	$\kv^{\mu\nu} R_{\mu\nu}R$,	&	\multirow{2}{*}{none}	&	$\kv^{\mu\nu} \cO_{\mu\nu}$,	&	\multirow{2}{*}{none}	&	\multirow{2}{*}{none}	&	\multirow{2}{*}{none}	\\[-5pt]
	&	$\kv^{\mu\nu}=\et^{\mu\nu}$	&		&	$\kv^{\mu\nu}=\et^{\mu\nu}$	&		&		&		\\
\multirow{2}{*}{ELLV, DI}	&	\multirow{2}{*}{none}	&	\multirow{2}{*}{none}	&	\multirow{2}{*}{none}	&	\multirow{2}{*}{none}	&	\multirow{2}{*}{none}	&	$\kb^a \cO_a$,	\\[-5pt]
	&		&		&		&		&		&	$\kb^a$ constant	\\
\multirow{2}{*}{LLI, EDV}	&	$\kb^{\mu\nu} R_{\mu\nu}R$,	&	$\kb^\mu \cO_\mu$,	&	$\kb^{\mu\nu} \cO_{\mu\nu}$,	&	 $\kb^\mu \cO_\mu$,	&	$\kb^{\mu\nu} R_{\mu\nu}R$,	&	$\kb^{\mu\nu} R_{\mu\nu}R$,	\\[-5pt]
	&	$\kb^{\mu\nu}=\et^{\mu\nu}$	&	$\kb^\mu$ constant	&	$\kb^{\mu\nu}=\et^{\mu\nu}$	&	$\kb^\mu$ nonconstant	&	 $\kb^{\mu\nu}$ nonconstant	&	 $\kb^{\mu\nu}$ constant	\\
\multirow{2}{*}{ELLV, EDV}	&	$\kb^{\mu a} \ivb\nu a R_{\mu\nu}R$,	&	$\kb^{\mu a} \cO_{\mu a}$,	&	$\kb^{\mu a} \cO_{\mu a}$,	&	$\kb^a \cO_a$,	&	$\kb^{\mu a} \ivb\nu a R_{\mu\nu} R$,	&	$\kb^{\mu a} \ivb\nu a R_{\mu\nu} R$,	\\[-5pt]
	&	$\kb^{\mu a}=\et^{\mu a}$	&	$\kb^{\mu a}$ constant	&	$\kb^{\mu a}=\et^{\mu a}$	&	$\kb^a$ nonconstant	&	$\kb^{\mu a}$ nonconstant 	&	$\kb^{\mu a}$ constant 	\\
\hline
\hline
\end{tabular}
\end{table*}

\begin{figure}[htp]
\centering
\includegraphics[width=0.48\textwidth]{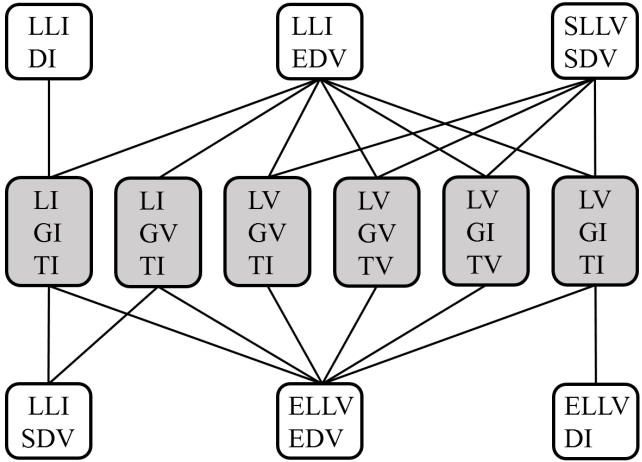}
\caption{Relationships between different types of effective terms
and their linearizations.}
\label{fig:linear}
\end{figure}

The correspondence between the spacetime symmetries
of a given term in the Lagrange density of a theory 
and the spacetime symmetries in the linearized limit
is depicted schematically in Fig.\ \ref{fig:linear}.
Each of the six boxes with a white background represents 
one of the classes of terms in Table \ref{tab:fulltypes},
labeled according to its properties 
under local Lorentz transformations and diffeomorphisms
using the abbreviations shown in the table.
Note that for simplicity 
we omit from the figure the two classes SLLV-EDV and ELLV-SDV
that involve mixed backgrounds.
Each of the six boxes with a gray background represents
one of the six nonempty classes of terms in Table \ref{tab:lineartypes},
labeled by its properties under Lorentz transformations,
gauge transformations, and translations. 
The lines specify the classes of linearized terms
that can arise from a given term in the original theory.
For example,
a term with spacetime symmetries in the SLLV-SDV class
can produce terms at the linearized level
lying in one of the four classes LV-GV-TI, LV-GV-TV, LV-GI-TV, and LV-GI-TI.
The figure applies to single terms in the original theory,
as combinations of terms can be associated with different classes.
It also assumes the term in the original theory
contributes in the linearized limit,
hence excluding certain possible operators 
such as cubic products of curvatures.

Table \ref{tab:relation} provides specific examples 
of the connections displayed in Fig.\ \ref{fig:linear}
between different types of terms and their linearizations.
The six entries in the first column
list the symmetry properties of the six types of terms
represented by the boxes with a white background in the figure. 
The other six columns in the table
are labeled with the symmetry properties 
of the six classes of linearized terms 
denoted by the boxes with a gray background in the figure.
Each of the 36 entries contained in these five columns
matches a particular line in the figure
and thereby provides a specific example of a term 
with the corresponding symmetries,
except for the 16 entries for which no such term exists. 
These 16 cases are excluded by the constraints 
\rf{lemma}, \rf{digi}, \rf{diti}, and \rf{tvlv}.
The backgrounds $k$ are generic unless otherwise indicated
and the operators $\cO$ represent appropriate dyamical fields.
The entries are only representative examples,
and additional possibilities exist.

The existence of certain invariances in a theory 
has implications for the observability 
of the fluctuations $h_{\mu\nu}$ and $\ch_{\mu a}$
introduced in Eq.\ \rf{eq:linear}.
Consider first a theory with local Lorentz invariance
described by a Lagrange density $\cL$.
Under an infinitesimal local Lorentz transformation 
with $\La^a{}_b(x)=\et^a{}_b+\ep^a{}_b(x)$,
the fluctuation $h_{\mu\nu}$ is invariant
while $\ch_{\mu a}$ changes at leading order,
\bea
h_{\mu\nu}(x) &\to& h_{\mu\nu}(x),
\nn\\ 
\ch_{\mu a}(x) &\to& \ch_{\mu a}(x)-\ep_{\mu a}(x).
\eea
Since $\ch_{\mu a}$ and $\ep_{\mu a}$ are both antisymmetric,
it follows that the $\ch_{\mu a}$ modes
can be transformed to zero 
using a suitable local Lorentz transformation.
Other dynamical fields change to new expressions
under this transformation.
For example,
the spinor field $\ps(x)$ changes according to 
\beq
\ps (x) \to \ps\pr(x)=\exp[\quar i\ep_{ab}(x)\si^{ab}]\ps(x).
\label{spinex}
\eeq
Writing the theory in terms of transformed fields
amounts to implementing field redefinitions
and so produces a Lagrange density equivalent to $\cL$
but lacking the $\ch_{\mu a}$ modes.
This confirms that the $\ch_{\mu a}$ modes
are unobservable in a theory with local Lorentz invariance
\cite{kt11}.
In particular,
it implies that the $\ch_{\mu a}$ modes play no physical role
in theories of the LLI-DI, LLI-SDV, and LLI-EDV types
shown in Fig.\ \ref{fig:linear}
and in the first, third, and fifth rows of Table \ref{tab:relation}.
Note,
however,
that the $\ch_{\mu a}$ modes can be physical 
in theories with Lorentz invariance in approximately flat spacetime,
as these can arise as linearized limits 
of theories with local Lorentz violation. 

Similarly,
some components of the metric and vierbein
are unobservable in theories with gauge invariance
in approximately flat spacetime.
Under an infinitesimal gauge transformation 
with $x^\mu\to x^\mu+\xi^\mu(x)$,
the fluctuations $h_{\mu\nu}$ and $\ch_{\mu a}$ change according to
Eq.\ \rf{eq:gaugetrans}.
It follows that four degrees of freedom in the metric and vierbein
are associated with gauge transformations
and hence are unobservable in theories with gauge invariance.
As in the case of the $\ch_{\mu a}$ modes discussed above,
field redefinitions can remove these $\xi_\mu$ modes
by generating a physically equivalent Lagrange density 
in which they are absent.
This establishes that the $\xi_\mu$ modes are unphysical
in linearized theories of the LI-GI-TI, LV-GI-TV, and LV-GI-TI classes
shown in Fig.\ \ref{fig:linear}
and in the corresponding columns of Table \ref{tab:relation}.

\subsection{No-go constraints}
\label{No-go constraints}

In any model based on Riemann geometry or its extensions
to include torsion and nonmetricity,
the fields must satisfy the Bianchi identities,
which are intrinsically imposed by the geometric structure
\cite{texts,hhk76}.
The Bianchi identities hold both on and off shell,
and their compatibility with the variational principle
imposes constraints that must be satisfied for consistency of the model.
In GR,
for example,
the Bianchi identity implies the on-shell conservation 
of the energy-momentum tensor, 
$D_\mu T^{\mu\nu}=0$,
which is compatible with the dynamics and symmetries of the theory
obtained by variation of the action.
Similarly,
in a model with spontaneous violation of one or more spacetime symmetries,
compatibility with the Bianchi identities is maintained
because the variational procedure is standard.
However, 
explicit violation of a spacetime symmetry
requires the presence in the action of 
one or more nondynamical background fields 
$\kb^{\mu\ldots}{}_{\nu\ldots}{}^{a\ldots}(x)$,
which behave unconventionally under variations.
The variational results can then become incompatible 
with implications from the Bianchi identities
and hence can render problematic a model containing explicit violation
\cite{ak04}.
This can,
for instance,
induce outright inconsistencies in the model
or impose unnatural requirements 
such as fine tuning of the explicit background.
The potential constraints on a model with explicit violation 
of spacetime symmetries are called no-go constraints.
Their role has been the subject of extensive recent investigation
by Bluhm and collaborators
\cite{rb15,15rb2,bs16,bbw19}.

As described in Sec.\ \ref{Backgrounds},
a spontaneous background $\kv$ 
can be viewed as a nondynamical background $\kb$ that is on shell
and that comes with dynamical fluctuations
including Nambu-Goldstone and massive modes.
In this context,
incompatibilities in a model with explicit breaking 
can be interpreted as due to the absence of dynamical fluctuations
\cite{ak04}.
Within the St\"uckelberg approach
\cite{stueckelberg,ags03},
the missing Nambu-Goldstone modes 
correspond to extra dynamical scalar fields
that can be added to restore the explicitly broken symmetry
\cite{bbw19}.
For explicit diffeomorphism violation,
the no-go constraints can be identified
with the Noether identities
\cite{noether18} 
arising from the requirement of general coordinate invariance of the model 
\cite{rb15}.
Under suitable circumstances,
the constraints can be satisfied
by appropriately fixing the $\xi_\mu$ modes. 
Similar results hold for local Lorentz violation
and the corresponding $\ch_{\mu a}$ modes 
\cite{bs16},
as well as in the presence of matter-gravity couplings 
\cite{bbw19}.
However,
in some scenarios the no-go constraints cannot be satisfied
for any choice of the $\xi_\mu$ or $\ch_{\mu a}$ modes
\cite{rb15}.

The present subsection contributes to the ongoing discussion 
of this topic by demonstrating that 
an explicit background $\kb^{\mu\ldots}{}_{\nu\ldots}{}^{a\ldots}(x)$
typically cannot satisfy the no-go constraints
in a model that modifies the physics only perturbatively.
This result holds in the matter-gravity sector
as well as in the pure-gravity one,
and it thus extends the practical impact
of the no-go constraints on the space of possible realistic models
that seek to describe small deviations from known physics.
Some comments about models based on non-Riemann geometry
are also presented.

Consider first a model with an explicit background $\bk_{\mu_1\cdots\mu_n}$
carrying $n$ covariant spacetime indices.
The associated current can be defined as usual by variation of the action,
$J^{\mu_1\cdots\mu_n}\equiv \de S/\de \bk_{\mu_1\cdots\mu_n}$.
Following the calculational procedure in Ref.\ \cite{ak04}
reveals that the Bianchi identity implies
\bea
D_\mu T^\mu_{\pt{\mu}\nu}&=&
J^{\mu_1\cdots\mu_n}D_\nu \bk_{\mu_1\cdots\mu_n}
-D_{\mu_1}(J^{\mu_1\cdots\mu_n}\bk_{\nu\mu_2\cdots\mu_n})
\nn\\
&&
\hskip20pt
-\ldots-D_{\mu_n}(J^{\mu_1\cdots\mu_n}\bk_{\mu_1\cdots\mu_{n-1}\nu})
\nn\\
&=&
0,
\label{covconst}
\eea
where the covariant derivatives are combinations 
of partial derivatives and Levi-Civita connections.
This result matches Eq.\ (27) of Ref.\ \cite{bs16},
and it reduces to Eq.\ (A.17) of Ref.\ \cite{bbw19}
in the limit of a single spacetime index.
It represents four no-go constraints 
that must be obeyed by the model for internal consistency.

In practical applications 
relevant for laboratory and solar-system experiments,
gravity is weak and the spacetime is approximately flat.
The metric can then be expanded as in Eq.\ \rf{eq:linear},
with the dynamics determined by the metric fluctuation $h_{\mu\nu}$.
Linearizing Eq.\ \rf{covconst} in $h_{\mu\nu}$ 
produces an equation of the schematic form
\beq
J(\prt k + k\prt h) + (\prt J + J \prt h)k = 0,
\eeq
where we suppress all indices and factors.
The no-go constraints thus correspond to conditions of the schematic form
\beq
\frac{\prt \bk}{\bk} + \frac{\prt J}{J} + \prt h = 0. 
\label{partialh}
\eeq
Note that the metric fluctuation $h$ appearing in this expression
includes both the conventional modes
appearing in GR and the $\xi_\mu$ modes.
The latter are physical in some scenarios 
but are unobservable in GR and models with gauge invariance,
as discussed at the end of the previous subsection.

We can consider the implications of the no-go constraints \rf{covconst} 
and the schematic condition \rf{partialh}
for different models with an explicit background $\bk_{\mu_1\cdots\mu_n}$.
Suppose first that the background $\bk_{\mu_1\cdots\mu_n}$ appears 
only in the pure-gravity sector of a model,
so that the current $J^{\mu_1\cdots\mu_n}$
is composed of $h_{\mu\nu}$.
The no-go constraints can then be impossible to satisfy.
A simple example is a model 
with a fixed but nonconstant cosmological term $\ov{\La}(x)$.
This is incompatible with the condition \rf{covconst}
required by the Bianchi identity,
which demands $\prt_\mu\ov{\La}=0$
\cite{rb15}.
Another example is the LLI-EDV model with action
\beq
S \propto \int d^4x\ e(R+\kb^{\mu\nu}g_{\mu\nu}),
\label{lliedvmodel}
\eeq
containing a two-index symmetric prescribed background $\kb^{\mu\nu}$.
At zeroth order in $h_{\mu\nu}$,
the condition \rf{covconst} reduces to the constraint
\beq
\prt_\mu \kb^{\mu\nu} 
+ \half \et_{\al\be}\et^{\mu\nu}\prt_\nu \kb^{\al\be} = 0,
\label{specialk}
\eeq
showing that only special choices of backgrounds $\kb^{\mu\nu}$ 
can be admissible.
Generic backgrounds $\kb^{\mu\nu}$ in this model 
are therefore perturbatively incompatible with the no-go constraints
independently of the behavior of $h_{\mu\nu}$ and its $\xi_\mu$ modes,
and so arbitrary explicit diffeomorphism violation is excluded.
Moreover,
at first order in $h$ the schematic condition \rf{partialh} 
in this model reduces to $\prt\kb/\kb \sim \prt h/h$.
Since $\prt h/h$ is tiny near the Earth
and since $\kb$ is perturbative by construction,
it follows that $\kb$ must be almost constant.
Thus,
even a restricted background $\kb^{\mu\nu}$ satisfying \rf{specialk}
must have a fine-tuned structure to satisfy the no-go constraints.

In the above models, 
the no-go constraints generate direct restrictions on explicit backgrounds
without involvement of $h_{\mu\nu}$ or $\xi_\mu$ at leading order
because the background terms in the action are linear in $h_{\mu\nu}$.
However,
backgrounds in more involved models typically also conflict
with the no-go constraints and the perturbative assumption.
Consider,
for example,
a model with action
\beq
S \propto \int d^4x\ e(R+ \kb(x) h_{\mu\nu}h^{\mu\nu}),
\label{mgmod}
\eeq
which can be viewed as incorporating 
a constant two-index background $\ov{\et}_{\mu\nu}$ 
such that $h_{\mu\nu}=g_{\mu\nu}-\ov{\et}_{\mu\nu}$
and $h^{\mu\nu}\equiv \ov{\et}^{\mu\al}\ov{\et}^{\nu\be}h_{\al\be}$,
along with a background function $\kb(x)$.
At leading order in small quantities,
the no-go constraints \rf{covconst} for this model
reduce to the schematic form $\prt \kb/\kb \sim \prt h/h$,
in accordance with the result \rf{partialh}.
As before,
this shows that the structure of $\kb(x)$ 
must be fine tuned to be nearly constant in the vicinity of the Earth
for the perturbative assumption to be valid.
In the particular special case that
$\kb(x)=m^2/2$ is a positive constant,
the action \rf{mgmod} describes a simple model for massive gravity.
At leading order,
the no-go constraints then collapse to $\prt_\mu h^{\mu\nu}=0$,
which is analogous to enforcing a particular gauge-fixing condition
and imposes a corresponding form for the $\xi_\mu$ modes.
In this limit,
the model \rf{mgmod} is therefore compatible with the no-go constraints
at least to first order in $h_{\mu\nu}$.
However,
even constant backgrounds may be insufficient to insure
compatibility with the no-go constraints in many models. 
For example,
no useful post-Newton expansion exists in pure-gravity models 
with constant-background $d=4$ terms of the form
$\ov s{}^{\mu\nu} R_{\mu\nu}$
or $\ov t{}^{\ka\la\mu\nu} R_{\ka\la\mu\nu}$
because the $\xi_\mu$ modes decouple at leading order
and hence cannot be used to satisfy the no-go constraints,
leading to severe restrictions on acceptable linearized curvatures 
\cite{rb15}.
As before,
the schematic constraint \rf{partialh} for these models
shows that the only potential backgrounds that are admissible 
must be almost constant,
so generic explicit backgrounds are excluded. 

Next,
suppose instead that the explicit background $\bk$ appears 
in the matter-gravity sector,
so that the current $J$ includes matter fields.
If we insist that $\bk$ modifies the physics only perturbatively,
as required in an effective field theory,
then we can show that the model is typically incompatible
with the no-go constraints.
Consider,
for example,
a matter sector involving a spinor field $\ps$.
The fermion bilinears can involve Dirac matrices
and associated couplings to the vierbein $\vb \mu a$,
which in approximately flat spacetime
can be expanded according to Eq.\ \rf{eq:linear}.
Linearizing the current $J$ in the metric and vierbein fluctuations
produces an expression of the schematic form
\beq
J\sim \psb\ps
+\mathcal{O}(h)\psb\ps +\mathcal{O}(\ch)\psb\ps
+\ldots,
\label{psiJ}
\eeq
where for simplicity we suppress all indices, factors, and 
structures involving Dirac matrices.
At leading order,
we thus find $\prt J/J \sim \psb\prt\ps/\psb\ps$.
However,
the size of $\psb\prt\ps/\psb\ps$ in laboratory experiments 
is much larger than the unperturbed GR contributions from $\prt h$,
which are dominated by the local gravitational acceleration.
For instance,
neutrons bound in the Earth's gravitational field
have values of $\psb\prt\ps/\psb\ps$ of order $10^{-12}$ GeV,
whereas the gravitational acceleration 
on the Earth's surface is of order $10^{-32}$ GeV.
The condition \rf{partialh} from the no-go constraints
thus typically cannot be satisfied
unless 
any physical effects from the $\xi_\mu$ modes or from ${\prt \bk}/{\bk}$ 
are also much larger than the gravitational acceleration,
both of which are excluded for perturbative modifications to GR. 

This line of reasoning can be applied to most models
with explicit backgrounds in the matter-gravity sector.
The argument can be evaded for certain cases 
in which 
the background $\kb$ is constant or nearly so,
$\prt \kb \sim 0$,
and the current $J$ happens to have a special form
that is conserved or almost conserved, 
$\prt J \sim 0$,
since then both $\bk$ and the $\xi_\mu$ modes 
can yield perturbative contributions 
to the usual gravitational acceleration
while still satisfying the condition \rf{partialh} 
from the no-go constraints.
For example,
one-loop radiative corrections in certain models of massive gravity 
yield a $\ov{c}$-type coefficient in the matter-gravity sector
\cite{rhr15,bbw19},
which amounts to a background $\kb$ with $\prt \kb \sim 0$ 
and a current $J$ with $\prt J \sim 0$.
As another example,
consider a term of the form 
$\cL \supset -\ab_\mu e^\mu{}_a \psb \ga^a \ps$
involving an explicit background $\ab_\mu$,
which can produce nontrivial physical effects 
when $\ab_\mu$ differs from a gradient, 
$\ab_\mu \neq \prt_\mu \ov{f}$
\cite{ak04}.
This term produces a current $J$ that is conserved, 
$\prt J =0$,
by virtue of a global U(1) symmetry.
The condition \rf{partialh} then reduces to $\prt \ab /\ab \sim \prt h$,
which is compatible with perturbative behavior.
In this instance,
further insight can be gleaned 
from the analogous term for spontaneous symmetry breaking,
$\cL \supset -a_\mu e^\mu{}_a \psb \ga^a \ps$,
where $a_\mu$ now contains a background $\av_\mu$ 
along with dynamical fluctuations
insuring compatibility between the Bianchi identity
and the variational procedure.
This situation is comparatively simple because $\prt J = 0$
and so the condition \rf{partialh} implies $\prt a \sim a \prt h$,
which is congruent with the solution (87) for the fluctuation modes
given in Ref.\ \cite{kt11}.
For a more complicated matter-gravity term with a spontaneous background $k$
that lacks a conservation law for $J$,
satisfying \rf{partialh} requires solving $\prt k \sim k \prt J/J$.
This typically is challenging to perform in detail 
and remains an open problem of definite interest.
In contrast,
for most explicit backgrounds, 
nonperturbative solutions that are compatible with the no-go constraints
may be possible in principle
but typically are incompatible with existing experiments.

Consider next a model with an explicit background $\bk{}^{\mu_1\cdots\mu_n}$ 
carrying $n$ contravariant spacetime indices.
This faces challenges similar to those 
for the background $\bk_{\mu_1\cdots\mu_n}$.
For this case,
the no-go constraints take the form 
\bea
D_\mu T^\mu_{\pt{\mu}\nu}&=&
J_{\mu_1\cdots\mu_n}D_\nu \bk{}^{\mu_1\cdots\mu_n}
+D_{\mu_1}(\bk{}^{\mu_1\cdots\mu_n}J_{\nu\mu_2\cdots\mu_n})
\nn\\ 
&&
\hskip20pt
+\ldots
+D_{\mu_n}(\bk{}^{\mu_1\cdots\mu_n}J_{\mu_1\cdots\mu_{n-1}\nu})
\nn\\
&=&0.
\eea
The solution to this equation
has the same schematic form as Eq.\ \rf{partialh}
and so exhibits the same problems,
with only fine-tuned backgrounds offering the potential for consistency.
While nonperturbative solutions of the no-go constraints may exist
for arbitrary backgrounds,
they typically are excluded by existing experiments.

Finally,
we discuss a model containing a background
$\bk_a{}^{b_1\cdots b_n}$ with $n+1$ local indices.
The corresponding no-go constraints now involve 
the antisymmetric part of the energy-momentum tensor.
They take the form
\bea
&&T^{\mu\nu}-T^{\nu\mu}
\nn\\
&&
\hskip10pt
=-e^{\mu a}e^{\nu b}[(J_{ac_1\cdots c_n}\bk_b{}^{c_1\cdots c_n}
+J_{c_1ac_2\cdots c_n}\bk{}^{c_1}{}_b{}^{c_2\cdots c_n}
\nn\\
&&
\hskip60pt
+\ldots+J_{c_1\cdots c_na}\bk{}^{c_1\cdots c_n}_{\pt{c_1\cdots c_n}b})
-(a\leftrightarrow b)]
\nn\\
&&
\hskip10pt
=0,
\label{antiT}
\eea
where
$J_{a b_1\cdots b_n}\equiv \de S/\de \bk{}^{a b_1\cdots b_n}$
is the relevant current.
These constraints amount to the requirement of the vanishing
of a generalized cross product between the current and the background.
As before,
the obstacle for this case can be understood in the perturbative limit,
where the modes of relevance are $h_{\mu\nu}$ and $\ch_{\mu a}$.

Consider first a background $\bk_a{}^{b_1\cdots b_n}$ 
appearing in the matter-gravity sector,
with the current $J$ expanded as in Eq.\ \rf{psiJ}.
Since $J$ cannot be significantly modified 
by perturbative $h_{\mu\nu}$ or $\ch_{\mu a}$,
the no-go constraints \rf{antiT} can be seen to require the vanishing 
of a linear combination of the currents determined by the background.
This represents a strong restriction 
on the structure of the background $\bk_a{}^{b_1\cdots b_n}$,
with generic backgrounds being inadmissible.

Next, 
suppose $\bk_a{}^{b_1\cdots b_n}$ appears instead in the pure-gravity sector.
For most terms in the corresponding Lagrange density,
including ones containing factors of the curvature,
the contribution to the current at leading order in small quantities
cannot contain the $\ch_{\mu a}$ modes. 
This is because the local indices on the background
must ultimately be contracted with factors of the vierbein,
which contains the $\ch_{\mu a}$ modes only at subleading order
and which has vanishing covariant derivative.
The no-go constraints \rf{antiT} then reduce to the requirement 
that a certain linear combination of derivatives of $h_{\mu\nu}$ must vanish,
which is problematic in the Newton limit.
Typical backgrounds obeying the no-go constraints
are thus incompatible with known perturbative physics.

A comparatively simple example illustrating a few of these features 
is given by the term 
$\cL = \kb^a{}_\mu e^\mu{}_a$,
for which the constraints \rf{antiT} reduce to
$\kb^a{}_\mu e^{\mu b} - \kb^b{}_\mu e^{\mu a} = 0$
and thus exclude an arbitrary explicit background $\kb^a{}_\mu$.
A suitable nonperturbative $\kb^a{}_\mu$ 
can nonetheless yield a condition determining $\ch_{\mu a}$
that provides consistency with the no-go constraints.
For example,
demanding that $\kb^a{}_\mu$ takes the form of a background vierbein 
yields an equation for $\ch_{\mu a}$ analogous to a gauge condition
\cite{bs16}.
Overall,
the above discussions confirm 
that the no-go constraints strongly restrict models 
with a generic background $\bk_a{}^{b_1\cdots b_n}$.

Note that the extension of Riemann geometry
to include torsion and nonmetricity
typically cannot satisfy the no-go constraints either.
If the torsion and nonmetricity are dynamical,
then they are fully determined by the equations of motion,
and so no extra degrees of freedom are available
to insure compatibility 
of the geometry with the variational principle.
The situation with dynamical torsion is discussed in Ref.\ \cite{ak04}.
If instead the torsion and nonmetricity are nondynamical,
then they are predetermined 
and hence can be interpreted as fixed background fields.
They therefore are also subject to no-go constraints,
and the same arguments apply.
Consistent solutions to the equations of motion typically are nonperturbative 
and hence incompatible with existing experiments.

The no-go constraints thus imply that a generic perturbative model 
with explicit violation of spacetime symmetries 
cannot be based on Riemann geometry
or its extensions to include torsion and nonmetricity.
Conceivably,
such a model might be formulated instead 
within the context of some other geometry,
or it might even be a nongeometric theory lacking a smooth manifold.
Investigations of these possibilities are of definite interest 
but lie beyond our present scope.
However,
it must be possible to approximate 
the infrared limit of any complete and consistent realistic model 
using an effective field theory based on GR and the SM.
The framework studied in this work
or its extension to include torsion and nonmetricity 
includes all possible backgrounds in the Lagrange density,
so it suffices as a low-energy approximation 
of any complete and realistic model.
Although this approximation may well violate the no-go constraints
from Riemann geometry,
the complete model must satisfy any corresponding constraints
arising from the underlying geometry. 

One option for a geometry compatible with explicit breaking 
is Finsler geometry,
which can be viewed as a generalization of Riemann geometry
with the role of the metric in determining geometric features
supplemented by other quantities prescribed on the manifold
\cite{rf,bcs00}.
With the latter quantities identified as explicit background fields,
Finsler geometry has been conjectured as a possible route 
to escape the no-go constraints in Riemann geometry
\cite{ak04}.
Investigating this conjecture in detail is hampered 
by the lack of a satisfactory definition for Lorentz-Finsler geometry,
which is currently the subject of active research
\cite{ak11,ek18,bjs20,ps18,em17,tbgm18,ilp18,cs18,rt18,xl18,sma19,sv20,rl20,fhpv20}. 
Support for the conjecture includes the demonstration that 
the trajectory of a fermion or scalar particle
in the present of explicit backgrounds
corresponds to a geodesic in a Riemann-Finsler space
\cite{ak11,ek18,ms19,dc17,fl15,nr15}.

\begin{figure}[htp]
\centering
\includegraphics[width=0.45\textwidth]{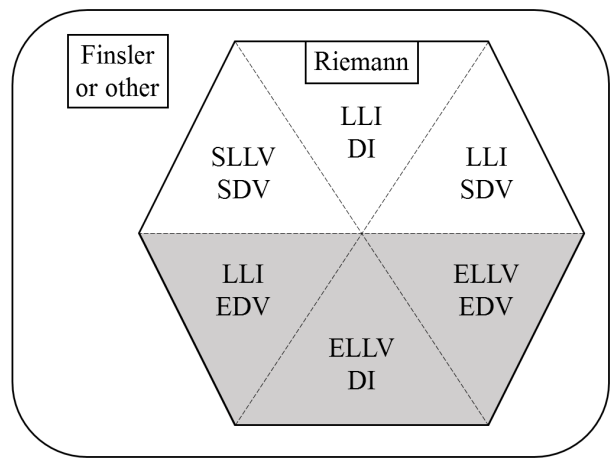}
\caption{Pictorial classification of background terms.}
\label{fig:geometry}
\end{figure}

Figure \ref{fig:geometry} provides a pictorial representation
of the different options for theories with backgrounds,
omitting as before the two classes SLLV-EDV and ELLV-SDV
that involve mixed backgrounds.
Terms within the hexagon are built on Riemann geometry.
Each of the six triangles in the hexagon
corresponds to one of the six classes of theories
contained in the rows of Table \ref{tab:relation}.
Theories in the LLI-DI, SLLV-SDV, and LLI-SDV classes
satisfy the no-go constraints and are compatible with Riemann geometry.
Theories in the LLI-EDV, ELLV-DI, and ELLV-EDV classes 
listed in the last three rows of Table \ref{tab:relation}
all involve explicit violation and are depicted by shaded triangles.
When they incorporate only perturbative deviations from GR coupled to the SM,
these models typically are inconsistent 
or incompatible with experiments.
Theories with explicit violation
that represent perturbative deviations from GR coupled to the SM
generically lie outside the hexagon,
so they must be constructed from some other geometry such as Finsler geometry
or have a nongeometrical basis.
Attempting to express them in terms of effective field theory
based on Riemann geometry is an approximation,
and it typically implies incompatibilities 
with the no-go constraints from Riemann geometry.
 
\section{Effective field theory}
\label{Effective gravity}

In this section,
we develop a methodology for the construction
of a realistic effective field theory involving gravity and matter 
in the presence of arbitrary backgrounds.
This enables the explicit derivation of all desired terms in the action,
including ones in the pure-gravity sector
and those involving matter-gravity couplings 
to gauge fields, fermions, and scalars.
It also yields the terms describing the dynamics of the background.

The methodology is initiated in Sec.\ \ref{Dynamical operators},
which describes the procedure for building dynamical operators 
with appropriate spacetime and gauge properties.
It is convenient to separate the action \rf{action} into four sectors,
\beq
S=\int d^4 x ~e(\cL_g + \cL_A + \cL_\ps + \cL_\ph),
\eeq
where $\cL_g$ contains pure-gravity terms
and any background dynamics,
$\cL_A$ describes gauge fields and their gravity couplings,
$\cL_\ps$ involves fermions including their gravity and gauge couplings, 
and $\cL_\ph$ contains all terms with scalars.
The pure-gravity sector is addressed in Sec.\ \ref{Pure-gravity sector},
along with the dynamics of the background.
We consider the matter-gravity sector in Sec.\ \ref{Matter sector},
starting with the pure-gauge sector
and then adding fermion and scalar terms.
For definiteness,
we work in Sec.\ \ref{Matter sector} 
with a Dirac fermion or complex scalar 
in a single representation of a gauge group.
Applications involving more general types of fermions and scalars
in realistic scenarios,
including the construction of 
the effective field theory based on GR coupled to the SM,
are provided in Sec.\ \ref{Applications}.

Any single effective term in the Lagrange density $\cL$
takes the form of a contraction between a dynamical operator $\cO$
and a background $k$ or its derivatives,
as illustrated in Eq.\ \rf{genlagterm}.
A specific operator $\cO$ may be contracted
directly to one or more backgrounds $k$ or their derivatives,
or may be contracted instead 
via combinations of the vierbein, metric, and Levi-Civita tensor.
It is convenient to adopt a compact notation
for these various types of backgrounds and contractions,
thereby simplifying expressions in the Lagrange density.
The idea is to introduce a quantity 
$\uk^{\mu\cdots}{}_{\nu\cdots}{}^{a\cdots}$ 
that is a linear combination of all terms
formed from background fields, vierbeins, metrics, and the Levi-Civita tensor.
Note that multiple vierbein and metric factors may appear in a given term,
but at most one Levi-Civita factor is needed
because products of the Levi-Civita tensor
reduce to products of vierbeins or metrics.
Contracting the combination $\uk^{\mu\cdots}{}_{\nu\cdots}{}^{a\cdots}$,
with any specific operator $\cO_{\mu\ldots}{}^{\nu\ldots}{}_{a\ldots}$
then produces a single expression in the Lagrange density $\cL$ of the form
$\cL \supset$
$\uk^{\mu\ldots}{}_{\nu\ldots}{}^{a\ldots}(x)$
$\cO_{\mu\ldots}{}^{\nu\ldots}{}_{a\ldots}(x)$, 
which expands into many individual terms of the form \rf{genlagterm}.
Terms involving contractions between dynamical operators 
and derivatives of backgrounds 
can also be combined in this way by using derivatives of  
$\uk^{\mu\cdots}{}_{\nu\cdots}{}^{a\cdots}$.

As an example,
consider the simplest case where $\uk$ has no indices,
so that the corresponding term $\cL\supset \uk\cO$
in the Lagrange density 
involves a dynamical operator $\cO$ without indices.
We can expand the quantity $\uk$ to display the component backgrounds, 
\bea
\uk &=&
k + k^{\mu\nu}g_{\mu\nu} + k_{\mu\nu}g^{\mu\nu} 
+ k_\mu{}^a e^\mu{}_a + k^{\mu a}e_{\mu a} + \ldots
\nn\\
&&
+ k^{\ka\la\mu\nu}g_{\ka\la}g_{\mu\nu} + \ldots
+ k^{\ka\la\mu\nu}\ep_{\ka\la\mu\nu} + \ldots,
\label{expansion}
\eea
which explicitly reveals the dependence on the gravitational degrees of freedom
and illustrates the compactness of the expression $\uk\cO$.
This expansion is also appropriate for terms 
such as $\cL\supset (D_\mu \uk) \cO^\mu$,
which involve the contraction of a dynamical operator 
with the derivative of a combination of backgrounds.
As another example,
consider the case where $\uk^\mu$ has a single contravariant spacetime index.
A term $\cL\supset \uk^\mu \cO_\mu$ in the Lagrange density
can then be expanded using
\beq
\uk^\mu = k^\mu + k_\nu g^{\mu\nu} + k^a e^\mu{}_a + \ldots .
\eeq
Similarly,
the term $\cL\supset \uk^a \cO_a$ 
involving a dynamical operator with a local index
can be expanded using
\beq
\uk^{a} = k^a + k^\mu e_\mu{}^a + k_\mu e^{\mu a} + \ldots . 
\eeq

Depending on the hypotheses of a specific theory,
the various backgrounds 
$k^{\mu\ldots}{}_{\nu\ldots}{}^{a\ldots}(x)$ 
combined in the above expansions 
may be partially or wholly related to each other.
For instance, 
a given theory containing only a single background $k^\mu$
might nonetheless have a Lagrange density with terms 
involving a two-index background $k^{\mu\nu}$ formed as 
$k^{\mu\nu} \propto k^\mu k^\nu$.
A given quantity $\uk^{\mu\cdots}{}_{\nu\cdots}{}^{a\cdots}$ 
may therefore be nonlinear in the backgrounds 
$k^{\mu\cdots}{}_{\nu\cdots}{}^{a\cdots}$
that specify a particular theory.
Since the mass dimensionality of each
$k^{\mu\cdots}{}_{\nu\cdots}{}^{a\cdots}$
is determined by the operator structure of the Lagrange density,
any nonlinear relationships may also involve dimensionful factors
that insure a definite mass dimensionality 
for $\uk^{\mu\cdots}{}_{\nu\cdots}{}^{a\cdots}$.

Expressed using the above notation,
the Lagrange density for the effective field theory 
can be used for spontaneous or explicit violation 
of spacetime symmetries,
with the various combinations of backgrounds
$k^{\mu\ldots}{}_{\nu\ldots}{}^{a\ldots}$
contained in the quantities $\uk^{\mu\cdots}{}_{\nu\cdots}{}^{a\cdots}$
correspondingly understood to be spontaneous or explicit. 
For spontaneous breaking,
the backgrounds $\kv^{\mu\ldots}{}_{\nu\ldots}{}^{a\ldots}$
are understood 
to come with concomitant dynamical fluctuations,
as described in Sec.\ \ref{Backgrounds}.
For explicit breaking,
the backgrounds $\kb^{\mu\ldots}{}_{\nu\ldots}{}^{a\ldots}$
are nondynamical,
and typically the underlying theory cannot be based on Riemann geometry
for reasons outlined in Sec.\ \ref{No-go constraints}.

\subsection{Dynamical operators}
\label{Dynamical operators}

To construct terms in the Lagrange density $\cL$,
we require a procedure to build suitable dynamical operators 
$\cO_{\mu\ldots}{}^{\nu\ldots}{}_{a\ldots}$.
For effective field theory based on GR and gauge theory, 
the terms must be independent 
of observer general coordinate transformations 
and be locally gauge invariant.
In Minkowski spacetime,
a procedure to construct gauge-covariant operators has recently been developed,
from which all gauge-invariant terms in the action can be built 
\cite{kl19}.
In this subsection,
we expand this procedure to construct 
gauge-covariant spacetime-tensor operators,
from which all terms with observer independence 
and gauge invariance can be found. 

For the gauge symmetry,
consider first the scenario in Minkowski spacetime
with a Dirac fermion $\ps$ in a representation $U$ of the gauge group
\cite{kl19}.
Then,
$\ps \to U \ps$ under a gauge transformation,
while the Dirac conjugate transforms as $\ol\ps \to \ol\ps U^\dagger$.
The gauge-covariant derivative acting on $\ps$ can be written as
$D_\mu\ps = \prt_\mu \ps - i g A_\mu \ps$,
where $g$ is the gauge coupling and $A_\mu$ is the gauge field
in the $U$ representation,
and it transforms as $D_\mu \to U D_\mu U^\dagger$.
The gauge field strength $F_{\mu\nu}$ in the $U$ representation
is generated by the commutator $[D_\mu, D_\nu] = -ig F_{\mu\nu}$.
By definition,
an operator $\cO$ formed from gauge fields 
is called gauge covariant if $\cO \to U\cO U^\dagger$.
Given gauge-covariant operators $\cO$ and $\cO^\prime$,
two kinds of gauge-invariant operators
can be constructed,
$\tr(\cO)$ and $(\ov{\cO\ps})\Ga \cO\pr\ps$,
where $\Ga$ represents the 16 matrices
$\{ 1, i \ga_5, \ga_\mu, \ga_5 \ga_\mu, \simn/2 \}$
spanning the spinor space.
These gauge-invariant operators are the desired objects
from which to build terms in the Lagrange density
for the effective field theory in Minkowski spacetime.  
More details about this construction and its implications can be found
in Sec.\ II of Ref.\ \cite{kl19}.

To generalize this construction to curved spacetime,
we can work with spacetime-tensor fields
and covariant derivatives extended to include an appropriate connection.
Relevant spacetime-tensor fields include the metric $g_{\mu\nu}$,
the curvature tensor $R_{\ka\la\mu\nu}$, 
the gauge field strength $F_{\mu\nu}$,
the spinor bilinears $\psb \Ga \ps$,
scalars $\ph$,
and combinations.
For this purpose,
the quantities $\ga_\mu$ appearing in $\Ga$ 
are now understood as $\ga_\mu = e_{\mu a} \ga^a$,
where the Dirac matrices $\ga^a$ are defined in the local frame.
In combinations,
gauge-invariant operators are placed inside fermion bilinears.
All these spacetime-tensor operators are also gauge covariant.
The covariant derivative $D_\mu$ acting on $\ps$ or on $DD\cdots D\ps$
can be expressed as
\beq
D_\mu=\overset{\star}{D}_\mu + \quar \nsc \mu ab \si_{ab} -ig A_\mu,
\eeq
where $\overset{\star}{D}_\mu$ is the usual covariant derivative of GR
containing the partial derivative $\prt_\mu$ 
and the appropriate connection term formed using Christoffel symbols,
and where $\nsc\mu ab$ is the spin connection.
For explicit derivations,
we adopt the conventions of Ref.\ \cite{ak04}.
Direct calculation shows that
any mixture of covariant derivatives 
and the spacetime-tensor operators
also forms a gauge-covariant spacetime-tensor operator.

Building the Lagrange density $\cL$
using all possible gauge-covariant spacetime-tensor operators
would introduce many redundancies
due to relationships between various mixtures of operators.
It is therefore useful to work instead with a standard basis set
that has no or controlled redundancies. 
The key result here is that any mixture 
of $g$, $R$, $F$, $\Ga$, and $D$
can be written in the standard form
\beq
[(D_{(n_1)}R) \cdots (D_{(n_t)}R)]
[(D_{(m_1)}F) \cdots (D_{(m_s)}F)] 
\Ga D_{(l)},
\label{form}
\eeq
where all indices on spacetime-tensor fields are omitted for simplicity.
In this expression,
we introduce the notation
\beq
D_{(n)} \equiv \frac 1 {n!} \sum \Da1 \Da2 \cdots D_{\al_n}
\eeq
as a symmetric sum over the $n$ indices.
Note that explicit factors of the metric tensor $g_{\mu\nu}$ 
can safely be disregarded in the form \rf{form},
as $g_{\mu\nu}$ commutes with covariant derivatives
and all other operators.

To prove the result \rf{form},
we follow a path similar to the proof of Eq.\ (2)
in Ref.\ \cite{kl19}.
It suffices to consider the case that the operator \rf{form} 
acts on a Dirac fermion field $\ps$,
as other cases are both similar and simpler.
Using the product rule for covariant derivatives,
any mixture of $D$, $R$, $F$, and $\Ga$ can be expressed in block form as
\bea
&&
\co = \sum (D D \cdots D R) \cdots (D D \cdots D R)
\nn\\
&&
\hskip35pt
\times (D D \cdots D F) \cdots (D D \cdots D F) \Ga D D \cdots D,
\qquad
\eea
where we use $D\Ga=0$,
which follows from Eq.\ \rf{dga}.
It therefore suffices to prove that operators of the form
$D_{\al_1} \Da2 \cdots D_{\al_n}$
can be expressed as linear combinations of the basis \rf{form}.
This can be achieved by mathematical induction.
The case $n=1$ follows directly from the definition of $D_{(n)}$.
Suppose the proposition holds for $n\leq k$.
Then,
we can decompose $D_{\al_1} \Da2 \cdots D_{\al_{k+1}}$ using Young tableaux,
\beq
\newtableau{4}{1}
\newbox{$\scriptstyle 1$} \newbox{$\scriptstyle 2$} 
\newbox{$\scriptstyle \cdots$} \newbox{$\scriptstyle k$}
\endtableau
\otimes
\newtableau{1}{1}
\newbox{$\scriptstyle k\!+\!1$}
\endtableau
=
\newtableau{5}{1}
\newbox{$\scriptstyle 1$} \newbox{$\scriptstyle 2$} 
\newbox{$\scriptstyle \cdots$} 
\newbox{$\scriptstyle k$}\newbox{$\scriptstyle k\!+\!1$}
\endtableau
\oplus
\newtableau{4}{2}
\newbox{$\scriptstyle 1$} \newbox{$\scriptstyle 2$} 
\newbox{$\scriptstyle \cdots$} \newbox{$\scriptstyle k$}
\newrow \newbox{$\scriptstyle k\!+\!1$}
\endtableau .
\label{young}
\eeq
The first term on the right-hand side is of the form $D_{(k+1)}$.
The second term contains at least one commutator 
of two covariant derivatives,
as proved in Ref.\ \cite{kl19}.
Direct calculation shows that
\beq
[D_\mu, D_\nu]=[\overset{\star}{D}_\mu,\overset{\star}{D}_\nu]
+\tfrac 14 i \lR \ka\la\mu\nu \uvb \ka a \uvb \la b \si_{ab}-ig F_{\mu\nu},
\eeq
where the commutator on the right-hand side 
generates a combination of curvature tensors,
so the commutator of covariant derivatives 
yields factors of $R$, $F$, and $\Ga$.
The remaining part of the second Young tableau \rf{young} 
thus contains at most $k-1<k$ covariant derivatives.
The proposition therefore holds for $n=k+1$,
concluding the proof.

\renewcommand\arraystretch{1.6}
\begin{table*}
\caption{
\label{tab:gravity}
Schematic structure of terms with $d\leq 6$ 
in the Lagrange density $\cL_g$.}
\setlength{\tabcolsep}{6pt}
\begin{tabular}{cl}
\hline
\hline
Component & Expression \\
\hline
$	\cL^{(2)}_g	$	&	$		\uk	$	\\
$	\cL^{(3)}_g	$	&	$		\cL_{\xx}^{(3)}		
					+	\uk(Dk)	$	\\
$	\cL^{(4)}_g	$	&	$		\uk R		
					+	\uk(Dk)(Dk)	$	\\
$	\cL^{(5)}_g	$	&	$		\uk DR		
					+	\cL_{\xx}^{(5)}		
					+	\uk (Dk)R		
					+	\uk(Dk)(Dk)(Dk)		
					+	\uk(Dk)(D_{(2)}k)	$	\\
$	\cL^{(6)}_g	$	&	$		\uk RR		
					+	\uk D_{(2)}R		
					+	\uk(Dk)(Dk)R		
					+	\uk(D_{(2)}k)R		
					+	\uk(Dk)(Dk)(Dk)(Dk)		
					+	\uk(Dk)(Dk)(D_{(2)}k)		
					+	\uk(D_{(2)}k)(D_{(2)}k)	$	\\
\hline
\hline
\end{tabular}
\end{table*}

Next,
we consider the linear independence of the operators \rf{form}.
Note first that the operators
$\Dn i R$ and $\Dn j R$ are linearly independent when $n_i \neq n_j$ 
because they have different mass dimensions.
Since $\Dn i R$ commutes with $\Dn j R$,
we can impose $n_1\leq \cdots \leq n_t$ on the basis \rf{form}.
Similarly,
$\Dn i F$ and $\Dn j F$ are linearly independent when $n_i \neq n_j$,
but they commute only for abelian gauge field theory.
In the abelian case,
we can therefore choose the linearly independent standard basis as
\bea
&& \Big\{ [(D_{(n_1)}R) \cdots (D_{(n_t)}R)]
[(D_{(m_1)}F) \cdots (D_{(m_s)}F)]
\Ga D_{(l)} \nn\\
&& \hskip 85pt
| n_1 \leq \cdots \leq n_t, m_1 \leq \cdots \leq m_s \Big \}.
\qquad
\label{qedform}
\eea
However,
in a nonabelian gauge theory,
we choose instead the basis
\bea
&& \Big\{ [(D_{(n_1)}R) \cdots (D_{(n_t)}R)]
[(D_{(m_1)}F) \cdots (D_{(m_s)}F)]
\Ga D_{(l)} \nn\\
&& \hskip 85pt
| n_1 \leq \cdots \leq n_t\Big \}.
\qquad
\eea
This basis is linearly independent in some cases
and is almost linearly dependent in others,
depending on the structure of the gauge group
\cite{kl19}.

In building a hermitian Lagrange density,
it is useful to have explicit expressions
for the hermitian conjugations of the operators \rf{form}.
Including fermions to form bilinears,
we find 
\bea
&&
\hskip -12pt
[ (\ov{D_{(m_0)} \ps}) (D_{(m_1)} F_{\be_1 \ga_1})
\scdots (D_{(m_s)} F_{\be_s \ga_s})\Ga (D_{(m_{s+1})}\ps) ]\da 
\nn\\
&=&
(\ov{D_{(m_{s+1})} \ps}) (D_{(m_s)} F_{\be_s \ga_s})
\scdots (D_{(m_1)} F_{\be_1 \ga_1})\Ga (D_{(m_0)}\ps).
\nn\\
\label{hermitian}
\eea
Any pieces involving $D_{(n)}R$ can be omitted
because they are independently real 
and commute with all other operators.
The result \rf{hermitian} takes the same form
as Eq.\ (13) in Ref.\ \cite{kl19},
but the covariant derivatives now include also
terms involving the spacetime and spin connections.
The proof of Eq.\ \rf{hermitian} follows the path
given in Ref.\ \cite{kl19}.

\subsection{Pure-gravity and background sector}
\label{Pure-gravity sector}

With the construction of generic gauge-invariant spacetime-tensor operators
in hand,
we can address specific sectors of the effective field theory in turn.
In this subsection,
we consider operators involving pure-gravity fields. 

For the pure-gravity sector,
it is convenient to distinguish terms in the Lagrange density $\cL_g$
according to mass dimension.
We therefore write
\beq
\cL_g = 
\frac 1 {2\ka}
(\cL_{g0}
+ \cL_g^{(2)}
+ \cL_g^{(3)}
+ \cL_g^{(4)}
+ \cL_g^{(5)}
+ \cL_g^{(6)}
+ \ldots),
\label{gravitylag}
\eeq
where $1/2\ka \equiv 1/16\pi G_N \simeq 3\times 10^{36}$ GeV${}^2$
is the gravitational coupling constant
formed from the Newton gravitational constant $G_N$.
The term $\cL_{g0}= R - 2 \La$ 
is the conventional Einstein-Hilbert expression with cosmological constant,
while the terms $\cL_g^{(d)}$ 
represent contributions to the effective field theory.
Note that each individual component $\cL_g^{(d)}$ has mass dimension two,
but by convention the superscript $d$ 
represents the mass dimension of the dynamical operator in $\cL_g^{(d)}$ 
including the factor of the gravitational coupling constant.
For example,
$\cL_g^{(4)}$ includes terms with the Riemann tensor as dynamical operator,
which in this convention is of mass dimension two.

Table \ref{tab:gravity} displays all terms in $\cL_g^{(d)}$ 
with $d\leq 6$ in schematic form.
The first column lists each component Lagrange density 
given in Eq.\ \rf{gravitylag},
while the second column specifies the corresponding operator structures
that can appear.
In this schematic notation,
each instance of $k$ can represent a distinct background
even when occurring in a single term,
and the various quantities $\uk$ may also be distinct. 
In each row,
terms without background derivatives $Dk$ are listed first.
All terms are invariant under general coordinate transformations
except for $\cL_{\xx}^{(d)}$ with $d=3$ or 5,
which transform into a total derivative
and hence maintain invariance of the action rather than the Lagrange density. 

Throughout Table \ref{tab:gravity},
all indices and any numerical or dimensionful factors are omitted 
for simplicity.
In each term,
we can absorb into $\uk$ any metric, vierbein, or Levi-Civita factors,
so all indices on $D_{(n)}\bk$ and $D_{(n)}R$ 
can be assumed distinct and contracted with indices on $\uk$.
To maintain hermiticity of the Lagrange density,
all backgrounds are taken to be real.
The mass dimensionality of a given background $k$ or $\uk$ 
is fixed by the operator structure of the Lagrange density.
As examples,
the quantity $\uk$ in $\cL_g^{(2)}$ has mass dimension $d=2$,
while $\uk$ in the term $\uk R$ is dimensionless.
The mass dimension of $k$ may be different in different theories,
so for convenience and simplicity
we adopt the convention that the mass dimension of $k$
is disregarded in assigning a given term
to a particular component $\cL_g^{(d)}$.
For instance,
the term $\uk (Dk)$ is assigned to $\cL_g^{(3)}$
irrespective of the actual mass dimension of $k$.
The number of derivatives $D$
is instead the relevant factor in defining the mass dimension
of terms involving background derivatives.

Since the commutator of two covariant derivatives generates curvature factors,
and since the product of two backgrounds 
can schematically be viewed as a single background,
certain terms can be omitted from the table without loss of generality.
For example,
the term $\uk [D,D] k$ is schematically equivalent to the term $\uk R$,
which itself is typically included in $\uk(Dk)(Dk)$ up to a surface term.
The only exception arises when the background is index free,
with $\uk R$ representing the term $k(x) g^{\al\ga}g^{\be\de}R_{\al\be\ga\de}$. 

\renewcommand\arraystretch{1.6}
\begin{table*}
\caption{
\label{tab:curvature}
Terms with $d\leq 8$ and without background derivatives
in the Lagrange density $\cL_g$.}
\setlength{\tabcolsep}{8pt}
\begin{tabular}{cl}
\hline
\hline
Component & Expression \\
\hline
$	\cL_{g0}	$	&		$		R-2\La	$	\\
$	\cL^{(2)}_g	$	&		$		\uk^{(2)}	$	\\
$	\cL^{(3)}_g	$	&		$		(\uk^{(3)}_{\Ga})^{\mu} \LC \al\mu\al	$	\\
$	\cL^{(4)}_g	$	&		$		(\uk^{(4)}_R)^{\al\be\ga\de}R_{\al\be\ga\de}	$	\\
$	\cL^{(5)}_g	$	&		$		(\uk_D^{(5)})^{\al\be\ga\de\ka}D_\ka R_{\al\be\ga\de}		
						+	(\uk^{(5)}_{{\rm CS}, 1})_\ka \ep^{\ka\la\mu\nu}\et_{ac}\et_{bd}(\nsc\la a b \prt_\mu \nsc \nu c d		
						+	\tfrac 23 \nsc \la a b \nsc \mu c e \llusc \nu e d)	$	\\
			&	\hskip20pt	$	+	(\uk^{(5)}_{{\rm CS}, 2})_\ka \ep^{\ka\la\mu\nu}\ep_{abcd}(\nsc\la a b \prt_\mu \nsc \nu c d		
						+	\tfrac 23 \nsc \la a b \nsc \mu c e \llusc \nu e d)	$	\\
$	\cL^{(6)}_g	$	&		$		(\uk_D^{(6)})^{\al\be\ga\de\ka\la}D_{(\ka}D_{\la)}R_{\al\be\ga\de}		
						+	(\uk_R^{(6)})^{\abcd1\abcd2}R_{\abcd1}R_{\abcd2}	$	\\
$	\cL^{(7)}_g	$	&		$		(\uk_D^{(7)})^{\al\be\ga\de\ka\la\mu}D_{(\ka}D_\la D_{\mu)}R_{\al\be\ga\de}		
						+	(\uk_{DR}^{(7)})^{\abcd1\abcd2\ka}R_{\abcd1}D_\ka R_{\abcd2}	$	\\
$	\cL^{(8)}_g	$	&		$		(\uk_D^{(8)})^{\al\be\ga\de\ka\la\mu\nu}D_{(\ka}D_\la D_\mu D_{\nu)}R_{\al\be\ga\de}		
						+	(\uk_{DR,1}^{(8)})^{\abcd1\abcd2\ka\la} R_{\abcd1} D_{(\ka}D_{\la)} R_{\abcd2}	$	\\
			&	\hskip20pt	$	+	(\uk_{DR,2}^{(8)})^{\abcd1\abcd2\ka\la}(D_\ka R_{\abcd1})(D_\la R_{\abcd2})	$	\\
			&	\hskip20pt	$	+	(\uk_R^{(8)})^{\abcd1\abcd2\abcd3} R_{\abcd1}R_{\abcd2}R_{\abcd3}	$	\\
\hline
\hline
\end{tabular}
\end{table*}

More insight can be obtained from the explicit forms
of the components of the Lagrange density $\cL_g$. 
First,
consider terms without background derivatives $Dk$.
The curvature plays a central role in most of these terms.
Table \ref{tab:curvature} displays the explicit form 
of all terms in $\cL_g$ for $d\leq 8$.
In this table,
unlike others in this section,
we include operators with $d=7$ and 8
because they are comparatively few in number
and because certain $d=8$ backgrounds 
have recently been constrained by observational data 
\cite{qb16,19shaobailey}.
Each row of the table lists a component of $\cL_g$ 
constructed from operators of a given mass dimension,
followed by the explicit form of the corresponding terms. 
In each term,
the combination $\uk$ of backgrounds is understood to be real
and to inherit from the dynamical operator 
the appropriate symmetry under permutation of the indices.
Most of the operators can be constructed by inspection
from Table \ref{tab:gravity}.
The components $\cL_{\xx}^{(d)}$ with $d=3$ and 5 
appearing in Table \ref{tab:gravity},
which are general coordinate invariant only up to a surface term,
are represented in Table \ref{tab:curvature}
by the three terms with backgrounds 
$(\uk_{\Ga}^{(3)})^\mu$,
$(\uk^{(5)}_{{\rm CS},1})_\mu$,
and $(\uk^{(5)}_{{\rm CS},2})_\mu$.
The operators associated with the latter two
are Chern-Simons terms expressed using the vierbein and spin connection.
To preserve general coordinate invariance of the action,
these three backgrounds must obey 
\bea
D_\mu(\uk_{\Ga}^{(3)})^\mu &=& 0,
\nn\\
D_\mu (\uk^{(5)}_{{\rm CS},1})_\nu-D_\nu (\uk^{(5)}_{{\rm CS},1})_\mu &=&0,
\nn\\
D_\mu (\uk^{(5)}_{{\rm CS},2})_\nu-D_\nu (\uk^{(5)}_{{\rm CS},2})_\mu &=&0.
\eea
The first equation implies $\prt_\mu [(\uk_{\Ga}^{(3)})^\mu /e]=0$
in any coordinate frame.
The second equation implies 
$\prt_\mu (\uk^{(5)}_{{\rm CS},1})_\nu
- \prt_\nu (\uk^{(5)}_{{\rm CS},1})_\mu = 0$,
so if the topology is trivial then we can write 
$(\uk^{(5)}_{{\rm CS},1})_\mu = \prt_\mu k$
for some scalar background $k(x)$.
A similar result holds for the third equation.

Next,
consider the terms in Table \ref{tab:gravity}
involving one or more factors of $Dk$
that cannot be moved onto the dynamical operators
via integration by parts.
Since different backgrounds and hence distinct $Dk$ factors 
can appear in a given term,
an explicit listing of all such terms is impractical in the general case.
We can,
however,
gain useful insight about generic terms with background derivatives $Dk$
by considering ones constructed explicitly
using a one-index background
because any given background $k^{\mu\ldots}{}_{\nu\ldots}{}^{a\ldots}$
can be expressed as a linear combination of products
of backgrounds with only one index.
For example,
a two-index background $k^{\mu\nu}$ can be viewed
as a linear combination of products of eight one-index backgrounds
as follows. 
Given $k^{\mu\nu}$ in a specific coordinate system,
we can define four one-index backgrounds in these coordinates
via $(k_\mu)^\nu \equiv k^{\mu\nu}$ 
and another four via $(k^\prime_\mu)^\nu \equiv \de_\mu{}^\nu$.
In the chosen coordinate system,
it then follows that
$k^{\mu\nu}=\sum_\la (k_\la)^\nu (k^\prime_\la)^\mu$.
Provided we define the eight one-index backgrounds to transform as 4-vectors
under general coordinate transformations,
this expression is a tensor expression
and hence is valid in any coordinate system,
as required.

\begin{table}
\caption{
\label{Dk}
Terms with $d\leq 6$ involving derivatives of a single background $k_\mu$.}
\setlength{\tabcolsep}{4pt}
\begin{tabular}{cl}
\hline
\hline
Component & Expression \\
\hline
$	\cL^{(3)}_k	$	&	$	(\uk^{(3)}_{Dk})^{\ka\la}(D_\ka k_\la)		$	\\
$	\cL^{(4)}_k	$	&	$	(\uk^{(4)}_{Dk})^{\ka\la\mu\nu}(D_\ka k_\la)(D_\mu k_\nu)		$	\\
$	\cL^{(5)}_k	$	&	$	(\uk^{(5)}_{Dk})^{\ka\la\mu\nu\rh\si}(D_\ka k_\la)(D_\mu k_\nu)(D_\rh k_\si)	,	$	\\
$		$	&	$	(\uk^{(5)}_{DDk})^{\ka\la\mu\nu\rh} (D_\ka k_\la)(D_{(\mu} D_{\nu)} k_\rh)	,	$	\\
$		$	&	$	(\uk^{(5)}_{DkR})^{\ka\la\mu\nu\rh\si} (D_\ka k_\la)R_{\mu\nu\rh\si}		$	\\
$	\cL^{(6)}_k	$	&	$	(\uk^{(6)}_{Dk})^{\ka\la\mu\nu\rh\si\ta\up}(D_\ka k_\la)(D_\mu k_\nu)(D_\rh k_\si)(D_\ta k_\up)	,	$	\\
$		$	&	$	(\uk^{(6)}_{DDk})^{\ka\la\mu\nu\rh\si\ta}(D_\ka k_\la)(D_\mu k_\nu)(D_{(\rh} D_{\si)} k_\ta)	,	$	\\
$		$	&	$	(\uk^{(6)}_{DDkDDk})^{\ka\la\mu\nu\rh\si} (D_{(\ka} D_{\la)} k_\mu)(D_{(\nu} D_{\rh)} k_\si)	,	$	\\
$		$	&	$	(\uk^{(6)}_{DkR})^{\ka\la\mu\nu\rh\si\ta\up} (D_\ka k_\la)(D_\mu k_\nu)R_{\rh\si\ta\up}	,	$	\\
$		$	&	$	(\uk^{(6)}_{DDkR})^{\ka\la\mu\nu\rh\si\ta}(D_{(\ka} D_{\la)} k_\mu)R_{\nu\rh\si\ta}		$	\\
\hline
\hline
\end{tabular}
\end{table}

\renewcommand\arraystretch{1.6}
\begin{table*}
\caption{
\label{fullgauge}
Schematic structure of terms with $d\leq 6$ 
in the gauge Lagrange density $\cL_A$.}
\setlength{\tabcolsep}{4pt}
\begin{tabular}{cl}
\hline
\hline
Component & Expression \\
\hline
$	\cL_A^{(1)}	$	&		$		\cL_{\xxa}^{(1)}	$	\\
$	\cL_A^{(3)}	$	&		$		\cL_{\xxa}^{(3)}		
						+	\uk (Dk) \tr(F)	$	\\
$	\cL_A^{(4)}	$	&		$		\uk \tr(FF)		
						+	\uk \tr(D_{(2)}F)		
						+	\uk R\tr(F)		
						+	\uk (Dk) (Dk) \tr(F)		
						+	\uk (D_{(2)}k) \tr(F)	$	\\
$	\cL_A^{(5)}	$	&		$		\uk \tr(FDF)		
						+	\uk \tr(D_{(3)}F)		
						+	\uk R \tr(DF)		
						+	\uk (DR) \tr(F)	$	\\
			&	\hskip20pt	$	+	\uk (Dk) R \tr(F)		
						+	\uk (Dk) (Dk) (Dk) \tr(F)		
						+	\uk (Dk) (D_{(2)}k) \tr(F)		
						+	\uk (D_{(3)}k) \tr(F)	$	\\
$	\cL_A^{(6)}	$	&		$		\uk \tr(FD_{(2)}F)		
						+	\uk \tr\big((DF)(DF)\big)		
						+	\uk \tr(D_{(4)}F)		
						+	[\uk \tr(FFF)+\hc]	$	\\
			&	\hskip20pt	$	+	\uk R\tr(FF)		
						+	\uk R \tr(D_{(2)}F)		
						+	\uk (D_{(2)}R) \tr(F)		
						+	\uk (DR) \tr(DF)		
						+	\uk RR \tr(F)	$	\\
			&	\hskip20pt	$	+	\uk (Dk) \tr(FDF)		
						+	\uk (Dk) (Dk) \tr(FF)		
						+	\uk (D_{(2)}k) \tr(FF)		
						+	\uk (D_{(4)}k) \tr(F)	$	\\
			&	\hskip20pt	$	+	\uk (Dk) (Dk) (Dk) (Dk) \tr(F)		
						+	\uk (Dk) (Dk) (D_{(2)}k) \tr(F)		
						+	\uk (D_{(2)}k) (D_{(2)}k) \tr(F)		
						+	\uk (Dk) (D_{(3)}k) \tr(F)	$	\\
			&	\hskip20pt	$	+	\uk (Dk) (DR) \tr(F)		
						+	\uk (Dk) (Dk) R \tr(F)		
						+	\uk (D_{(2)}k) R \tr(F)	$	\\
\hline
\hline
\end{tabular}
\end{table*}

Taking advantage of this result,
we can consider a scenario involving a single one-index background $k_\mu$
as an example to gain insight.
Table \ref{Dk} displays possible operators of mass dimension $d\leq 6$ 
involving derivatives of a background $k_\mu$ 
with one covariant spacetime index.
The first column lists $\cL_k^{(d)}$ for different values of $d$,
while the second column displays 
the corresponding possible terms involving $D_\mu k_\nu$. 
As before,
each combination $\uk$ of backgrounds is taken as real
and inherits the symmetry under index permutation of the operator,
and each operator can be constructed using Table \ref{tab:gravity}
as a guide.
Following the convention described above,
the assigned value of the mass dimension $d$ of a particular term 
is definied for convenience by the number of derivatives $D$ involved. 
As a result,
the various combinations $\uk^{(d)}$ at given $d$  
may have different mass dimensions
and may incorporate different powers of $k_\mu$.
Notice also that the combinations $\uk^{(d)}$ with even numbers of indices
cannot contain terms linear in $k_\mu$.
For example,
the leading-order $k_\mu$-dependent term in $(\uk^{(3)}_{Dk})^{\ka\la}$ is
$(\uk^{(3)}_{Dk})^{\ka\la}
\propto
k_\mu k_\nu g^{\mu \ka} g^{\nu \la}$,
which produces a cubic coupling in $k_\mu$ 
for the component $\cL_k^{(3)}$.

If the breaking is spontaneous,
the type of index carried by the background
has no effect on the physics,
as discussed in Sec.\ \ref{Backgrounds}.
In this scenario, 
Table \ref{Dk} also encompasses operators of mass dimension $d\leq 6$ 
constructed from a single background $k^\mu$ or $k^a$,
provided each instance of $k_\mu$ in the table
is replaced with
$g_{\mu\nu}k^\nu$ or  
$e_{\mu a} k^a$,
respectively.
However,
if the breaking is explicit,
then each of the three types of single-index background 
$k_\mu$, $k^\mu$, and $k^a$
can lead to different physics.
If all three are present,
then we can write
\beq
\cL_k^{(3)}=(\uk^{(3)}_{Dk,1})^{\mu\nu} D_\mu\bk_\nu 
+(\uk^{(3)}_{Dk,2})^{\mu}_{\pt{\mu}\nu} D_\mu \bk{}^\nu
+(\uk^{(3)}_{Dk,3})^{\mu a}D_\mu \bk_a,
\eeq
with distinct combinations $\uk$ for each term. 

Taken together,
the results in this subsection provide 
the explicit form of all terms at $d\leq 6$
in the expansion (57) of Ref.\ \cite{ak04}
in the zero-torsion limit.
Our construction covers both the pure-gravity sector
and any background dynamics,
allowing for either spontaneous or explicit violations
of spacetime symmetries. 
It opens the way to broadening theoretical explorations
of many issues including,
for example,
the $t$ puzzle
\cite{bkx15,yb15},
the relevance of spontaneous violation
\cite{ms18},
and the influence of gravitational effects on the matter sector
\cite{eps19}.
It also has potential applications
to the numerous ongoing experimental searches for backgrounds.

\subsection{Matter sector}
\label{Matter sector}

Next,
we turn attention to the matter sector,
including pure matter, matter-gravity, and matter-background terms.
It is convenient to separate the discussion
according to the type of matter involved.
In what follows,
we first consider the inclusion of gauge fields,
then Dirac spinors,
and finally complex scalar fields.

\subsubsection{Gauge fields}
\label{sec:gauge}

In the gauge sector,
we consider for definiteness a Lagrange density $\cL_A$
determining the behavior of a gauge field $A_\mu$ 
with gauge field strength $F_{\mu\nu}$
in the $U$ representation of the gauge group.
This follows the setup adopted in Sec.\ II of Ref.\ \cite{kl19}
and summarized in Sec.\ \ref{Dynamical operators} above.
Couplings of $A_\mu$ to gravity and backgrounds are included
in this sector. 
It is convenient to split $\cL_A$ into pieces
according to the mass dimension $d$ of the dynamical operator, 
\beq
\cL_A=
\cL_{A0} +\cL^{(1)}_A+\cL^{(2)}_A+\cL^{(3)}_A+\cL^{(4)}_A+\cL^{(5)}_A+\ldots.
\eeq
The first piece $\cL_{A0} = \tr(F_{\mu\nu}F^{\mu\nu})/2$
is the usual Yang-Mills term in curved spacetime
expressed with a trace taken in the gauge space.
On each of the other pieces,
the superscript indicates the value of $d$.

\renewcommand\arraystretch{1.6}
\begin{table*}
\caption{
\label{tab:gauge}
Terms with $d\leq 6$ and without background derivatives 
in the gauge Lagrange density $\cL_A$.}
\setlength{\tabcolsep}{8pt}
\begin{tabular}{cl}
\hline
\hline
Component & Expression \\
\hline
$	\cL_{A0}	$	&		$	-	\half\tr(F_{\mu\nu}F^{\mu\nu})	$	\\
$	\cL_A^{(1)}	$	&		$	-	(\uk^{(1)})^{\mu} \tr(A_\mu)	$	\\
$	\cL_A^{(3)}	$	&		$		(\uk^{(3)}_{\rm{CS}})_{\ka} \ep^{\ka\la\mu\nu} \tr (A_\la F_{\mu\nu} +\tfrac{2}{3} ig A_\la A_\mu A_\nu)		
						+	(\uk^{(3)}_{DF})^{\al\mu\nu}\tr(D_\al F_{\mu\nu})	$	\\
$	\cL_A^{(4)}	$	&		$	-	\half (\uk^{(4)}_F)^{\ka\la\mu\nu} \tr (F_{\ka\la} F_{\mu\nu})		
						-	(\uk^{(4)}_{DF})^{\al\be\mu\nu} \tr (D_{(\al}D_{\be)} F_{\mu\nu})		
						-	(\uk_{RF}^{(4)})^{\al\be\ga\de\mu\nu}R_{\al\be\ga\de}\tr(F_{\mu\nu})	$	\\
$	\cL_A^{(5)}	$	&		$		\half (\uk^{(5)}_D)^{\al\ka\la\mu\nu} \tr (F_{\ka\la} D_\al F_{\mu\nu})		
						+	(\uk^{(5)}_{DF})^{\al\be\ga\mu\nu} \tr (D_{(\al}D_\be D_{\ga)} F_{\mu\nu})	$	\\
			&	\hskip20pt	$	+	(\uk_{RDF}^{(5)})^{\al\be\ga\de\ep\mu\nu}R_{\al\be\ga\de}\tr(D_\ep F_{\mu\nu})		
						+	(\uk_{DRF}^{(5)})^{\al\be\ga\de\ep\mu\nu}(D_\ep R_{\al\be\ga\de})\tr(F_{\mu\nu})	$	\\
$	\cL_A^{(6)}	$	&		$	-	\half (\uk_D^{(6)})^{\al\be\ka\la\mu\nu} \tr (F_{\ka\la} D_{(\al}D_{\be)} F_{\mu\nu})		
						-	\half (\uk_{DFDF}^{(6)})^{\al\be\ka\la\mu\nu} \tr \big((D_\al F_{\ka\la}) (D_\be F_{\mu\nu})\big)	$	\\
			&	\hskip20pt	$	-	(\uk^{(6)}_{DF})^{\al\be\ga\de\mu\nu} \tr (D_{(\al}D_\be D_\ga D_{\de)} F_{\mu\nu})		
						-	\tfrac1{12}[(\uk_{F}^{(6)})^{\ka\la\mu\nu\rh\si}\tr(F_{\ka\la}F_{\mu\nu}F_{\rh\si})+\hc]	$	\\
			&	\hskip20pt	$	-	\half (\uk_{RFF}^{(6)})^{\al\be\ga\de\ka\la\mu\nu}R_{\al\be\ga\de}\tr(F_{\ka\la}F_{\mu\nu})		
						-	(\uk_{RDDF}^{(6)})^{\al\be\ga\de\ep\ze\mu\nu}R_{\al\be\ga\de}\tr(D_{(\ep}D_{\ze)}F_{\mu\nu})	$	\\
			&	\hskip20pt	$	-	(\uk_{DDRF}^{(6)})^{\al\be\ga\de\ep\ze\mu\nu}(D_{(\ep}D_{\ze)}R_{\al\be\ga\de})\tr(F_{\mu\nu})		
						-	(\uk_{DRDF}^{(6)})^{\al\be\ga\de\ep\ze\mu\nu}(D_\ep R_{\al\be\ga\de})\tr(D_{\ze}F_{\mu\nu})	$	\\
			&	\hskip20pt	$	-	(\uk_{RRF}^{(6)})^{\abcd1\abcd2\mu\nu}R_{\abcd1}R_{\abcd2}\tr(F_{\mu\nu})	$	\\
\hline
\hline
\end{tabular}
\end{table*}

Table \ref{fullgauge} lists the schematic form 
of all terms in $\cL_A^{(d)}$ with mass dimension $d\leq6$.
The first entry in each row shows a component of $\cL_A$ with fixed $d$,
while the second entry lists the schematic forms 
of the corresponding operators. 
In this notation,
each occurrence of $k$ in a given term 
can represent a distinct background,
and the combinations $\uk$ appearing in different terms 
can also be distinct. 
Terms in the pure gauge sector are shown first,
followed by terms with gauge-curvature couplings,
and then by terms with background derivatives $Dk$.
All terms in the Lagrange density are invariant under gauge transformations
except $\cL_{\xxa}^{(d)}$ with $d=1$ and 3,
which become total derivatives and thus leave the action invariant instead.  
The table omits all indices and factors,
and all backgrounds can be taken as real
except for the one involving the operator $\tr(FFF)$.
Note that the components $\cL_A^{(d)}$ have mass dimension four,
unlike the components $\cL_g^ {(d)}$ in Eq.\ \rf{gravitylag}.
Other properties of the backgrounds $k$ and the combinations $\uk$
follow those described for Table \ref{tab:gravity}.

The explicit form of terms in $\cL_A$ without background derivatives 
are displayed in Table \ref{tab:gauge} for $d\leq 6$.
The structure of the table follows that of Table \ref{fullgauge},
and the various expressions can be derived using the latter as a guide.
Each combination $\uk$ appearing in Table \ref{tab:gauge}
is understood to have index symmetry determined 
by the index structure of the corresponding dynamical operator.
All combinations $\uk$ are real 
except for the one controlling the dynamical operator 
$\tr(F_{\ka\la}F_{\mu\nu}F_{\rh\si})$.
The terms $\cL_{\xxa}^{(1)}$ and $\cL_{\xxa}^{(3)}$ 
in Table \ref{fullgauge} 
are explicitly given in Table \ref{tab:gauge}
as the terms with backgrounds 
$(\uk^{(1)})^\mu$ and $(\uk^{(3)}_{\rm{CS}})_{\ka}$,
with the latter governing the nonabelian Chern-Simons operator.
To maintain gauge symmetry,
these backgrounds must satisfy the conditions
\bea
D_\mu(\uk^{(1)})^\mu &=& 0,
\nn\\
D_\mu (\uk_{\rm{CS}}^{(3)})_\nu-D_\nu (\uk_{\rm{CS}}^{(3)})_\mu &=&0.
\eea
The first of these expressions implies 
$\prt_\mu [(\uk^{(1)})^\mu /e]=0$,
while the second implies 
$\prt_\mu (\uk_{\rm{CS}}^{(3)})_\mu-\prt_\nu (\uk_{\rm{CS}}^{(3)})_\nu=0$.
If the topology is trivial,
a scalar background field $k(x)$ can be found
such that $(\uk_{\rm{CS}}^{(3)})_\mu=\prt_\mu k$,
in which case its contribution to $\cL_A^{(3)}$
is equivalent under partial integration to a $d=4$ term 
$k \ep^{\ka\la\mu\nu} F_{\ka\la}F_{\mu\nu}$.
If the topology is nontrivial,
however,
the Chern-Simons contribution to $\cL_A^{(3)}$ 
can produce independent physical effects.
Note that the putative $d=2$ term
$\uk^{\mu\nu}\tr(F_{\mu\nu})=
\uk^{\mu\nu}\tr(\prt_\mu A_\nu-\prt_\nu A_\mu)$
involves a combination of partial derivatives of $A_\mu$,
so it is included in $\cL^{(1)}_A$ up to a surface term.
We also omit terms such as $(\tr F)(\tr (\cdots F))$ 
because these terms either vanish if the gauge group is SU($N$)
or are incorporated in terms like $\tr(F\cdots F)$ if the gauge group is U(1),
but their inclusion may be appropriate for other gauge groups.

\renewcommand\arraystretch{1.6}
\begin{table*}
\caption{
\label{tab:fullspinor}
Schematic structure of terms with $d\leq 6$ 
in the fermion Lagrange density $\cL_\ps$.}
\setlength{\tabcolsep}{8pt}
\begin{tabular}{cl}
\hline
\hline
Component & Expression \\
\hline
$	\cL^{(3)}_\ps	$	&		$		\uk \psb \Ga \ps			$	\\
$	\cL^{(4)}_\ps	$	&		$		\uk \psb\Ga i D\ps				
						+	\uk (Dk) \psb \Ga \ps	+	\hc	$	\\
$	\cL^{(5)}_\ps	$	&		$		\uk \psb \Ga i^2 D_{(2)} \ps				
						+	\uk \psb \Ga F\ps				
						+	\uk R\psb \Ga\ps				
						+	\uk (Dk)\psb \Ga iD\ps				
						+	\uk (Dk) (Dk) \psb \Ga \ps				
						+	\uk (\D2k) \psb \Ga \ps	+	\hc	$	\\
$	\cL^{(6)}_\ps	$	&		$		\uk \psb \Ga i^3 D_{(3)} \ps				
						+	\uk R \psb \Ga iD \ps				
						+	\uk (DR) \psb \Ga \ps				
						+	\uk \psb \Ga F iD \ps				
						+	\uk \psb \Ga (DF) \ps				
						+	\uk (\psb \Ga \ps) (\psb \Ga \ps)			$	\\
			&	\hskip20pt	$	+	\uk (Dk) \psb \Ga i^2 \D2 \ps				
						+	\uk (Dk) R \psb \Ga \ps				
						+	\uk (Dk) \psb \Ga F \ps				
						+	\uk (\D2k) \psb \Ga iD \ps				
						+	\uk (Dk) (Dk) \psb \Ga iD \ps			$	\\
			&	\hskip20pt	$	+	\uk (Dk)(\D2k) \psb \Ga\ps				
						+	\uk (Dk)(Dk)(Dk)\psb\Ga\ps				
						+	\uk (\D3k)\psb\Ga\ps	+	\hc	$	\\
\hline
\hline
\end{tabular}
\end{table*}

\renewcommand\arraystretch{1.6}
\begin{table*}
\caption{
\label{tab:spinor2}
Terms with $d\leq 6$ and without background derivatives 
in the fermion Lagrange density $\cL_\ps$.}
\setlength{\tabcolsep}{8pt}
\begin{tabular}{cl}
\hline
\hline
Component & Expression \\
\hline
$	\cL_{\ps0}	$	&		$		\half \psb (\ivb \mu a\ga^a i D_\mu - m_\ps ) \ps	+	\hc	$	\\
$	\cL^{(3)}_\ps	$	&		$	-	\um\pr \psb \ps 				
						-	i \um_5 \psb \ga_5\ps 				
						-	\ua_\ka \ivb\ka a\psb\ga^a\ps				
						-	\ub_\ka\ivb\ka a\psb\gamma_5\ga^a\ps				
						-	\half \uH_{\ka\la}\ivb\ka a\ivb\la b\psb\si^{ab} \ps			$	\\
$	\cL^{(4)}_{\ps D}	$	&		$	-	\half \uc_{\ka\mu} \ivb\ka a \ivb\mu b e^{\nu b} \psb\ga^a i D_\nu \ps				
						-	\half \ud_{\ka\mu}\ivb\ka a\ivb\mu b e^{\nu b} \psb \ga_5 \ga^a i D_\nu \ps			$	\\
			&	\hskip20pt	$	-	\half \ue_\mu \ivb\mu b e^{\nu b} \psb iD_\nu \ps				
						-	\half i \uf_\mu \ivb\mu b e^{\nu b}  \psb \ga_5 iD_\mu \ps				
						-	\quar \ug_{\ka\la\mu}\ivb\ka a \ivb\la b \ivb\mu b e^{\nu b} \psb \si^{ab} iD_\nu \ps	+	\hc	$	\\
$	\cL_{\ps D}^{(5)}	$	&		$	-	\half (\um^{(5)})^{\mu\nu}\psb iD_{(\mu} iD_{\nu)}\ps				
						-	\half i (\um^{(5)}_{5})^{\mu\nu} \psb \ga_5 iD_{(\mu} iD_{\nu)} \ps			$	\\
			&	\hskip20pt	$	+	\half (\ua^{(5)})^{\ka\mu\nu} \lvb\ka a\psb \ga^a iD_{(\mu} iD_{\nu)} \ps				
						+	\half (\ub^{(5)})^{\ka\mu\nu}\lvb\ka a\psb \ga_5 \ga^a iD_{(\mu} iD_{\nu)} \ps			$	\\
			&	\hskip20pt	$	-	\quar (\uH^{(5)})^{\ka\la\mu\nu} \lvb\ka a\lvb\la b\psb \si^{ab} iD_{(\mu} iD_{\nu)} \ps	+	\hc	$	\\
$	\cL_{\ps R}^{(5)}	$	&		$	-	(\um^{(5)}_{R})^{\mu\nu\rh\si} R_{\mu\nu\rh\si}\psb\ps				
						-	i(\um^{(5)}_{5R})^{\mu\nu\rh\si} R_{\mu\nu\rh\si} \psb\ga_5\ps			$	\\
			&	\hskip20pt	$	-	(\ua^{(5)}_{R})^{\ka\mu\nu\rh\si}\lvb\ka a R_{\mu\nu\rh\si} \psb \ga^a \ps				
						-	(\ub^{(5)}_{R})^{\ka\mu\nu\rh\si}\lvb\ka a R_{\mu\nu\rh\si} \psb \ga_5 \ga^a \ps				
						-	\half (\uH^{(5)}_{R})^{\ka\la\mu\nu\rh\si}\lvb\ka a\lvb\la b R_{\mu\nu\rh\si} \psb \si^{ab} \ps			$	\\
$	\cL_{\ps F}^{(5)}	$	&		$	-	\half (\um^{(5)}_{F})^{\mu\nu} \psb F_{\mu\nu} \ps				
						-	\half i(\um^{(5)}_{5F})^{\mu\nu} \psb \ga_5 F_{\mu\nu} \ps			$	\\
			&	\hskip20pt	$	+	\half (\ua^{(5)}_{F})^{\ka\mu\nu}\lvb\ka a \psb \ga^a F_{\mu\nu} \ps				
						+	\half (\ub^{(5)}_{F})^{\ka\mu\nu}\lvb\ka a \psb \ga_5 \ga^a F_{\mu\nu} \ps				
						-	\quar (\uH^{(5)}_{F})^{\ka\la\mu\nu}\lvb\ka a\lvb\la b \psb \si^{ab} F_{\mu\nu} \ps			$	\\
$	\cL_{\ps D}^{(6)}	$	&		$	-	\half (\uc^{(6)})^{\ka\mu\nu\rh} \lvb\ka a\psb \ga^a iD_{(\mu} iD_\nu iD_{\rh)} \ps				
						-	\half (\ud^{(6)})^{\ka\mu\nu\rh} \lvb\ka a\psb \ga_5 \ga^a iD_{(\mu} iD_\nu iD_{\rh)} \ps			$	\\
			&	\hskip20pt	$	+	\half (\ue^{(6)})^{\mu\nu\rh} \psb iD_{(\mu} iD_\nu iD_{\rh)} \ps				
						+	\half i (\uf^{(6)})^{\mu\nu\rh} \psb \ga_5 iD_{(\mu} iD_\nu iD_{\rh)} \ps			$	\\
			&	\hskip20pt	$	+	\quar (\ug^{(6)})^{\ka\la\mu\nu\rh}\lvb\ka a\lvb\la b \psb \si^{ab} iD_{(\mu} iD_\nu iD_{\rh)} \ps	+	\hc	$	\\
$	\cL_{\ps R}^{(6)}	$	&		$	-	\half (\uc^{(6)}_{R})^{\ka\mu\nu\rh\si\ta}\lvb \ka a R_{\nu\rh\si\ta}\psb \ga^a  iD_\mu \ps				
						-	\half (\ud^{(6)}_{R})^{\ka\mu\nu\rh\si\ta}\lvb\ka a R_{\nu\rh\si\ta} \psb \ga_5 \ga^a iD_\mu \ps			$	\\
			&	\hskip20pt	$	-	\half (\ue^{(6)}_{R})^{\mu\nu\rh\si\ta}R_{\nu\rh\si\ta} \psb iD_\mu \ps				
						-	\half i(\uf^{(6)}_{R})^{\mu\nu\rh\si\ta} R_{\nu\rh\si\ta} \psb \ga_5 iD_\mu \ps			$	\\
			&	\hskip20pt	$	-	\quar (\ug^{(6)}_{R})^{\ka\la\mu\nu\rh\si\ta}\lvb\ka a\lvb\la b R_{\nu\rh\si\ta} \psb \si^{ab} iD_\mu \ps	+	\hc	$	\\
$	\cL_{\ps DR}^{(6)}	$	&		$	-	(\um^{(6)}_{DR})^{\mu\nu\rh\si\ta}(D_\mu R_{\nu\rh\si\ta}) \psb \ps				
						-	i (\um^{(6)}_{5DR})^{\mu\nu\rh\si\ta}(D_\mu R_{\nu\rh\si\ta}) \psb \ga_5 \ps			$	\\
			&	\hskip20pt	$	-	(\ua^{(6)}_{DR})^{\ka\mu\nu\rh\si\ta} \lvb\ka a (D_\mu R_{\nu\rh\si\ta})\psb \ga^a\ps				
						-	(\ub^{(6)}_{DR})^{\ka\mu\nu\rh\si\ta}\lvb\ka a (D_\mu R_{\nu\rh\si\ta}) \psb \ga_5 \ga^a\ps			$	\\
			&	\hskip20pt	$	-	\half (\uH^{(6)}_{DR})^{\ka\la\mu\nu\rh\si\ta} \lvb\ka a\lvb\la b(D_\mu R_{\nu\rh\si\ta})\psb \si^{ab}\ps			$	\\
$	\cL_{\ps F}^{(6)}	$	&		$	-	\quar (\uc^{(6)}_{F})^{\ka\mu\nu\rh}\lvb\ka a \psb \ga^a F_{\nu\rh} iD_\mu \ps				
						-	\quar (\ud^{(6)}_{F})^{\ka\mu\nu\rh}\lvb\ka a \psb \ga_5 \ga^a F_{\nu\rh} iD_\mu \ps			$	\\
			&	\hskip20pt	$	+	\quar (\ue^{(6)}_{F})^{\mu\nu\rh} \psb F_{\nu\rh} iD_\mu \ps				
						+	\quar i(\uf^{(6)}_{F})^{\mu\nu\rh} \psb \ga_5 F_{\nu\rh} iD_\mu \ps				
						+	\eigh (\ug^{(6)}_{F})^{\ka\la\mu\nu\rh}\lvb\ka a\lvb\la b \psb \si^{ab} F_{\nu\rh} iD_\mu \ps	+	\hc	$	\\
$	\cL_{\ps DF}^{(6)}	$	&		$	-	\half (\um^{(6)}_{DF})^{\mu\nu\rh} \psb (D_\mu F_{\nu\rh})\ps				
						-	\half i (\um^{(6)}_{5DF})^{\mu\nu\rh} \psb \ga_5 (D_\mu F_{\nu\rh})\ps			$	\\
			&	\hskip20pt	$	+	\half (\ua^{(6)}_{DF})^{\ka\mu\nu\rh}\lvb\ka a \psb \ga^a (D_\mu F_{\nu\rh})\ps				
						+	\half (\ub^{(6)}_{DF})^{\ka\mu\nu\rh} \lvb\ka a\psb \ga_5 \ga^a(D_\mu F_{\nu\rh})\ps			$	\\
			&	\hskip20pt	$	-	\quar (\uH^{(6)}_{DF})^{\ka\la\mu\nu\rh}\lvb\ka a\lvb\la b \psb \si^{ab}(D_\mu F_{\nu\rh})\ps			$	\\
$	\cL_{\ps\ps}^{(6)}	$	&		$		\uk_{SS}(\psb\ps)(\psb\ps)				
						-	\uk_{PP}(\psb\ga_5\ps)(\psb\ga_5\ps)				
						+	i\uk_{SP}(\psb\ps)(\psb\ga_5\ps)			$	\\
			&	\hskip20pt	$	-	(\uk_{SV})^\ka \lvb\ka a(\psb\ps)(\psb\ga^a\ps)				
						-	(\uk_{SA})^\ka\lvb\ka a (\psb\ps) (\psb\ga_5\ga^a\ps)				
						+	\half (\uk_{ST})^{\ka\la}\lvb\ka a\lvb\la b(\psb\ps)(\psb\si^{ab}\ps)			$	\\
			&	\hskip20pt	$	-	i(\uk_{PV})^\ka\lvb\ka a (\psb\ga_5\ps) (\psb\ga^a\ps)				
						-	i(\uk_{PA})^\ka\lvb\ka a (\psb\ga_5\ps) (\psb\ga_5\ga^a\ps)				
						+	\half i(\uk_{PT})^{\ka\la} \lvb\ka a\lvb\la b(\psb\ga_5\ps) (\psb\si^{ab}\ps)			$	\\
			&	\hskip20pt	$	+	\half (\uk_{VV})^{\ka\la}\lvb\ka a\lvb\la b(\psb\ga^a\ps)(\psb\ga^b\ps)				
						+	\half (\uk_{AA})^{\ka\la}\lvb\ka a\lvb\la b(\psb\ga_5\ga^a\ps)(\psb\ga_5\ga^b\ps)			$	\\
			&	\hskip20pt	$	-	\half (\uk_{VT})^{\ka\la\mu} \lvb\ka a\lvb\la b\lvb\mu c(\psb \ga^a \ps) (\psb \si^{bc} \ps)				
						-	\half (\uk_{AT})^{\ka\la\mu} \lvb\ka a\lvb\la b\lvb\mu c(\psb \ga_5\ga^a \ps) (\psb \si^{bc} \ps)			$	\\
			&	\hskip20pt	$	+	(\uk_{V\!A})^{\ka\la}\lvb\ka a\lvb\la b(\psb\ga^a\ps) (\psb \ga_5\ga^b \ps)				
						+	\eigh (\uk_{TT})^{\ka\la\mu\nu}\lvb\ka a\lvb\la b\lvb\mu c\lvb\nu d(\psb\si^{ab}\ps)(\psb\si^{cd}\ps)			$	\\
\hline
\hline
\end{tabular}
\end{table*}

\subsubsection{Fermions}
\label{Fermions}

Consider next the fermion sector,
including couplings to gravity and to gauge fields. 
For definiteness,
we adopt the scenario described in Sec.\ II of Ref.\ \cite{kl19}
and outlined in Sec.\ \ref{Dynamical operators} above,
with a Dirac fermion $\ps$ of mass $m_\ps$
lying in the $U$ representation of the gauge group.
The Lagrange density $\cL_\ps$
can be split into components 
containing dynamical operators with definite mass dimension $d$,
\beq
\cL_\ps =
\cL_{\ps0} +\cL^{(3)}_\ps+\cL^{(4)}_\ps+\cL^{(5)}_\ps+\cL^{(6)}_\ps+\ldots.
\eeq
The term $\cL_{\ps0}$ is the usual Dirac Lagrange density
$\cL_{\ps0} =\psb (\ivb \mu a\ga^a i D_\mu - m_\ps ) \ps/2 + \hc$,
which includes minimal couplings to gravity and to the nonabelian field. 
All components $\cL^{(d)}_\ps$ have overall mass dimension four,
with the superscript $d$ indicating the mass dimension 
of the associated dynamical operator.

Table \ref{tab:fullspinor} provides the schematic structure
of terms in $\cL_\ps^{(d)}$ of mass dimension $d\leq 6$.
The first column lists the components of $\cL_\ps$,
while the second column displays the corresponding operators. 
In this schematic context,
the backgrounds $k$ or combinations $\uk$ can all be distinct,
even within a single term.
In a given row,
terms involving fermions and covariant derivatives are listed first,
then ones involving curvature couplings and gauge couplings,
and finally ones with background derivatives $Dk$.
All indices and constants are omitted for simplicity,
and all structures involving hermitian combinations of Dirac matrices 
are represented by $\Ga$.
The factor $\hc$ appearing at the end of most expressions 
for a given component $\cL_\ps^{(d)}$ 
represents the addition of the hermitian conjugate
of all terms explicitly written for that component. 
In principle,
any background $k$ or combination $\uk$ can be taken as complex,
but its imaginary part is contained up to surface terms
in the real parts of other backgrounds.
For example,
the imaginary part of $\uk$ in $\uk\psb\Ga iD\ps$
is contained in the real part of $\uk(D\bk)\psb\Ga\ps$,
and the imaginary part of $\uk$ in $\uk R\psb\Ga iD\ps$
is contained in the real parts of 
$\uk(DR)\psb\Ga\ps$ and $\uk(D\bk)R\psb\Ga\ps$.
To avoid redundancy,
all background fields in this schematic notation
can therefore be taken as real.
Further properties of the backgrounds $k$ and the combinations $\uk$
discussed for Table \ref{tab:gravity} also hold here.

To offer more insight into the content of $\cL_\ps$,
we provide in Table \ref{tab:spinor2} 
the explicit form of all terms in $\cL_\ps$ 
with operators of mass dimension $d\leq 6$
and without background derivatives,
omitting total-derivative terms.
The structure of this table matches that of Table \ref{tab:fullspinor}.
The notation for the combinations $\uk$ appearing in this table 
is chosen to match standard conventions in the literature,
with different symbols distinguishing the spin and CPT properties
of the various dynamical operators as usual.
All backgrounds can be taken real.
The index symmetry of each background $\uk$
is understood to match the index symmetry
of the associated dynamical operator. 
The position of the indices is also chosen
to match conventions in the literature.
In particular,
the notation in $\cL^{(3)}_\ps$ and $\cL^{(4)}_{\ps D}$ 
involves covariant indices on the backgrounds 
in agreement with Ref.\ \cite{ak04}
and reducing in Minkowski spacetime to Ref.\ \cite{ck97},
while the usage of contravariant background indices 
in $\cL^{(d)}$ with $d\geq 5$
is compatible with that in published works 
discussing the nonminimal fermion sector,
such as Refs.\ \cite{dk16,kl19}.

The reader is reminded that for spontaneous breaking 
the index position on the background has no effect on the physics,
as discussed in Sec.\ \ref{Backgrounds},
while for explicit breaking the choice of index position
establishes a definition of the corresponding physical effects.
For example,
in explicit breaking the physical effects of the two backgrounds $\ol b_\mu$
and $\ol b^{\prime \mu}$ can be different,
with these two possibilities contained among others 
in $\ub_\mu$ as $\ol b_\mu$ and $g_{\mu\nu}\ol b^{\prime\nu}$.
To achieve an unambiguous statement,
experiments measuring a given background component
must therefore report results on the background 
using a specific convention.
A separate confusion can arise
because raising or lowering indices on a given background component
can introduce spurious signs and coordinate dependences.
This issue exists already in Minkowski spacetime
but can be particularly acute in gravitational experiments.
It could, for example, be problematic 
to report results for an quantity $\ol k^\mu$ defined as
$\ol k^\mu \equiv g^{\mu\nu}\ol b_\nu$
using data from an experiment sensitive to a background $\ol b_\mu$,
given that $g^{\mu\nu}$ itself is a nontrivial object.

\renewcommand\arraystretch{1.6}
\begin{table*}
\caption{
\label{tab:fullscalar}
Schematic structure of terms with $d\leq 6$ 
in the scalar Lagrange density $\cL_\ph$.}
\setlength{\tabcolsep}{4pt}
\begin{tabular}{cl}
\hline
\hline
Component & Expression \\
\hline
$	\cL^{(2)}_\ph	$	&		$		\uk	\V2		
						+	\uk(Dk)	\V1	$	\\
$	\cL^{(3)}_\ph	$	&		$		\uk	\V3		
						+	\uk	R\V1		
						+	\uk	F\V1		
						+	\uk(Dk)	\V2		
						+	\uk(Dk)(Dk)	\V1		
						+	\uk(\D2k)	\V1	$	\\
$	\cL^{(4)}_\ph	$	&		$		\uk	\V4		
						+	\uk	R\V2		
						+	\uk	F\V2		
						+	\uk	(DR)\V1		
						+	\uk	(DF)\V1		
						+	\uk(Dk)	\V3		
						+	\uk(Dk)	R\V1		
						+	\uk(Dk)	F\V1	$	\\
			&	\hskip20pt	$	+	\uk(Dk)(Dk)	\V2		
						+	\uk(\D2k)	\V2		
						+	\uk(Dk)(Dk)(Dk)	\V1		
						+	\uk(Dk)(\D2k)	\V1		
						+	\uk(\D3k)	\V1	$	\\
$	\cL^{(5)}_\ph	$	&		$		\uk	\V5		
						+	\uk	R\V3		
						+	\uk	F\V3		
						+	\uk	(DR)\V2		
						+	\uk	(DF)\V2		
						+	\uk	(\psb\Ga\ps)\V2		
						+	\uk	(\D2R)\V1		
						+	\uk	(\D2F)\V1	$	\\
			&	\hskip20pt	$	+	\uk	RR\V1		
						+	\uk	FF\V1		
						+	\uk	RF\V1		
						+	\uk	[(\psb\Ga D\ps)\V1 +\hc]		
						+	\uk(Dk)	\V4		
						+	\uk(Dk)	R\V2	$	\\
			&	\hskip20pt	$	+	\uk(Dk)	F\V2		
						+	\uk(Dk)	(DR)\V1		
						+	\uk(Dk)	(DF)\V1		
						+	\uk(Dk)(Dk)	\V3		
						+	\uk(Dk)(Dk)	R\V1	$	\\
			&	\hskip20pt	$	+	\uk(Dk)(Dk)	F\V1		
						+	\uk(\D2k)	\V3		
						+	\uk(\D2k)	R\V1		
						+	\uk(\D2k)	F\V1		
						+	\uk(Dk)(Dk)(Dk)	\V2	$	\\
			&	\hskip20pt	$	+	\uk(Dk)(\D2k)	\V2		
						+	\uk(\D3k)	\V2		
						+	\uk(Dk)(Dk)(Dk)(Dk)	\V1		
						+	\uk(Dk)(Dk)(\D2 k)	\V1	$	\\
			&	\hskip20pt	$	+	\uk(\D2 k)(\D2 k)	\V1		
						+	\uk(Dk)(\D3 k)	\V1		
						+	\uk(\D4 k)	\V1	$	\\
$	\cL^{(6)}_\ph	$	&		$		\uk	\V6		
						+	\uk	R\V4		
						+	\uk	F\V4		
						+	\uk	(DR)\V3		
						+	\uk	(DF)\V3		
						+	\uk	(\psb\Ga\ps)\V3		
						+	\uk	(\D2 R)\V2		
						+	\uk	(\D2 F)\V2	$	\\
			&	\hskip20pt	$	+	\uk	RR\V2		
						+	\uk	FF\V2		
						+	\uk	RF\V2		
						+	\uk	[(\psb\Ga D\ps)\V2 +\hc]		
						+	\uk	(\D3R)\V1		
						+	\uk	(\D3F)\V1	$	\\
			&	\hskip20pt	$	+	\uk	R(\D1R)\V1		
						+	\uk	F(\D1R)\V1		
						+	\uk	F(\D1F)\V1		
						+	\uk	R(\D1F)\V1		
						+	\uk	R(\psb\Ga\ps)\V1		
						+	\uk	F(\psb\Ga\ps)\V1	$	\\
			&	\hskip20pt	$	+	\uk	[(\psb\Ga\D2\ps)\V1+\hc]		
						+	\uk	(\ov{D\ps}\Ga D\ps)\V1		
						+	\uk(Dk)	\V5		
						+	\uk(Dk)	R\V3		
						+	\uk(Dk)	F\V3	$	\\
			&	\hskip20pt	$	+	\uk(Dk)	(DR)\V2		
						+	\uk(Dk)	(DF)\V2		
						+	\uk(Dk)	(\psb\Ga\ps)\V2		
						+	\uk(Dk)	(\D2R)\V1		
						+	\uk(Dk)	(\D2F)\V1	$	\\
			&	\hskip20pt	$	+	\uk(Dk)	RR\V1		
						+	\uk(Dk)	FF\V1		
						+	\uk(Dk)	RF\V1		
						+	\uk(Dk)	[(\psb\Ga D\ps)\V1 +\hc]		
						+	\uk(Dk)(Dk)	\V4	$	\\
			&	\hskip20pt	$	+	\uk(Dk)(Dk)	R\V2		
						+	\uk(Dk)(Dk)	F\V2		
						+	\uk(Dk)(Dk)	(DR)\V1		
						+	\uk(Dk)(Dk)	(DF)\V1		
						+	\uk(\D2k)	\V4	$	\\
			&	\hskip20pt	$	+	\uk(\D2k)	R\V2		
						+	\uk(\D2k)	F\V2		
						+	\uk(\D2k)	(DR)\V1		
						+	\uk(\D2k)	(DF)\V1		
						+	\uk(Dk)(Dk)(Dk)	\V3	$	\\
			&	\hskip20pt	$	+	\uk(Dk)(Dk)(Dk)	R\V1		
						+	\uk(Dk)(Dk)(Dk)	F\V1		
						+	\uk(Dk)(\D2k)	\V3		
						+	\uk(Dk)(\D2k)	R\V1	$	\\
			&	\hskip20pt	$	+	\uk(Dk)(\D2k)	F\V1		
						+	\uk(\D3k)	\V3		
						+	\uk(\D3k)	R\V1		
						+	\uk(\D3k)	F\V1		
						+	\uk(Dk)(Dk)(Dk)(Dk)	\V2	$	\\
			&	\hskip20pt	$	+	\uk(Dk)(Dk)(\D2 k)	\V2		
						+	\uk(\D2 k)(\D2 k)	\V2		
						+	\uk(Dk)(\D3 k)	\V2		
						+	\uk(\D4 k)	\V2	$	\\
			&	\hskip20pt	$	+	\uk(Dk)(Dk)(Dk)(Dk)(Dk)	\V1		
						+	\uk(Dk)(Dk)(Dk)(\D2k)	\V1		
						+	\uk(Dk)(Dk)(\D3k)	\V1	$	\\
			&	\hskip20pt	$	+	\uk(Dk)(\D2k)(\D2k)	\V1		
						+	\uk(Dk)(\D4k)	\V1		
						+	\uk(\D2k)(\D3k)	\V1		
						+	\uk(\D5k)	\V1	$	\\
\hline
\hline
\end{tabular}
\end{table*}

For any $d$,
the structure of the couplings involving gravity, gauge, and fermion fields 
can be written in schematic form as
\bea
\cL\pr
&\supset&
\uk(Dk)\cdots(Dk) (DR)\cdots (DR) 
\nn\\
&&\hskip5pt
\times(\tr[(DF)\cdots(DF)])\cdots (\tr[(DF)\cdots(DF)])
\nn\\
&&\hskip10pt
\times[(\ov{D\ps})(DF)\cdots(DF)\Ga(D\ps)]\cdots
\nn\\
&&\hskip20pt
\times\cdots[(\ov{D\ps})(DF)\cdots(DF)\Ga(D\ps)]+\hc,
\qquad
\label{ggf}
\eea
where $D$ here denotes any symmetric combination of derivatives.
Inspection reveals that the only terms of this form 
absent from Table \ref{tab:spinor2} involve the operators
$\tr(F)\psb\Ga\ps$, $\tr(F)\psb\Ga iD\ps$, and $\tr(DF)\psb\Ga\ps$.
These vanish for gauge groups with traceless adjoint generators,
including SU($N$),
but they reduce to terms in Table \ref{tab:spinor2} 
if the gauge group is abelian.
Otherwise,
the operators of lowest mass dimension that involve traces of powers of $F$ 
take the schematic form $\tr(FF)\psb\Ga\ps$,
which have $d=7$.
Although this value of $d$ lies outside the range of Table \ref{tab:spinor2},
the unique structure of these terms may be of interest 
for theoretical and experimental studies.
 
\renewcommand\arraystretch{1.6}
\begin{table*}
\caption{
\label{tab:scalar}
Terms with $d\leq 6$ and without background derivatives 
in the scalar Lagrange density $\cL_\ph$.}
\setlength{\tabcolsep}{4pt}
\begin{tabular}{cl}
\hline
\hline
Component & Expression \\
\hline
$	\cL_{\ph 0}	$	&		$	-	(D_\mu \ph)\da D^\mu\ph				
						\pm	\mu^2\ph\da\ph				
						-	\tfrac 1 6 \la(\ph\da\ph)^2			$	\\
$	\cL^{(2)}_\ph	$	&		$		\uk^{(2)} \ph\da\ph			$	\\
$	\cL^{(3)}_\ph	$	&		$	-	\half(\uk^{(3)})^\mu \ph\da iD_\mu\ph	+	\hc	$	\\
$	\cL^{(4)}_\ph	$	&		$		[\half(\uk^{(4)})^{\mu\nu} \ph\da  iD_{(\mu} iD_{\nu)} \ph	+	\hc]		
						-	(\uk^{(4)}_{R\ph})^{\mu\nu\rh\si} R_{\mu\nu\rh\si} \ph\da \ph				
						-	\half(\uk^{(4)}_{F\ph})^{\mu\nu} \ph\da  F_{\mu\nu} \ph				
						-	\tfrac 1 6 \uk^{(4)}_{\ph\ph} (\ph\da  \ph)^2			$	\\
$	\cL^{(5)}_\ph	$	&		$	[-	\half(\uk^{(5)})^{\mu\nu\rh} \ph\da  iD_{(\mu} iD_\nu iD_{\rh)} \ph				
						+	\half(\uk^{(5)}_{RD\ph})^{\mu\nu\rh\si\ta} R_{\nu\rh\si\ta} \ph\da  iD_\mu \ph				
						+	\quar(\uk^{(5)}_{FD\ph})^{\mu\nu\rh} \ph\da  F_{\nu\rh} iD_\mu \ph			$	\\
			&	\hskip20pt	$	+	\half(\uk^{(5)}_{\ph D\ph})^\mu (\ph\da\ph) (\ph\da iD_\mu\ph)	+	\hc]		
						+	\half(\uk^{(5)}_{DR\ph})^{\mu\nu\rh\si\ta} (D_\mu R_{\nu\rh\si\ta}) \ph\da  \ph				
						+	\half(\uk^{(5)}_{DF\ph})^{\mu\nu\rh} \ph\da  (D_\mu F_{\nu\rh}) \ph			$	\\
			&	\hskip20pt	$	+	\uk^{(5)}_{S\ph}(\psb\ps)(\ph\da\ph)				
						+	i\uk^{(5)}_{P\ph}(\psb\ga_5\ps)(\ph\da\ph)				
						+	(\uk^{(5)}_{V\ph})^\ka\lvb\ka a(\psb\ga^a\ps)(\ph\da\ph)				
						+	(\uk^{(5)}_{A\ph})^\ka\lvb\ka a(\psb\ga_5\ga^a\ps)(\ph\da\ph)			$	\\
			&	\hskip20pt	$	+	\half(\uk^{(5)}_{T\ph})^{\ka\la} \lvb\ka a \lvb\la b(\psb\si^{ab}\ps)(\ph\da\ph)			$	\\
$	\cL^{(6)}_\ph	$	&		$		[\half(\uk^{(6)})^{\mu\nu\rh\si} \ph\da iD_{(\mu}iD_{\nu}iD_{\rh}iD_{\si)}\ph				
						+	\half(\uk^{(6)}_{RDD\ph})^{\mu\nu\rh\si\ta\up}R_{\rh\si\ta\up}\ph\da  iD_{(\mu}iD_{\nu)}\ph				
						+	\quar(\uk^{(6)}_{FDD\ph})^{\mu\nu\rh\si}\ph\da F_{\rh\si} iD_{(\mu}iD_{\nu)}\ph			$	\\
			&	\hskip20pt	$	+	\half(\uk^{(6)}_{DRD\ph})^{\mu\nu\rh\si\ta\up} (D_\nu R_{\rh\si\ta\up}) \ph\da iD_\mu\ph				
						+	\quar(\uk^{(6)}_{DFD\ph})^{\mu\nu\rh\si} \ph\da(D_\nu F_{\rh\si})iD_\mu\ph				
						+	\half(\uk^{(6)}_{\ph DD\ph})^{\mu\nu} (\ph\da\ph)(\ph\da iD_{(\mu}iD_{\nu)}\ph)			$	\\
			&	\hskip20pt	$	+	\half(\uk^{(6)}_{D\ph D\ph})^{\mu\nu} (\ph\da iD_\mu\ph)(\ph\da iD_\nu\ph)				
						+	\half(\uk^{(6)}_{D\ph\da D\ph})^{\mu\nu} \big((iD_\mu\ph)\da\ph\big)(\ph\da iD_\nu\ph)				
						+	\half(\uk^{(6)}_{SD\ph})^\mu (\psb\ps)(\ph\da iD_\mu\ph)			$	\\
			&	\hskip20pt	$	+	\half i(\uk^{(6)}_{PD\ph})^\mu (\psb\ga^5\ps)(\ph\da iD_\mu\ph)				
						+	\half (\uk^{(6)}_{VD\ph})^{\ka\mu} e_{\ka a} (\psb\ga^a\ps)(\ph\da iD_\mu\ph)				
						+	\half (\uk^{(6)}_{AD\ph})^{\ka\mu} e_{\ka a} (\psb\ga_5\ga^a\ps)(\ph\da iD_\mu\ph)			$	\\
			&	\hskip20pt	$	+	\quar (\uk^{(6)}_{TD\ph})^{\ka\la\mu} e_{\ka a}e_{\la b} (\psb\si^{ab}\ps)(\ph\da iD_\mu\ph)				
						+	\half(\uk^{(6)}_{DS\ph})^\mu (\psb iD_\mu\ps)(\ph\da\ph)				
						+	\half i(\uk^{(6)}_{DP\ph})^\mu (\psb\ga^5 iD_\mu\ps)(\ph\da\ph)			$	\\
			&	\hskip20pt	$	+	\half (\uk^{(6)}_{DV\ph})^{\ka\mu} e_{\ka a} (\psb\ga^a iD_\mu\ps)(\ph\da\ph)				
						+	\half (\uk^{(6)}_{DA\ph})^{\ka\mu} e_{\ka a} (\psb\ga_5\ga^a iD_\mu\ps)(\ph\da\ph)			$	\\
			&	\hskip20pt	$	+	\quar (\uk^{(6)}_{DT\ph})^{\ka\la\mu} e_{\ka a}e_{\la b} (\psb\si^{ab} iD_\mu\ps)(\ph\da\ph)	+	\hc]		
						+	(\uk^{(6)}_{DDR\ph})^{\mu\nu\rh\si\ta\up} (D_{(\mu}D_{\nu)} R_{\rh\si\ta\up}) \ph\da\ph			$	\\
			&	\hskip20pt	$	+	\half(\uk^{(6)}_{DDF\ph})^{\mu\nu\rh\si} \ph\da (D_{(\mu}D_{\nu)}F_{\rh\si})\ph				
						+	(\uk^{(6)}_{RR\ph})^{\mu\nu\rh\si\ta\up\ch\om} R_{\mu\nu\rh\si}R_{\ta\up\ch\om} \ph\da\ph				
						+	\half(\uk^{(6)}_{RF\ph})^{\mu\nu\rh\si\ta\up} R_{\rh\si\ta\up} \ph\da F_{\mu\nu}\ph			$	\\
			&	\hskip20pt	$	+	\quar(\uk^{(6)}_{FF\ph,1})^{\mu\nu\rh\si} \ph\da F_{\mu\nu}F_{\rh\si}\ph				
						+	\quar(\uk^{(6)}_{FF\ph,2})^{\mu\nu\rh\si} \tr(F_{\mu\nu}F_{\rh\si}) \ph\da\ph				
						+	\quar(\uk^{(6)}_{R\ph\ph})^{\mu\nu\rh\si} R_{\mu\nu\rh\si} (\ph\da\ph)^2			$	\\
			&	\hskip20pt	$	+	\quar(\uk^{(6)}_{F\ph\ph})^{\mu\nu} (\ph\da\ph)(\ph\da F_{\mu\nu}\ph)				
						+	\tfrac 1 {120} \uk^{(6)}_{\ph\ph\ph} (\ph\da\ph)^3			$	\\
\hline
\hline
\end{tabular}
\end{table*}

\subsubsection{Scalars}
\label{Scalars}

In the scalar sector,
we consider first a scenario with a scalar field $\ph$ 
lying in any representation of the gauge group.
Depending on the model of interest and the gauge group involved,
the representation may be complex, real, or pseudoreal,
and it may be different from the representations
of the fermion and gauge fields.
The Lagrange density $\cL_\ph$ involves all couplings of $\ph$,
including to gravity, gauge, and fermion fields
as well as self couplings.
It can be decomposed into a sum of pieces,
\beq
\cL_\ph = \cL_{\ph0}+\cL^{(2)}_\ph+\cL^{(3)}_\ph
+\cL^{(4)}_\ph+\cL^{(5)}_\ph+\cL^{(6)}_\ph+\ldots,
\eeq
where $\cL_{\ph 0}$ is a conventional 
renormalizable Lagrange density for $\ph$
that is invariant under gauge transformations,
local Lorentz transformations, and diffeomorphisms.
Typically,
$\cL_{\ph 0}$ incorporates a kinetic term
quadratic in the covariant derivative of $\ph$
along with a potential term
involving a globally stable polynomial in $\ph$.
According to its sign,
the term quadratic in $\ph$ may represent a scalar mass 
or may trigger spontaneous breaking of the gauge symmetry.
Where compatible with gauge invariance,
$\cL_{\ph 0}$ also includes conventional couplings to other sectors.
The components $\cL^{(d)}_\ph$
have overall mass dimension four,
and they represent corrections in the effective field theory 
involving dynamical operators of mass dimension $d$.

In constructing possible contributions to the terms $\cL^{(d)}_\ph$
in the effective field theory,
we seek gauge-invariant operators
that are polynomials in $\ph$ and in covariant derivatives of $\ph$.
Any given scalar operator of this type 
can be characterized by its mass dimension $d$
and assigned to a corresponding set $V^{(d)}$.
For instance,
$V^{(1)}$ contains the operator $\ph$
and any independent conjugates,
$V^{(2)}$
contains 
$\ph\ph$, $D_\mu\ph$, and various conjugates,
and
$V^{(3)}$
contains
$\ph\ph\ph$,
$\ph D_\mu \ph$,
$D_\mu D_\nu \ph$,
and various conjugates.
Note that any given operator in the set $V^{(d)}$
may carry up to $d-1$ spacetime indices.
The properties of operators quadratic in $\ph$ at any $d$
are studied in Ref.\ \cite{ek18}.

Table \ref{tab:fullscalar}
displays the terms in $\cL^{(d)}_\ph$ with $d\leq 6$ in schematic form.
For each $\cL^{(d)}_\ph$ with given $d$ shown in the first column,
the possible corresponding operator structures 
are presented in the second column.
In this notation,
a factor $V^{(d)}$ represents any element 
of the associated set of scalar operators,
so an expression containing $V^{(d)}$
is a compact description of several distinct terms.
Operators without background derivatives are listed first, 
with pure-scalar terms preceding ones 
containing gravity, gauge, and fermion couplings.
These are followed by operators involving 
increasing numbers of background derivatives $Dk$.
Indices and all numerical or dimensional factors are omitted,
and any appearance of a hermitian combination of Dirac matrices
is represented by $\Ga$.
A few terms appear with hermitian conjugates,
which are denoted by $\hc$ in the table.
All appearances of the backgrounds $k$ and combinations $\uk$
can be distinct,
even within a single term,
and the backgrounds are assumed real.
Since all terms are required to be gauge invariant,
the choice of representation for $\ph$ restricts 
the appearance of some operator structures.
For example,
the term $\uk (Dk) V^{(1)}$ in $\cL_\ph^{(2)}$
is gauge invariant only when the scalar is in the singlet
representation. 

To offer further insight,
the explicit terms with $d\leq 6$
in the Lagrange density for a special model
are provided in Table \ref{tab:scalar}.
For simplicity and definiteness,
the complex scalar $\ph$ in this model 
is assumed to be in the same $U$ representation 
of the gauge group as the fermion field and the gauge field strength,
only terms invariant by virtue of the combination $U^{\dagger} U = I$ 
are listed,
and total-derivative terms are omitted.
This excludes,
for instance, 
singlet terms arising from products of $U$ representations
that are specific to a particular gauge group.
For example,
when the gauge group is SU(3)
and the scalar is in the octet representation,
the Lagrange density $\cL_\ph^{(6)}$ contains an additional term
\beq
\cL_\ph^{(6)} \supset
(\uk^{(6)}_{D\ph DD\ph})^{\mu\nu\rh}\ep_{abc}\ph_a (iD_\mu\ph)_b 
(iD_{(\mu}iD_{\nu)}\ph)_c + \hc,
\eeq
where $a, b, c$ are the gauge indices in the adjoint representation.

The format of Table \ref{tab:scalar}
parallels that of Table \ref{tab:fullscalar},
and the contents of the former can be deduced
by inspection of the latter.
In Table \ref{tab:scalar},
the mass parameter $\pm \mu^2$ in the conventional piece $\cL_{\ph 0}$ 
of the Lagrange density can take either sign,
with the lower sign triggering spontaneous breaking of the gauge symmetry.
In other terms,
each background $\uk$ is understood to have index symmetry
inherited from the index symmetry of the corresponding operator.
All backgrounds can be assumed real.
The results in the table emphasize
the rapid growth with $d$ of terms involving scalar couplings
already visible in the schematic notation of Table \ref{tab:fullscalar},
even in the context of a specific and comparatively simple model.

\section{Applications}
\label{Applications}

In this section,
we consider some applications of the formalism
to several cases of practical relevance.
These include effective field theories extending 
Einstein-Maxwell theory,
GR coupled to the SM,
and conventional models involving only scalar coupling constants.

\renewcommand\arraystretch{1.6}
\begin{table*}
\caption{
\label{tab:photon}
Terms with $d\leq 6$ 
in the photon sector of the Einstein-Maxwell effective field theory.}
\setlength{\tabcolsep}{4pt}
\begin{tabular}{cl}
\hline
\hline
Component & Expression \\
\hline
$	\cL_{A0}	$	&		$	-	\quar F_{\mu\nu}F^{\mu\nu}	$	\\
$	\cL_A^{(1)}	$	&		$	-	(\uk_A)_{\mu} A^\mu	$	\\
$	\cL_A^{(3)}	$	&		$		\half(\uk_{AF})^{\ka} \ep_{\ka\la\mu\nu} A^\la F^{\mu\nu} 		
						+	\half(\uk^{(3)}_{DF})^{\al\mu\nu}D_\al F_{\mu\nu}	$	\\
$	\cL_A^{(4)}	$	&		$	-	\quar (\uk_F)_{\ka\la\mu\nu} F^{\ka\la} F^{\mu\nu}		
						-	\half(\uk^{(4)}_{DF})^{\al\be\mu\nu} D_{(\al}D_{\be)} F_{\mu\nu}		
						-	\half(\uk_{RF})^{\al\be\ga\de\mu\nu}R_{\al\be\ga\de} F_{\mu\nu}	$	\\
$	\cL_A^{(5)}	$	&		$		\quar (\uk^{(5)}_D)^{\al\ka\la\mu\nu}F_{\ka\la} D_\al F_{\mu\nu}		
						+	\half(\uk^{(5)}_{DF})^{\al\be\ga\mu\nu} D_{(\al}D_\be D_{\ga)} F_{\mu\nu}	$	\\
			&	\hskip20pt	$	+	\half (\uk_{RDF}^{(5)})^{\al\be\ga\de\ep\mu\nu}R_{\al\be\ga\de} D_\ep F_{\mu\nu}		
						+	\half(\uk_{DRF}^{(5)})^{\al\be\ga\de\ep\mu\nu}(D_\ep R_{\al\be\ga\de}) F_{\mu\nu}	$	\\
$	\cL_A^{(6)}	$	&		$	-	\quar (\uk_D^{(6)})^{\al\be\ka\la\mu\nu} F_{\ka\la} D_{(\al}D_{\be)} F_{\mu\nu}		
						-	\half(\uk^{(6)}_{DF})^{\al\be\ga\de\mu\nu} D_{(\al}D_\be D_\ga D_{\de)} F_{\mu\nu}		
						-	\quar(\uk^{(6)}_{DFDF})^{\al\be\ka\la\mu\nu} (D_\al F_{\ka\la}) (D_\be F_{\mu\nu})	$	\\
			&	\hskip20pt	$	-	\tfrac1{12}(\uk_{F}^{(6)})^{\ka\la\mu\nu\rh\si} F_{\ka\la}F_{\mu\nu}F_{\rh\si}		
						-	\quar (\uk_{RFF}^{(6)})^{\al\be\ga\de\ka\la\mu\nu}R_{\al\be\ga\de} F_{\ka\la}F_{\mu\nu}		
						-	\half(\uk_{RDDF}^{(6)})^{\al\be\ga\de\ep\ze\mu\nu}R_{\al\be\ga\de} D_{(\ep}D_{\ze)}F_{\mu\nu}	$	\\
			&	\hskip20pt	$	-	\half(\uk_{DDRF}^{(6)})^{\al\be\ga\de\ep\ze\mu\nu}(D_{(\ep}D_{\ze)}R_{\al\be\ga\de}) F_{\mu\nu}		
						-	\half(\uk_{DRDF}^{(6)})^{\al\be\ga\de\ep\ze\mu\nu}(D_\ep R_{\al\be\ga\de}) D_{\ze}F_{\mu\nu}	$	\\
			&	\hskip20pt	$	-	\half(\uk_{RRF}^{(6)})^{\abcd1\abcd2\mu\nu}R_{\abcd1}R_{\abcd2} F_{\mu\nu}	$	\\
\hline
\hline
\end{tabular}
\end{table*}

\renewcommand\arraystretch{1.6}
\begin{table*}
\caption{
\label{tab:realscalar}
Terms with $d\leq 6$ for an uncharged scalar coupled 
to the Einstein-Maxwell effective field theory.}
\setlength{\tabcolsep}{4pt}
\begin{tabular}{cl}
\hline
\hline
Component & Expression \\
\hline
$	\cL_{\ph 0}	$	&		$	-	\half(D_\mu \ph) D^\mu\ph		
						\pm	\half\mu^2\ph^2		
						+	\tfrac{1}{6}g\ph^3		
						-	\tfrac{1}{24}\la\ph^4	$	\\
$	\cL^{(2)}_\ph	$	&		$		\half\uk^{(2)} \ph^2	$	\\
$	\cL^{(3)}_\ph	$	&		$		\tfrac{1}{6} \uk^{(3)}_\ph \ph^3		
						+	(\uk^{(3)}_{R\ph})^{\mu\nu\rh\si}R_{\mu\nu\rh\si}\ph		
						+	\half(\uk^{(3)}_{F\ph})^{\mu\nu} F_{\mu\nu}\ph	$	\\
$	\cL^{(4)}_\ph	$	&		$		\half(\uk^{(4)})^{\mu\nu} \ph  iD_{(\mu} iD_{\nu)} \ph		
						-	\tfrac{1}{24} \uk^{(4)}_{\ph} \ph^4		
						-	\half(\uk^{(4)}_{R\ph})^{\mu\nu\rh\si} R_{\mu\nu\rh\si} \ph^2		
						-	\quar(\uk^{(4)}_{F\ph})^{\mu\nu} F_{\mu\nu} \ph^2		
						-	(\uk^{(4)}_{DR\ph})^{\mu\nu\rh\si\ta} (D_\mu R_{\nu\rh\si\ta})\ph	$	\\
			&	\hskip20pt	$	-	\half(\uk^{(4)}_{DF\ph})^{\mu\nu\rh} (D_\mu F_{\nu\rh})\ph	$	\\
$	\cL^{(5)}_\ph	$	&		$		\tfrac{1}{6}(\uk^{(5)}_{DD\ph})^{\mu\nu} \ph^2 iD_{(\mu}iD_{\nu)}\ph		
						+	\tfrac 1 6 (\uk^{(5)}_{D\ph D\ph})^{\mu\nu} \ph (iD_\mu\ph)(iD_\nu\ph)		
						+	\tfrac{1}{120}\uk^{(5)}_\ph \ph^5		
						+	\tfrac{1}{6}(\uk^{(5)}_{R\ph})^{\mu\nu\rh\si} R_{\mu\nu\rh\si}\ph^3	$	\\
			&	\hskip20pt	$	+	\half(\uk^{(5)}_{DR\ph})^{\mu\nu\rh\si\ta} (D_\mu R_{\nu\rh\si\ta}) \ph^2		
						+	(\uk^{(5)}_{DDR\ph})^{\mu\nu\rh\si\ta\up}(D_{(\mu}D_{\nu)}R_{\rh\si\ta\up}) \ph		
						+	\tfrac{1}{12} (\uk^{(5)}_{F\ph})^{\mu\nu} F_{\mu\nu}\ph^3	$	\\
			&	\hskip20pt	$	+	\quar(\uk^{(5)}_{DF\ph})^{\mu\nu\rh} (D_\mu F_{\nu\rh}) \ph^2		
						+	\half(\uk^{(5)}_{DDF\ph})^{\mu\nu\rh\si} (D_{(\mu}D_{\nu)}F_{\rh\si})\ph		
						+	(\uk^{(5)}_{RR\ph})^{\mu\nu\rh\si\ta\up\ch\om} R_{\mu\nu\rh\si}R_{\ta\up\ch\om}\ph	$	\\
			&	\hskip20pt	$	+	\quar(\uk^{(5)}_{FF\ph})^{\mu\nu\rh\si} F_{\mu\nu}F_{\rh\si}\ph		
						+	\half(\uk^{(5)}_{RF\ph})^{\mu\nu\rh\si\ta\up} R_{\mu\nu\rh\si}F_{\ta\up}\ph	$	\\
$	\cL^{(6)}_\ph	$	&		$		\half(\uk^{(6)})^{\mu\nu\rh\si} \ph iD_{(\mu}iD_{\nu}iD_{\rh}iD_{\si)}\ph		
						+	\tfrac{1}{24} (\uk^{(6)}_{DD\ph})^{\mu\nu} \ph^3 iD_{(\mu}iD_{\nu)}\ph		
						+	\tfrac{1}{24} (\uk^{(6)}_{D\ph D\ph})^{\mu\nu} \ph^2(iD_\mu\ph)(iD_\nu\ph)		
						+	\tfrac{1}{720} \uk^{(6)}_\ph \ph^6	$	\\
			&	\hskip20pt	$	+	\tfrac{1}{24} (\uk^{(6)}_{R\ph})^{\mu\nu\rh\si} R_{\mu\nu\rh\si}\ph^4		
						+	\tfrac{1}{48} (\uk^{(6)}_{F\ph})^{\mu\nu} F_{\mu\nu}\ph^4		
						+	\half(\uk^{(6)}_{RDD\ph})^{\mu\nu\rh\si\ta\up}R_{\rh\si\ta\up}\ph  iD_{(\mu}iD_{\nu)}\ph	$	\\
			&	\hskip20pt	$	+	\quar(\uk^{(6)}_{FDD\ph})^{\mu\nu\rh\si}F_{\rh\si}\ph iD_{(\mu}iD_{\nu)}\ph		
						+	\tfrac 1 {6} (\uk^{(6)}_{DR\ph})^{\mu\nu\rh\si\ta} (D_\mu R_{\nu\rh\si\ta}) \ph^3		
						+	\tfrac 1 {12} (\uk^{(6)}_{DF\ph})^{\mu\nu\rh} (D_\mu F_{\nu\rh}) \ph^3	$	\\
			&	\hskip20pt	$	+	\half(\uk^{(6)}_{DDR\ph})^{\mu\nu\rh\si\ta\up} (D_{(\mu}D_{\nu)} R_{\rh\si\ta\up}) \ph^2		
						+	\quar(\uk^{(6)}_{DDF\ph})^{\mu\nu\rh\si} (D_{(\mu}D_{\nu)} F_{\rh\si}) \ph^2		
						+	\half(\uk^{(6)}_{RR\ph})^{\mu\nu\rh\si\ta\up\ch\om} R_{\mu\nu\rh\si}R_{\ta\up\ch\om} \ph^2	$	\\
			&	\hskip20pt	$	+	\quar(\uk^{(6)}_{RF\ph})^{\mu\nu\rh\si\ta\up} R_{\mu\nu\rh\si} F_{\ta\up}\ph^2		
						+	\eigh(\uk^{(6)}_{FF\ph})^{\mu\nu\rh\si} F_{\mu\nu}F_{\rh\si}\ph^2		
						+	(\uk^{(6)}_{DDDR\ph})^{\mu\nu\rh\si\ta\up\ch} (D_{(\mu}D_{\nu}D_{\rh)}R_{\si\ta\up\ch})\ph	$	\\
			&	\hskip20pt	$	+	\half (\uk^{(6)}_{DDDF\ph})^{\mu\nu\rh\si\ta} (D_{(\mu}D_{\nu}D_{\rh)}F_{\si\ta})\ph		
						+	(\uk^{(6)}_{RDR\ph})^{\la\mu\nu\rh\si\ta\up\ch\om} R_{\mu\nu\rh\si}(D_\la R_{\ta\up\ch\om})\ph	$	\\
			&	\hskip20pt	$	+	\half(\uk^{(6)}_{RDF\ph})^{\mu\nu\rh\si\ta\up\ch} R_{\nu\rh\si\ta}(D_\mu F_{\up\ch})\ph		
						+	\half(\uk^{(6)}_{DRF\ph})^{\mu\nu\rh\si\ta\up\ch} (D_\mu R_{\nu\rh\si\ta})F_{\up\ch}\ph		
						+	\quar(\uk^{(6)}_{FDF\ph})^{\mu\nu\rh\si\ta} F_{\nu\rh}(D_\mu F_{\si\ta})\ph	$	\\
\hline
\hline
\end{tabular}
\end{table*}

\subsection{Einstein-Maxwell theories}
\label{sec:abelian}

As a first application,
we consider effective field theories
based on the usual Einstein-Maxwell theory
coupling gravity to electrodynamics
and possibly additional matter fields.
In any specific scenario,
the Einstein-Maxwell Lagrange density $\cL_{\rm EM}$
can be separated as
\beq
\cL_{\rm EM} = \cL_g + \cL_A + \ldots,
\eeq
where $\cL_g$ contains terms 
in the pure-gravity and background sector,
$\cL_A$ describes the photon sector
including gravitational couplings,
and the ellipsis indicates any other component Lagrange densities 
for couplings to fermions and scalars. 

The general features of the Lagrange density $\cL_g$ for the gravity sector 
are treated in Sec.\ \ref{Pure-gravity sector}.
The discussion applies directly to the Einstein-Maxwell case.
In particular,
the explicit forms of the terms in $\cL_g$ 
without background derivatives
are listed in Table \ref{tab:curvature}.

For the photon sector,
the terms in $\cL_A$ can be extracted
from the results obtained for nonabelian gauge theories
in Sec.\ \ref{sec:gauge},
and in particular from Table \ref{tab:gauge}.
However,
the abelian nature of the gauge group
simplifies the structure of certain terms,
while some terms that are distinct in the nonabelian case
merge in the abelian limit. 
For clarity,
we display in Table \ref{tab:photon} 
all terms in $\cL_A$ with operators of mass dimension $d\leq6$
without background derivatives.
The first column lists the components $\cL_A^{(d)}$ of $\cL_A$,
while the second shows the terms they incorporate.
In this table,
each background can be taken to be real
and to have index symmetry inherited 
from that of the corresponding operator.
The index positions and the labels on the backgrounds
match existing conventions in the literature.

\renewcommand\arraystretch{1.6}
\begin{table*}
\caption{
\label{tab:SME-gauge}
Terms with $d\leq 6$ 
in the gravity and gauge sectors $\cL_{\rm gravity}$ and $\cL_{\rm gauge}$.}
\setlength{\tabcolsep}{4pt}
\begin{tabular}{cl}
\hline
\hline
Component & Expression \\
\hline
$	\cL_{{\rm gravity},0}+\cL_{{\rm gauge},0}	$	&		$		\tfrac{1}{2\ka}(R-2\La)		
						-	\half\tr(G_{\mu\nu}G^{\mu\nu})		
						-	\half\tr(W_{\mu\nu}W^{\mu\nu})		
						-	\quar B_{\mu\nu}B^{\mu\nu}	$	\\
$	\cL^{(2)}_{\rm gravity}	$	&		$		\tfrac{1}{2\ka}\uk^{(2)}	$	\\
$	\cL^{(3)}_{\rm gravity}	$	&		$		\tfrac{1}{2\ka}(\uk^{(3)}_{\Ga})^{\mu} \LC \al\mu\al	$	\\
$	\cL^{(4)}_{\rm gravity}	$	&		$		\tfrac{1}{2\ka}(\uk^{(4)}_R)^{\al\be\ga\de}R_{\al\be\ga\de}	$	\\
$	\cL^{(5)}_{\rm gravity}	$	&		$		\tfrac{1}{2\ka}\big[(\uk_D^{(5)})^{\al\be\ga\de\ka}D_\ka R_{\al\be\ga\de}		
						+	(\uk^{(5)}_{{\rm CS}, 1})_\ka \ep^{\ka\la\mu\nu}\et_{ac}\et_{bd}(\nsc\la a b \prt_\mu \nsc \nu c d		
						+	\tfrac 23 \nsc \la a b \nsc \mu c e \llusc \nu e d)	$	\\
			&	\hskip20pt	$	+	(\uk^{(5)}_{{\rm CS}, 2})_\ka \ep^{\ka\la\mu\nu}\ep_{abcd}(\nsc\la a b \prt_\mu \nsc \nu c d		
						+	\tfrac 23 \nsc \la a b \nsc \mu c e \llusc \nu e d)\big]	$	\\
$	\cL^{(6)}_{\rm gravity}	$	&		$		\tfrac{1}{2\ka}\big[(\uk_D^{(6)})^{\al\be\ga\de\ka\la}D_{(\ka}D_{\la)}R_{\al\be\ga\de}		
						+	(\uk_R^{(6)})^{\abcd1\abcd2}R_{\abcd1}R_{\abcd2}\big]	$	\\
$	\cL_{{\rm gauge}}^{(1)}	$	&		$		(\uk_0)_\ka B_\ka	$	\\
$	\cL_{{\rm gauge}}^{(3)}	$	&		$		(\uk_3)_{\ka} \ep^{\ka\la\mu\nu} \tr (G_\la G_{\mu\nu} +\tfrac{2}{3} ig_3 G_\la G_\mu G_\nu)		
						+	(\uk_2)_{\ka} \ep^{\ka\la\mu\nu} \tr (W_\la W_{\mu\nu} +\tfrac{2}{3} ig W_\la W_\mu W_\nu)		
						+	(\uk_1)_\ka \ep^{\ka\la\mu\nu} B_\la B_{\mu\nu}	$	\\
			&	\hskip20pt	$	+	\half(\uk^{(3)}_{DB})^{\al\ka\la} D_\al B_{\ka\la}	$	\\
$	\cL_{{\rm gauge}}^{(4)}	$	&		$	-	\half (\uk_G)_{\ka\la\mu\nu} \tr(G^{\ka\la} G^{\mu\nu})		
						-	\half (\uk_W)_{\ka\la\mu\nu} \tr(W^{\ka\la} W^{\mu\nu})		
						-	\quar (\uk_B)_{\ka\la\mu\nu} B^{\ka\la}B^{\mu\nu}		
						-	\half (\uk_{RB})^{\ka\la\mu\nu\rh\si}R_{\ka\la\mu\nu}B_{\rh\si}	$	\\
			&	\hskip20pt	$	-	\half(\uk^{(4)}_{DB})^{\al\be\ka\la} D_{(\al}D_{\be)} B_{\ka\la}	$	\\
$	\cL_{{\rm gauge}}^{(5)}	$	&		$	-	\half (\uk^{(5)}_{DG})^{\al\ka\la\mu\nu} \tr (G_{\ka\la} D_\al G_{\mu\nu})		
						-	\half (\uk^{(5)}_{DW})^{\al\ka\la\mu\nu} \tr (W_{\ka\la} D_\al W_{\mu\nu})		
						-	\quar(\uk^{(5)}_{DB})^{\al\ka\la\mu\nu} B_{\ka\la}D_\al B_{\mu\nu}	$	\\
			&	\hskip20pt	$	-	\half(\uk^{(5)}_{DRB})^{\al\ka\la\mu\nu\rh\si} (D_\al R_{\ka\la\mu\nu})B_{\rh\si}		
						-	\half(\uk^{(5)}_{RDB})^{\al\ka\la\mu\nu\rh\si} R_{\ka\la\mu\nu}D_\al B_{\rh\si}		
						-	\half(\uk^{(5)}_{DB})^{\al\be\ga\ka\la} D_{(\al}D_{\be}D_{\ga)} B_{\ka\la}	$	\\
$	\cL_{{\rm gauge}}^{(6)}	$	&		$	-	\half (\uk_{DDG}^{(6)})^{\al\be\ka\la\mu\nu} \tr(G_{\ka\la} D_{(\al}D_{\be)} G_{\mu\nu})		
						-	\half (\uk_{DDW}^{(6)})^{\al\be\ka\la\mu\nu} \tr(W_{\ka\la} D_{(\al}D_{\be)} W_{\mu\nu})	$	\\
			&	\hskip20pt	$	-	\quar(\uk^{(6)}_{DDB})^{\al\be\ka\la\mu\nu} B_{\ka\la}D_{(\al}D_{\be)}B_{\mu\nu}		
						-	\half (\uk_{DGDG}^{(6)})^{\al\be\ka\la\mu\nu} \tr\big((D_\al G_{\ka\la})(D_\be G_{\mu\nu})\big)	$	\\
			&	\hskip20pt	$	-	\half (\uk_{DWDW}^{(6)})^{\al\be\ka\la\mu\nu} \tr\big((D_\al W_{\ka\la})(D_\be W_{\mu\nu})\big)		
						-	\quar(\uk^{(6)}_{DBDB})^{\al\be\ka\la\mu\nu} (D_\al B_{\ka\la})(D_\be B_{\mu\nu})	$	\\
			&	\hskip20pt	$	-	\tfrac1{12}\big[(\uk_{G}^{(6)})^{\ka\la\mu\nu\rh\si}\tr(G_{\ka\la}G_{\mu\nu}G_{\rh\si})+\hc\big]		
						-	\tfrac1{12}\big[(\uk_{W}^{(6)})^{\ka\la\mu\nu\rh\si}\tr(W_{\ka\la}W_{\mu\nu}W_{\rh\si})+\hc\big]	$	\\
			&	\hskip20pt	$	-	\half(\uk_{RDDB}^{(6)})^{\al\be\ga\de\ep\ze\mu\nu}R_{\al\be\ga\de} D_{(\ep}D_{\ze)}B_{\mu\nu}		
						-	\half(\uk_{DDRB}^{(6)})^{\al\be\ga\de\ep\ze\mu\nu}(D_{(\ep}D_{\ze)}R_{\al\be\ga\de}) B_{\mu\nu}	$	\\
			&	\hskip20pt	$	-	\half(\uk_{DRDB}^{(6)})^{\al\be\ga\de\ep\ze\mu\nu}(D_\ep R_{\al\be\ga\de}) D_{\ze}B_{\mu\nu}		
						-	\tfrac1{12}(\uk_{B}^{(6)})^{\ka\la\mu\nu\rh\si}B_{\ka\la}B_{\mu\nu}B_{\rh\si}	$	\\
			&	\hskip20pt	$	-	\quar(\uk^{(6)}_{BGG})^{\ka\la\mu\nu\rh\si}B_{\ka\la}\tr(G_{\mu\nu}G_{\rh\si})		
						-	\quar(\uk^{(6)}_{BWW})^{\ka\la\mu\nu\rh\si}B_{\ka\la}\tr(W_{\mu\nu}W_{\rh\si})	$	\\
			&	\hskip20pt	$	-	\half (\uk_{RGG}^{(6)})^{\al\be\ga\de\ka\la\mu\nu}R_{\al\be\ga\de}\tr(G_{\ka\la}G_{\mu\nu})		
						-	\half (\uk_{RWW}^{(6)})^{\al\be\ga\de\ka\la\mu\nu}R_{\al\be\ga\de}\tr(W_{\ka\la}W_{\mu\nu})	$	\\
			&	\hskip20pt	$	-	\quar (\uk_{RBB}^{(6)})^{\al\be\ga\de\ka\la\mu\nu}R_{\al\be\ga\de}B_{\ka\la}B_{\mu\nu}		
						-	\half(\uk_{RRB}^{(6)})^{\abcd1\abcd2\ka\la}R_{\abcd1}R_{\abcd2}B_{\ka\la}	$	\\
\hline
\hline
\end{tabular}
\end{table*}

Augmenting the Einstein-Maxwell theory with matter fields 
implies a corresponding extension of the effective field theory.
The inclusion of a single Dirac field,
which may be charged under the U(1) gauge group,
yields an additional component $\cL_\ps$ in $\cL_{\rm EM}$.
The terms in $\cL_\ps$ 
involving operators of mass dimension $d\leq6$
and without background derivatives
can be extracted directly from Table \ref{tab:spinor2} 
in Sec.\ \ref{Fermions},
with the gauge field strength $F_{\mu\nu}$ taken as abelian.
Similarly,
adding a complex scalar with U(1) charge 
generates a component $\cL_\ph$ in $\cL_{\rm EM}$.
All contributions to $\cL_\ph$ 
containing operators of mass dimension $d\leq6$
without background derivatives
are given by Table \ref{tab:scalar} in Sec.\ \ref{Scalars},
with $F_{\mu\nu}$ understood to be abelian.

The inclusion of an uncharged real scalar
is more involved because additional terms can be constructed. 
For this case,
all terms in $\cL_\ph$ with operators of mass dimension $d\leq6$ 
are presented in Table \ref{tab:realscalar}.
The format of the table matches that of Table \ref{tab:scalar},
and most of the comments in Sec.\ \ref{Scalars} apply.
In some of the terms unique to this scenario, 
a factor of $i$ has been inserted to keep all backgrounds
and combinations real.

\renewcommand\arraystretch{1.5}
\begin{table*}
\caption{
\label{tab:leptonquark}
Terms with $d\leq 6$ in the fermion sector
$\cL_{\rm fermion}=\cL_{\rm lepton}+\cL_{\rm quark}$.}
\setlength{\tabcolsep}{3pt}
\begin{tabular}{cl}
\hline
\hline
Component & Expression \\
\hline
$	\cL_{{\rm fermion},0}	$	&		$		\half \ivb\mu a \Lb_A\ga^a iD_\mu L_A 				
						+	\half \ivb\mu a \Rb_A\ga^a iD_\mu R_A 				
						+	\half \ivb\mu a \Qb_A\ga^a iD_\mu Q_A 				
						+	\half \ivb\mu a \Ub_A\ga^a iD_\mu U_A 				
						+	\half \ivb\mu a \Db_A\ga^a iD_\mu D_A 	+	\hc	$	\\
$	\cL^{(3)}_{\rm lepton}	$	&		$	-	(\ua_L)_{\ka AB} \ivb\ka a \Lb_A\ga^a L_B				
						-	(\ua_R)_{\ka AB} \ivb\ka a \Rb_A\ga^a R_B			$	\\
$	\cL^{(4)}_{\rm {\rm lepton}}	$	&		$	-	\half (\uc_L)_{\ka\mu AB} \ivb\ka a \ivb\mu b e^{\nu b} \Lb_A\ga^a i D_\nu L_B				
						-	\half (\uc_R)_{\ka\mu AB} \ivb\ka a \ivb\mu b e^{\nu b} \Rb_A\ga^a i D_\nu R_B	+	\hc	$	\\
$	\cL_{{\rm lepton},D}^{(5)}	$	&		$		\half (\ua^{(5)}_L)^{\ka\mu\nu}_{AB} \lvb\ka a \Lb_A \ga^a iD_{(\mu} iD_{\nu)} L_B				
						+	\half (\ua^{(5)}_R)^{\ka\mu\nu}_{AB} \lvb\ka a \Rb_A \ga^a iD_{(\mu} iD_{\nu)} R_B	+	\hc	$	\\
$	\cL_{{\rm lepton},R}^{(5)}	$	&		$		(\ua^{(5)}_{R,L})^{\ka\mu\nu\rh\si}_{AB}\lvb\ka a R_{\mu\nu\rh\si} \Lb_A \ga^a L_B				
						+	(\ua^{(5)}_{R,R})^{\ka\mu\nu\rh\si}_{AB}\lvb\ka a R_{\mu\nu\rh\si} \Rb_A \ga^a R_B			$	\\
$	\cL_{{\rm lepton},W}^{(5)}	$	&		$		\half (\ua^{(5)}_{W,L})^{\ka\mu\nu}_{AB}\lvb\ka a \Lb_A W_{\mu\nu} \ga^a L_B			$	\\
$	\cL_{{\rm lepton},B}^{(5)}	$	&		$		\half (\ua^{(5)}_{B,L})^{\ka\mu\nu}_{AB}\lvb\ka a B_{\mu\nu} \Lb_A \ga^a L_B				
						+	\half (\ua^{(5)}_{B,R})^{\ka\mu\nu}_{AB}\lvb\ka a B_{\mu\nu} \Rb_A \ga^a R_B			$	\\
$	\cL^{(6)}_{{\rm lepton},D}	$	&		$	-	\half (\uc^{(6)}_L)^{\ka\mu\nu\rh}_{AB} \lvb\ka a \Lb_A\ga^a iD_{(\mu}iD_\nu iD_{\rh)} L_B				
						-	\half (\uc^{(6)}_R)^{\ka\mu\nu\rh}_{AB} \lvb\ka a \Rb_A\ga^a iD_{(\mu}iD_\nu iD_{\rh)} R_B	+	\hc	$	\\
$	\cL^{(6)}_{{\rm lepton},R}	$	&		$	-	\half (\uc^{(6)}_{R,L})^{\ka\mu\nu\rh\si\ta}_{AB} \lvb\ka a R_{\nu\rh\si\ta}\Lb_A\ga^a iD_\mu L_B				
						-	\half (\uc^{(6)}_{R,R})^{\ka\mu\nu\rh\si\ta}_{AB} \lvb\ka a R_{\nu\rh\si\ta}\Rb_A\ga^a iD_\mu R_B	+	\hc	$	\\
$	\cL^{(6)}_{{\rm lepton},DR}	$	&		$		(\uc^{(6)}_{DR,L})^{\ka\mu\nu\rh\si\ta}_{AB} \lvb\ka a (D_\mu R_{\nu\rh\si\ta})\Lb_A\ga^a L_B				
						+	(\uc^{(6)}_{DR,R})^{\ka\mu\nu\rh\si\ta}_{AB} \lvb\ka a (D_\mu R_{\nu\rh\si\ta})\Rb_A\ga^a R_B			$	\\
$	\cL^{(6)}_{{\rm lepton},W}	$	&		$	-	\quar (\uc^{(6)}_{W,L})^{\ka\mu\nu\rh}_{AB} \lvb\ka a \Lb_A\ga^a W_{\nu\rh} iD_\mu L_B	+	\hc	$	\\
$	\cL^{(6)}_{{\rm lepton},DW}	$	&		$		\half(\uc^{(6)}_{DW,L})^{\ka\mu\nu\rh}_{AB} \lvb\ka a \Lb_A\ga^a (D_\mu W_{\nu\rh}) L_B			$	\\
$	\cL^{(6)}_{{\rm lepton},B}	$	&		$	-	\quar (\uc^{(6)}_{B,L})^{\ka\mu\nu\rh}_{AB} \lvb\ka a B_{\nu\rh}\Lb_A\ga^a iD_\mu L_B				
						-	\quar (\uc^{(6)}_{B,R})^{\ka\mu\nu\rh}_{AB} \lvb\ka a B_{\nu\rh}\Rb_A\ga^a iD_\mu R_B	+	\hc	$	\\
$	\cL^{(6)}_{{\rm lepton},DB}	$	&		$		\half(\uc^{(6)}_{DB,L})^{\ka\mu\nu\rh}_{AB} \lvb\ka a (D_\mu B_{\nu\rh})\Lb_A\ga^a L_B				
						+	\half(\uc^{(6)}_{DB,R})^{\ka\mu\nu\rh}_{AB} \lvb\ka a (D_\mu B_{\nu\rh})\Rb_A\ga^a R_B			$	\\
$	\cL^{(6)}_{{\rm lepton},2}	$	&		$		\half(\uk^{(6)}_{LL})^{\ka\la}_{ABCD} \lvb\ka a \lvb\la b (\Lb_A\ga^a L_B)(\Lb_C\ga^b L_D)				
						+	\half(\uk^{(6)}_{RR})^{\ka\la}_{ABCD} \lvb\ka a \lvb\la b (\Rb_A\ga^a R_B)(\Rb_C\ga^b R_D)			$	\\
			&	\hskip20pt	$	+	(\uk^{(6)}_{LR})^{\ka\la}_{ABCD} \lvb\ka a \lvb\la b (\Lb_A\ga^a L_B)(\Rb_C\ga^b R_D)			$	\\
$	\cL^{(3)}_{\rm quark}	$	&		$	-	(\ua_Q)_{\ka AB} \ivb\ka a \Qb_A\ga^a Q_B				
						-	(\ua_U)_{\ka AB} \ivb\ka a \Ub_A\ga^a U_B				
						-	(\ua_D)_{\ka AB} \ivb\ka a \Db_A\ga^a D_B			$	\\
$	\cL^{(4)}_{\rm quark}	$	&		$	-	\half (\uc_Q)_{\ka\mu AB} \ivb\ka a \Qb_A\ga^a i D^\mu Q_B				
						-	\half (\uc_U)_{\ka\mu AB} \ivb\ka a \Ub_A\ga^a i D^\mu U_B				
						-	\half (\uc_D)_{\ka\mu AB} \ivb\ka a \Db_A\ga^a i D^\mu D_B	+	\hc	$	\\
$	\cL_{{\rm quark},D}^{(5)}	$	&		$		\half (\ua^{(5)}_Q)^{\ka\mu\nu}_{AB} \lvb\ka a \Qb_A \ga^a iD_{(\mu} iD_{\nu)} Q_B				
						+	\half (\ua^{(5)}_U)^{\ka\mu\nu}_{AB} \lvb\ka a \Ub_A \ga^a iD_{(\mu} iD_{\nu)} U_B				
						+	\half (\ua^{(5)}_D)^{\ka\mu\nu}_{AB} \lvb\ka a \Db_A \ga^a iD_{(\mu} iD_{\nu)} D_B	+	\hc	$	\\
$	\cL_{{\rm quark},R}^{(5)}	$	&		$		(\ua^{(5)}_{R,Q})^{\ka\mu\nu\rh\si}_{AB}\lvb\ka a R_{\mu\nu\rh\si} \Qb_A \ga^a Q_B				
						+	(\ua^{(5)}_{R,U})^{\ka\mu\nu\rh\si}_{AB}\lvb\ka a R_{\mu\nu\rh\si} \Ub_A \ga^a U_B				
						+	(\ua^{(5)}_{R,D})^{\ka\mu\nu\rh\si}_{AB}\lvb\ka a R_{\mu\nu\rh\si} \Db_A \ga^a D_B			$	\\
$	\cL_{{\rm quark},G}^{(5)}	$	&		$		\half (\ua^{(5)}_{G,Q})^{\ka\mu\nu}_{AB}\lvb\ka a \Qb_A G_{\mu\nu} \ga^a Q_B				
						+	\half (\ua^{(5)}_{G,U})^{\ka\mu\nu}_{AB}\lvb\ka a \Ub_A G_{\mu\nu} \ga^a U_B				
						+	\half (\ua^{(5)}_{G,D})^{\ka\mu\nu}_{AB}\lvb\ka a \Db_A G_{\mu\nu} \ga^a D_B			$	\\
$	\cL_{{\rm quark},W}^{(5)}	$	&		$		\half (\ua^{(5)}_{W,Q})^{\ka\mu\nu}_{AB}\lvb\ka a \Qb_A W_{\mu\nu} \ga^a Q_B			$	\\
$	\cL_{{\rm quark},B}^{(5)}	$	&		$		\half (\ua^{(5)}_{B,Q})^{\ka\mu\nu}_{AB}\lvb\ka a B_{\mu\nu} \Qb_A \ga^a Q_B				
						+	\half (\ua^{(5)}_{B,U})^{\ka\mu\nu}_{AB}\lvb\ka a B_{\mu\nu} \Ub_A \ga^a U_B				
						+	\half (\ua^{(5)}_{B,D})^{\ka\mu\nu}_{AB}\lvb\ka a B_{\mu\nu} \Db_A \ga^a D_B			$	\\
$	\cL^{(6)}_{{\rm quark},D}	$	&		$	-	\half (\uc^{(6)}_Q)^{\ka\mu\nu\rh}_{AB} \lvb\ka a \Qb_A\ga^a iD_{(\mu}iD_\nu iD_{\rh)} Q_B				
						-	\half (\uc^{(6)}_U)^{\ka\mu\nu\rh}_{AB} \lvb\ka a \Ub_A\ga^a iD_{(\mu}iD_\nu iD_{\rh)} U_B			$	\\
			&	\hskip20pt	$	-	\half (\uc^{(6)}_D)^{\ka\mu\nu\rh}_{AB} \lvb\ka a \Db_A\ga^a iD_{(\mu}iD_\nu iD_{\rh)} D_B	+	\hc	$	\\
$	\cL^{(6)}_{{\rm quark},R}	$	&		$	-	\half (\uc^{(6)}_{R,Q})^{\ka\mu\nu\rh\si\ta}_{AB} \lvb\ka a R_{\nu\rh\si\ta}\Qb_A\ga^a iD_\mu Q_B				
						-	\half (\uc^{(6)}_{R,U})^{\ka\mu\nu\rh\si\ta}_{AB} \lvb\ka a R_{\nu\rh\si\ta}\Ub_A\ga^a iD_\mu U_B			$	\\
			&	\hskip20pt	$	-	\half (\uc^{(6)}_{R,D})^{\ka\mu\nu\rh\si\ta}_{AB} \lvb\ka a R_{\nu\rh\si\ta}\Db_A\ga^a iD_\mu D_B	+	\hc	$	\\
$	\cL^{(6)}_{{\rm quark},DR}	$	&		$		(\uc^{(6)}_{DR,Q})^{\ka\mu\nu\rh\si\ta}_{AB} \lvb\ka a (D_\mu R_{\nu\rh\si\ta})\Qb_A\ga^a Q_B				
						+	(\uc^{(6)}_{DR,U})^{\ka\mu\nu\rh\si\ta}_{AB} \lvb\ka a (D_\mu R_{\nu\rh\si\ta})\Ub_A\ga^a U_B			$	\\
			&	\hskip20pt	$	+	(\uc^{(6)}_{DR,D})^{\ka\mu\nu\rh\si\ta}_{AB} \lvb\ka a (D_\mu R_{\nu\rh\si\ta})\Db_A\ga^a D_B			$	\\
$	\cL^{(6)}_{{\rm quark},G}	$	&		$	-	\quar (\uc^{(6)}_{G,Q})^{\ka\mu\nu\rh}_{AB} \lvb\ka a \Qb_A\ga^a G_{\nu\rh} iD_\mu Q_B				
						-	\quar (\uc^{(6)}_{G,U})^{\ka\mu\nu\rh}_{AB} \lvb\ka a \Ub_A\ga^a G_{\nu\rh} iD_\mu U_B				
						-	\quar (\uc^{(6)}_{G,D})^{\ka\mu\nu\rh}_{AB} \lvb\ka a \Db_A\ga^a G_{\nu\rh} iD_\mu D_B	+	\hc	$	\\
$	\cL^{(6)}_{{\rm quark},DG}	$	&		$		\half(\uc^{(6)}_{DG,Q})^{\ka\mu\nu\rh}_{AB} \lvb\ka a \Qb_A\ga^a (D_\mu G_{\nu\rh}) Q_B				
						+	\half(\uc^{(6)}_{DG,U})^{\ka\mu\nu\rh}_{AB} \lvb\ka a \Ub_A\ga^a (D_\mu G_{\nu\rh}) U_B				
						+	\half(\uc^{(6)}_{DG,D})^{\ka\mu\nu\rh}_{AB} \lvb\ka a \Db_A\ga^a (D_\mu G_{\nu\rh}) D_B			$	\\
$	\cL^{(6)}_{{\rm quark},W}	$	&		$	-	\quar (\uc^{(6)}_{W,Q})^{\ka\mu\nu\rh}_{AB} \lvb\ka a \Qb_A\ga^a W_{\nu\rh} iD_\mu Q_B	+	\hc	$	\\
$	\cL^{(6)}_{{\rm quark},DW}	$	&		$		\half(\uc^{(6)}_{DW,Q})^{\ka\mu\nu\rh}_{AB} \lvb\ka a \Qb_A\ga^a (D_\mu W_{\nu\rh}) Q_B			$	\\
$	\cL^{(6)}_{{\rm quark},B}	$	&		$	-	\quar (\uc^{(6)}_{B,Q})^{\ka\mu\nu\rh}_{AB} \lvb\ka a B_{\nu\rh}\Qb_A\ga^a iD_\mu Q_B				
						-	\quar (\uc^{(6)}_{B,U})^{\ka\mu\nu\rh}_{AB} \lvb\ka a B_{\nu\rh}\Ub_A\ga^a iD_\mu U_B				
						-	\quar (\uc^{(6)}_{B,D})^{\ka\mu\nu\rh}_{AB} \lvb\ka a B_{\nu\rh}\Db_A\ga^a iD_\mu D_B	+	\hc	$	\\
$	\cL^{(6)}_{{\rm quark},DB}	$	&		$		\half(\uc^{(6)}_{DB,Q})^{\ka\mu\nu\rh}_{AB} \lvb\ka a (D_\mu B_{\nu\rh})\Qb_A\ga^a Q_B				
						+	\half(\uc^{(6)}_{DB,U})^{\ka\mu\nu\rh}_{AB} \lvb\ka a (D_\mu B_{\nu\rh})\Ub_A\ga^a U_B				
						+	\half(\uc^{(6)}_{DB,D})^{\ka\mu\nu\rh}_{AB} \lvb\ka a (D_\mu B_{\nu\rh})\Db_A\ga^a D_B			$	\\
$	\cL^{(6)}_{{\rm quark},2}	$	&		$		\half(\uk^{(6)}_{QQ})^{\ka\la}_{ABCD} \lvb\ka a \lvb\la b (\Qb_A\ga^a Q_B)(\Qb_C\ga^b Q_D)				
						+	\half(\uk^{(6)}_{UU})^{\ka\la}_{ABCD} \lvb\ka a \lvb\la b (\Ub_A\ga^a U_B)(\Ub_C\ga^b U_D)			$	\\
			&	\hskip20pt	$	+	\half(\uk^{(6)}_{DD})^{\ka\la}_{ABCD} \lvb\ka a \lvb\la b (\Db_A\ga^a D_B)(\Db_C\ga^b D_D)				
						+	(\uk^{(6)}_{QU})^{\ka\la}_{ABCD} \lvb\ka a \lvb\la b (\Qb_A\ga^a Q_B)(\Ub_C\ga^b U_D)			$	\\
			&	\hskip20pt	$	+	(\uk^{(6)}_{QD})^{\ka\la}_{ABCD} \lvb\ka a \lvb\la b (\Qb_A\ga^a Q_B)(\Db_C\ga^b D_D)				
						+	(\uk^{(6)}_{UD})^{\ka\la}_{ABCD} \lvb\ka a \lvb\la b (\Ub_A\ga^a U_B)(\Db_C\ga^b D_D)			$	\\
$	\cL^{(6)}_{{\rm quark,lepton}}	$	&		$		(\uk^{(6)}_{QL})^{\ka\la}_{ABCD} \lvb\ka a \lvb\la b (\Qb_A\ga^a Q_B)(\Lb_C\ga^b L_D)				
						+	(\uk^{(6)}_{QR})^{\ka\la}_{ABCD} \lvb\ka a \lvb\la b (\Qb_A\ga^a Q_B)(\Rb_C\ga^b R_D)			$	\\
			&	\hskip20pt	$	+	(\uk^{(6)}_{UL})^{\ka\la}_{ABCD} \lvb\ka a \lvb\la b (\Ub_A\ga^a U_B)(\Lb_C\ga^b L_D)				
						+	(\uk^{(6)}_{UR})^{\ka\la}_{ABCD} \lvb\ka a \lvb\la b (\Ub_A\ga^a U_B)(\Rb_C\ga^b R_D)			$	\\
			&	\hskip20pt	$	+	(\uk^{(6)}_{DL})^{\ka\la}_{ABCD} \lvb\ka a \lvb\la b (\Db_A\ga^a D_B)(\Lb_C\ga^b L_D)				
						+	(\uk^{(6)}_{DR})^{\ka\la}_{ABCD} \lvb\ka a \lvb\la b (\Db_A\ga^a D_B)(\Rb_C\ga^b R_D)			$	\\
\hline
\hline
\end{tabular}
\end{table*}

\renewcommand\arraystretch{1.6}
\begin{table*}
\caption{
\label{tab:higgs}
Terms with $d\leq 6$ in the Higgs sector $\cL_{\rm Higgs}$.}
\setlength{\tabcolsep}{2pt}
\begin{tabular}{cl}
\hline
\hline
Component & Expression \\
\hline
$	\cL_{{\rm Higgs},0}	$	&		$	-	(D_\mu \ph)\da D^\mu\ph				
						+	\mu^2\ph\da\ph				
						-	\tfrac 1 6 \la(\ph\da\ph)^2			$	\\
$	\cL^{(2)}_{\rm Higgs}	$	&		$		\uk^{(2)}_\ph \ph\da\ph			$	\\
$	\cL^{(3)}_{\rm Higgs}	$	&		$		(\uk_\ph)^\mu \ph\da iD_\mu\ph	+	\hc	$	\\
$	\cL^{(4)}_{\rm Higgs}	$	&		$		[\half(\uk_{\ph\ph})^{\mu\nu} (D_\mu\ph)\da D_\nu\ph	+	\hc]		
						-	(\uk_{\ph R})^{\mu\nu\rh\si} R_{\mu\nu\rh\si} \ph\da \ph				
						-	\half(\uk_{\ph W})^{\mu\nu} \ph\da  W_{\mu\nu} \ph				
						-	\half(\uk_{\ph B})^{\mu\nu} B_{\mu\nu} \ph\da\ph				
						-	\tfrac 1 6 \uk^{(4)}_{\ph\ph} (\ph\da \ph)^2			$	\\
$	\cL^{(5)}_{\rm Higgs}	$	&		$		[\half(\uk^{(5)})^{\mu\nu\rh} \ph\da  iD_{(\mu} iD_\nu iD_{\rh)} \ph				
						+	\half(\uk^{(5)}_{RD\ph})^{\mu\nu\rh\si\ta} R_{\nu\rh\si\ta} \ph\da  iD_\mu \ph				
						+	\quar(\uk^{(5)}_{WD\ph})^{\mu\nu\rh} \ph\da  W_{\nu\rh} iD_\mu \ph				
						+	\quar(\uk^{(5)}_{BD\ph})^{\mu\nu\rh} B_{\nu\rh}\ph\da iD_\mu \ph			$	\\
			&	\hskip20pt	$	+	\half(\uk^{(5)}_{\ph D\ph})^\mu (\ph\da\ph) (\ph\da iD_\mu\ph)	+	\hc]		
						+	(\uk^{(5)}_{DR\ph})^{\mu\nu\rh\si\ta} (D_\mu R_{\nu\rh\si\ta}) \ph\da  \ph				
						+	\half(\uk^{(5)}_{DW\ph})^{\mu\nu\rh} \ph\da  (D_\mu W_{\nu\rh}) \ph			$	\\
			&	\hskip20pt	$	+	\half(\uk^{(5)}_{DB\ph})^{\mu\nu\rh}(D_\mu B_{\nu\rh})\ph\da \ph			$	\\
$	\cL^{(6)}_{\rm Higgs}	$	&		$		[\half(\uk^{(6)})^{\mu\nu\rh\si} \ph\da iD_{(\mu}iD_{\nu}iD_{\rh}iD_{\si)}\ph				
						+	\half(\uk^{(6)}_{RDD\ph})^{\mu\nu\rh\si\ta\up}R_{\rh\si\ta\up}\ph\da  iD_{(\mu}iD_{\nu)}\ph				
						+	\quar(\uk^{(6)}_{WDD\ph})^{\mu\nu\rh\si}\ph\da W_{\rh\si}iD_{(\mu}iD_{\nu)}\ph			$	\\
			&	\hskip20pt	$	+	\quar(\uk^{(6)}_{BDD\ph})^{\mu\nu\rh\si}B_{\rh\si}\ph\da iD_{(\mu}iD_{\nu)}\ph				
						+	\half(\uk^{(6)}_{DRD\ph})^{\mu\nu\rh\si\ta\up} (D_\nu R_{\rh\si\ta\up}) \ph\da iD_\mu\ph				
						+	\quar(\uk^{(6)}_{DWD\ph})^{\mu\nu\rh\si} \ph\da(D_\nu W_{\rh\si})iD_\mu\ph			$	\\
			&	\hskip20pt	$	+	\quar(\uk^{(6)}_{DBD\ph})^{\mu\nu\rh\si} (D_\nu B_{\rh\si})\ph\da iD_\mu\ph				
						+	\half(\uk^{(6)}_{\ph DD\ph})^{\mu\nu} (\ph\da\ph)(\ph\da iD_{(\mu}iD_{\nu)}\ph)				
						+	\half(\uk^{(6)}_{D\ph D\ph})^{\mu\nu} (\ph\da iD_\mu\ph)(\ph\da iD_\nu\ph)			$	\\
			&	\hskip20pt	$	+	\half(\uk^{(6)}_{D\ph\da D\ph})^{\mu\nu} \big((iD_\mu\ph)\da\ph\big)(\ph\da iD_\nu\ph)	+	\hc]		
						+	(\uk^{(6)}_{DDR\ph})^{\mu\nu\rh\si\ta\up} (D_{(\mu}D_{\nu)} R_{\rh\si\ta\up}) \ph\da\ph			$	\\
			&	\hskip20pt	$	+	\half(\uk^{(6)}_{DDW\ph})^{\mu\nu\rh\si} \ph\da (D_{(\mu}D_{\nu)}W_{\rh\si})\ph				
						+	\half(\uk^{(6)}_{DDB\ph})^{\mu\nu\rh\si} (D_{(\mu}D_{\nu)}B_{\rh\si}) \ph\da\ph				
						+	(\uk^{(6)}_{RR\ph})^{\mu\nu\rh\si\ta\up\ch\om} R_{\mu\nu\rh\si}R_{\ta\up\ch\om} \ph\da\ph			$	\\
			&	\hskip20pt	$	+	(\uk^{(6)}_{RW\ph})^{\mu\nu\rh\si\ta\up} R_{\rh\si\ta\up} \ph\da W_{\mu\nu}\ph				
						+	\half(\uk^{(6)}_{RB\ph})^{\mu\nu\rh\si\ta\up} R_{\rh\si\ta\up}B_{\mu\nu} \ph\da\ph				
						+	\quar(\uk^{(6)}_{WW\ph,1})^{\mu\nu\rh\si} \ph\da W_{\mu\nu}W_{\rh\si}\ph			$	\\
			&	\hskip20pt	$	+	\quar(\uk^{(6)}_{WW\ph,2})^{\mu\nu\rh\si} \tr(W_{\mu\nu}W_{\rh\si}) \ph\da\ph				
						+	\quar(\uk^{(6)}_{BB\ph})^{\mu\nu\rh\si} B_{\mu\nu}B_{\rh\si} \ph\da\ph				
						+	\quar(\uk^{(6)}_{BW\ph})^{\mu\nu\rh\si} B_{\mu\nu}\ph\da W_{\rh\si} \ph			$	\\
			&	\hskip20pt	$	+	\quar(\uk^{(6)}_{R\ph\ph})^{\mu\nu\rh\si} R_{\mu\nu\rh\si} (\ph\da\ph)^2				
						+	\quar(\uk^{(6)}_{W\ph\ph})^{\mu\nu} (\ph\da\ph)(\ph\da W_{\mu\nu}\ph)				
						+	\quar(\uk^{(6)}_{B\ph\ph})^{\mu\nu}B_{\mu\nu} (\ph\da\ph)^2				
						+	\tfrac 1 {120} \uk^{(6)}_{\ph\ph\ph} (\ph\da\ph)^3			$	\\
\hline
\hline
\end{tabular}
\end{table*}

\subsection{General Relativity and the Standard Model}
\label{sec:SME}

Next,
we turn attention to the effective field theory
constructed from GR coupled to the SM,
the gravitational Standard-Model Extension (SME).
The framework for this scenario is described in Ref.\ \cite{ak04},
which explicitly presents all minimal operators 
allowing for nonzero torsion in Riemann-Cartan spacetime.
Here,
we extend these results to include operators
of mass dimension $d\leq6$,
restricting attention to the zero-torsion limit
and to terms without background derivatives.

The Lagrange density $\cL_{\rm SME}$ for this theory
can be decomposed into a sum of terms assigned to sectors
for each type of field,
\bea
\cL_{\rm SME} &=&
\cL_{\rm gravity} 
+ \cL_{\rm gauge} 
+\cL_{\rm lepton} 
+ \cL_{\rm quark} 
\nonumber\\ &&
+ \cL_{\rm Higgs} 
+ \cL_{\rm Yukawa} . 
\eea
The gauge invariance of the corresponding action 
is the SM group SU(3)$\times$SU(2)$\times$U(1).
In what follows,
we present explicit terms in the effective field theory 
prior to the usual spontaneous breaking of the gauge group 
to SU(3)$\times$U(1),
adopting the conventions of Ref.\ \cite{ak04}.
In particular,
the covariant derivative $D_\mu$ 
is both spacetime and SU(3)$\times$SU(2)$\times$U(1) covariant.
The coupling constants for SU(3), SU(2), and U(1)
are denoted by $g_3$, $g$, and $g^\prime$,
respectively.
They are related to the charge $q$ for the electromagnetic U(1) group
and to the angle $\th_W$
via $q = g \sin \th_W = g^\prime \cos \th_W$.

The piece $\cL_{\rm gravity}$
can be identified with the Lagrange density $\cL_g$ 
discussed in Sec.\ \ref{Pure-gravity sector},
$\cL_{\rm gravity}=\cL_g$,
which has terms listed in Table \ref{tab:curvature}.
In the gauge sector,
which includes gauge couplings to gravitational fields,
the gauge fields for the strong interaction
are described by the hermitian SU(3) adjoint matrix $G_\mu$,
while the SU(2) gauge fields form a hermitian adjoint matrix $W_\mu$.
The hermitian singlet hypercharge gauge field is denoted $B_\mu$. 
The field strengths corresponding to these gauge fields are 
$G_{\mu\nu}$, $W_{\mu\nu}$, and $B_{\mu\nu}$.
The corresponding Lagrange density
$\cL_{\rm gauge}$
can be constructed using the techniques and results
presented in Sec.\ \ref{sec:gauge}
and Table \ref{tab:gauge},
along with the U(1) limit provided in Table \ref{tab:photon}.

Table \ref{tab:SME-gauge} displays the terms
in $\cL_{\rm gravity}$ and $\cL_{\rm gauge}$
without background derivatives
and containing operators of mass dimension $d\leq 6$. 
The first row of the table
provides the usual Lagrange density in the absence of backgrounds.
For the remainder of the table,
the first column lists the component
$\cL_{\rm gravity}^{(d)}$ or $\cL_{\rm gauge}^{(d)}$,
while the second column contains the corresponding terms.
The notation for the background combinations $\uk$
is introduced in Sec.\ \ref{Effective gravity}.
Each combination $\uk$ is real 
and has index symmetry derived 
from that of its associated dynamical operator.
For the terms with $d=3$ and 4,
the index positions and identifying subscripts on the combinations $\uk$ 
match those adopted in Ref.\ \cite{ak04}. 
For generality,
we include in this table 
all terms with dynamical operators that are total derivatives.
Note that 
the meaning of the trace operation $\tr(\cO)$ 
for an operator $\cO$ in the gauge sector 
depends on the context,
being taken in the appropriate representation for $\cO$. 
Since $\tr{G_\mu}$ and $\tr{W_\mu}$ both vanish,
only traces linear in the U(1) gauge fields appear in the table.

\renewcommand\arraystretch{1.6}
\begin{table*}
\caption{
\label{tab:yukawa}
Terms with $d\leq 6$ in the Yukawa sector $\cL_{\rm Yukawa}$.}
\setlength{\tabcolsep}{2pt}
\begin{tabular}{cl}
\hline
\hline
Component & Expression \\
\hline
$	\cL_{{\rm Yukawa},0}	$	&		$	-	(G_L)_{AB}\Lb_A\ph R_B				
						-	(G_U)_{AB}\Qb_A\ph^c U_B				
						-	(G_D)_{AB}\Qb_A\ph D_B	+	\hc	$	\\
$	\cL^{(4)}_{\rm Yukawa}	$	&		$	-	(\uG^{(4)}_L)_{AB}\Lb_A\ph R_B				
						-	(\uG^{(4)}_U)_{AB}\Qb_A\ph^c U_B				
						-	(\uG^{(4)}_D)_{AB}\Qb_A\ph D_B				
						-	\half(\uH_L)_{\ka\la AB}\ivb\ka a\ivb\la b\Lb_A\ph\si^{ab}R_B			$	\\
			&	\hskip20pt	$	-	\half(\uH_U)_{\ka\la AB}\ivb\ka a\ivb\la b\Qb_A\ph\si^{ab}U_B				
						-	\half(\uH_D)_{\ka\la AB}\ivb\ka a\ivb\la b\Qb_A\ph\si^{ab}D_B	+	\hc	$	\\
$	\cL^{(5)}_{\rm Yukawa}	$	&		$	-	(\uG^{(5)}_L)_{AB}^\mu \Lb_A\ph iD_\mu R_B				
						-	(\uG^{(5)}_U)_{AB}^\mu \Qb_A\ph^c iD_\mu U_B				
						-	(\uG^{(5)}_D)_{AB}^\mu \Qb_A\ph iD_\mu D_B				
						-	\half(\uH^{(5)}_L)^{\ka\la\mu}_{AB}\lvb\ka a\lvb\la b\Lb_A\ph\si^{ab}iD_\mu R_B			$	\\
			&	\hskip20pt	$	-	\half(\uH^{(5)}_U)^{\ka\la\mu}_{AB}\lvb\ka a\lvb\la b\Qb_A\ph^c \si^{ab}iD_\mu U_B				
						-	\half(\uH^{(5)}_D)^{\ka\la\mu}_{AB}\lvb\ka a\lvb\la b\Ub_A\ph\si^{ab}iD_\mu D_B				
						-	(\uG^{(5)}_{D\ph,L})_{AB}^\mu \Lb_A(iD_\mu\ph) R_B			$	\\
			&	\hskip20pt	$	-	(\uG^{(5)}_{D\ph,U})_{AB}^\mu \Qb_A(iD_\mu\ph)^c U_B				
						-	(\uG^{(5)}_{D\ph,D})_{AB}^\mu \Qb_A(iD_\mu\ph) D_B				
						-	\half(\uH^{(5)}_{D\ph,L})^{\ka\la\mu}_{AB}\lvb\ka a\lvb\la b\Lb_A(iD_\mu\ph)\si^{ab} R_B			$	\\
			&	\hskip20pt	$	-	\half(\uH^{(5)}_{D\ph,U})^{\ka\la\mu}_{AB}\lvb\ka a\lvb\la b\Qb_A(iD_\mu\ph)^c \si^{ab}U_B				
						-	\half(\uH^{(5)}_{D\ph,D})^{\ka\la\mu}_{AB}\lvb\ka a\lvb\la b\Ub_A(iD_\mu\ph) \si^{ab} D_B	+	\hc	$	\\
$	\cL^{(5)}_{{\rm Yukawa},\psi\ph}	$	&		$		(\uk^{(5)}_{L\ph})^\ka_{AB} \lvb\ka a(\Lb_A \ga^a L_B)(\ph\da\ph)				
						+	(\uk^{(5)}_{R\ph})^\ka_{AB} \lvb\ka a(\Rb_A \ga^a R_B)(\ph\da\ph)				
						+	(\uk^{(5)}_{Q\ph})^\ka_{AB} \lvb\ka a(\Qb_A \ga^a Q_B)(\ph\da\ph)			$	\\
			&	\hskip20pt	$	+	(\uk^{(5)}_{U\ph})^\ka_{AB} \lvb\ka a(\Ub_A \ga^a U_B)(\ph\da\ph)				
						+	(\uk^{(5)}_{D\ph})^\ka_{AB} \lvb\ka a(\Db_A \ga^a D_B)(\ph\da\ph)	+	\hc	$	\\
$	\cL^{(6)}_{{\rm Yukawa},D}	$	&		$	-	(\uG^{(6)}_{L})_{AB}^{\mu\nu} \Lb_A\ph iD_{(\mu}iD_{\nu)} R_B				
						-	(\uG^{(6)}_{U})_{AB}^{\mu\nu} \Qb_A\ph^c iD_{(\mu}iD_{\nu)} U_B				
						-	(\uG^{(6)}_{D})_{AB}^{\mu\nu} \Qb_A\ph iD_{(\mu}iD_{\nu)} D_B			$	\\
			&	\hskip20pt	$	-	\half(\uH^{(6)}_L)^{\ka\la\mu\nu}_{AB}\lvb\ka a\lvb\la b\Lb_A\ph\si^{ab}iD_{(\mu}iD_{\nu)} R_B				
						-	\half(\uH^{(6)}_U)^{\ka\la\mu\nu}_{AB}\lvb\ka a\lvb\la b\Qb_A\ph^c\si^{ab}iD_{(\mu}iD_{\nu)} U_B			$	\\
			&	\hskip20pt	$	-	\half(\uH^{(6)}_D)^{\ka\la\mu\nu}_{AB}\lvb\ka a\lvb\la b\Qb_A\ph\si^{ab}iD_{(\mu}iD_{\nu)} D_B				
						-	(\uG^{(6)}_{DD\ph,L})_{AB}^{\mu\nu} \Lb_A (iD_{(\mu}iD_{\nu)}\ph) R_B			$	\\
			&	\hskip20pt	$	-	(\uG^{(6)}_{DD\ph,U})_{AB}^{\mu\nu} \Qb_A (iD_{(\mu}iD_{\nu)}\ph)^c U_B				
						-	(\uG^{(6)}_{DD\ph,D})_{AB}^{\mu\nu} \Qb_A (iD_{(\mu}iD_{\nu)}\ph) D_B			$	\\
			&	\hskip20pt	$	-	\half(\uH^{(6)}_{DD\ph,L})^{\ka\la\mu\nu}_{AB}\lvb\ka a\lvb\la b\Lb_A(iD_{(\mu}iD_{\nu)}\ph)\si^{ab} R_B				
						-	\half(\uH^{(6)}_{DD\ph,U})^{\ka\la\mu\nu}_{AB}\lvb\ka a\lvb\la b\Qb_A(iD_{(\mu}iD_{\nu)}\ph)^c\si^{ab} U_B			$	\\
			&	\hskip20pt	$	-	\half(\uH^{(6)}_{DD\ph,D})^{\ka\la\mu\nu}_{AB}\lvb\ka a\lvb\la b\Qb_A(iD_{(\mu}iD_{\nu)}\ph)\si^{ab} D_B				
						-	(\uG^{(6)}_{D\ph,L})_{AB}^{\mu\nu} \Lb_A (iD_{\mu}\ph) iD_\nu R_B			$	\\
			&	\hskip20pt	$	-	(\uG^{(6)}_{D\ph,U})_{AB}^{\mu\nu} \Qb_A (iD_{\mu}\ph)^c iD_\nu U_B				
						-	(\uG^{(6)}_{D\ph,D})_{AB}^{\mu\nu} \Qb_A (iD_{\mu}\ph) iD_\nu D_B				
						-	\half(\uH^{(6)}_{D\ph,L})^{\ka\la\mu\nu}_{AB}\lvb\ka a\lvb\la b\Lb_A(iD_{\mu}\ph)\si^{ab} iD_\nu R_B			$	\\
			&	\hskip20pt	$	-	\half(\uH^{(6)}_{D\ph,U})^{\ka\la\mu\nu}_{AB}\lvb\ka a\lvb\la b\Qb_A(iD_{\mu}\ph)^c\si^{ab} iD_\nu U_B				
						-	\half(\uH^{(6)}_{D\ph,D})^{\ka\la\mu\nu}_{AB}\lvb\ka a\lvb\la b\Qb_A(iD_{\mu}\ph)\si^{ab} iD_\nu D_B	+	\hc	$	\\
$	\cL^{(6)}_{{\rm Yukawa},R}	$	&		$	-	(\uG^{(6)}_{R,L})_{AB}^{\mu\nu\rh\si}R_{\mu\nu\rh\si} \Lb_A\ph R_B				
						-	(\uG^{(6)}_{R,U})_{AB}^{\mu\nu\rh\si} R_{\mu\nu\rh\si}\Qb_A\ph^c U_B				
						-	(\uG^{(6)}_{R,D})_{AB}^{\mu\nu\rh\si} R_{\mu\nu\rh\si}\Qb_A\ph D_B			$	\\
			&	\hskip20pt	$	-	\half(\uH^{(6)}_L)^{\ka\la\mu\nu\rh\si}_{AB}\lvb\ka a\lvb\la b R_{\mu\nu\rh\si}\Lb_A\ph\si^{ab}R_B				
						-	\half(\uH^{(6)}_U)^{\ka\la\mu\nu\rh\si}_{AB}\lvb\ka a\lvb\la b R_{\mu\nu\rh\si}\Qb_A\ph^c\si^{ab}U_B			$	\\
			&	\hskip20pt	$	-	\half(\uH^{(6)}_D)^{\ka\la\mu\nu\rh\si}_{AB}\lvb\ka a\lvb\la b R_{\mu\nu\rh\si}\Qb_A\ph\si^{ab}D_B	+	\hc	$	\\
$	\cL^{(6)}_{{\rm Yukawa},G}	$	&		$	-	(\uG^{(6)}_{G,U})_{AB}^{\mu\nu} \Qb_A G_{\mu\nu}\ph^c U_B				
						-	(\uG^{(6)}_{G,D})_{AB}^{\mu\nu} \Qb_A G_{\mu\nu}\ph D_B				
						-	\half(\uH^{(6)}_{G,U})^{\ka\la\mu\nu}_{AB}\lvb\ka a\lvb\la b \Qb_A G_{\mu\nu}\ph^c\si^{ab}U_B			$	\\
			&	\hskip20pt	$	-	\half(\uH^{(6)}_{G,D})^{\ka\la\mu\nu}_{AB}\lvb\ka a\lvb\la b \Qb_A G_{\mu\nu}\ph\si^{ab}D_B	+	\hc	$	\\
$	\cL^{(6)}_{{\rm Yukawa},W}	$	&		$	-	(\uG^{(6)}_{W,L})_{AB}^{\mu\nu} \Lb_A W_{\mu\nu}\ph R_B				
						-	(\uG^{(6)}_{W,U})_{AB}^{\mu\nu} \Qb_A W_{\mu\nu}\ph^c U_B				
						-	(\uG^{(6)}_{W,D})_{AB}^{\mu\nu} \Qb_A W_{\mu\nu}\ph D_B			$	\\
			&	\hskip20pt	$	-	\half(\uH^{(6)}_{W,L})^{\ka\la\mu\nu}_{AB}\lvb\ka a\lvb\la b \Lb_A W_{\mu\nu}\ph\si^{ab}R_B				
						-	\half(\uH^{(6)}_{W,U})^{\ka\la\mu\nu}_{AB}\lvb\ka a\lvb\la b \Qb_A W_{\mu\nu}\ph^c\si^{ab}U_B			$	\\
			&	\hskip20pt	$	-	\half(\uH^{(6)}_{W,D})^{\ka\la\mu\nu}_{AB}\lvb\ka a\lvb\la b \Qb_A W_{\mu\nu}\ph\si^{ab}D_B	+	\hc	$	\\
$	\cL^{(6)}_{{\rm Yukawa},B}	$	&		$	-	(\uG^{(6)}_{B,L})_{AB}^{\mu\nu}B_{\mu\nu} \Lb_A \ph R_B				
						-	(\uG^{(6)}_{B,U})_{AB}^{\mu\nu} B_{\mu\nu}\Qb_A \ph^c U_B				
						-	(\uG^{(6)}_{B,D})_{AB}^{\mu\nu} B_{\mu\nu}\Qb_A \ph D_B			$	\\
			&	\hskip20pt	$	-	\half(\uH^{(6)}_{B,L})^{\ka\la\mu\nu}_{AB}\lvb\ka a\lvb\la b B_{\mu\nu}\Lb_A \ph\si^{ab}R_B				
						-	\half(\uH^{(6)}_{B,U})^{\ka\la\mu\nu}_{AB}\lvb\ka a\lvb\la b B_{\mu\nu}\Qb_A \ph^c\si^{ab}U_B			$	\\
			&	\hskip20pt	$	-	\half(\uH^{(6)}_{B,D})^{\ka\la\mu\nu}_{AB}\lvb\ka a\lvb\la b B_{\mu\nu}\Qb_A \ph\si^{ab}D_B	+	\hc	$	\\
$	\cL^{(6)}_{{\rm Yukawa},\ph}	$	&		$	-	(\uG^{(6)}_{\ph,L})_{AB} (\ph\da\ph) \Lb_A \ph R_B 				
						-	(\uG^{(6)}_{\ph,U})_{AB} (\ph\da\ph)\Qb_A \ph^c U_B				
						-	(\uG^{(6)}_{\ph,D})_{AB} (\ph\da\ph)\Qb_A \ph D_B			$	\\
			&	\hskip20pt	$	-	\half(\uH^{(6)}_{\ph,L})^{\ka\la}_{AB}\lvb\ka a\lvb\la b (\ph\da\ph)\Lb_A \ph\si^{ab}R_B				
						-	\half(\uH^{(6)}_{\ph,U})^{\ka\la}_{AB}\lvb\ka a\lvb\la b (\ph\da\ph)\Qb_A \ph^c\si^{ab}U_B			$	\\
			&	\hskip20pt	$	-	\half(\uH^{(6)}_{\ph,D})^{\ka\la}_{AB}\lvb\ka a\lvb\la b (\ph\da\ph)\Qb_A \ph\si^{ab}D_B	+	\hc	$	\\
$	\cL^{(6)}_{{\rm Yukawa},\psi\ph}	$	&		$		\half(\uk^{(6)}_{LD\ph})^{\ka\mu}_{AB} \lvb\ka a(\Lb_A \ga^a L_B)(\ph\da iD_\mu\ph)				
						+	\half(\uk^{(6)}_{RD\ph})^{\ka\mu}_{AB} \lvb\ka a(\Rb_A \ga^a R_B)(\ph\da iD_\mu\ph)			$	\\
			&	\hskip20pt	$	+	\half(\uk^{(6)}_{QD\ph})^{\ka\mu}_{AB} \lvb\ka a(\Qb_A \ga^a Q_B)(\ph\da iD_\mu\ph)				
						+	\half(\uk^{(6)}_{UD\ph})^{\ka\mu}_{AB} \lvb\ka a(\Ub_A \ga^a U_B)(\ph\da iD_\mu\ph)			$	\\
			&	\hskip20pt	$	+	\half(\uk^{(6)}_{DD\ph,1})^{\ka\mu}_{AB} \lvb\ka a(\Db_A \ga^a D_B)(\ph\da iD_\mu\ph)				
						+	\half(\uk^{(6)}_{DL\ph})^{\ka\mu}_{AB} \lvb\ka a(\Lb_A \ga^a  iD_\mu L_B)(\ph\da\ph)			$	\\
			&	\hskip20pt	$	+	\half(\uk^{(6)}_{DR\ph})^{\ka\mu}_{AB} \lvb\ka a(\Rb_A \ga^a  iD_\mu R_B)(\ph\da\ph)				
						+	\half(\uk^{(6)}_{DQ\ph})^{\ka\mu}_{AB} \lvb\ka a(\Qb_A \ga^a  iD_\mu Q_B)(\ph\da\ph)			$	\\
			&	\hskip20pt	$	+	\half(\uk^{(6)}_{DU\ph})^{\ka\mu}_{AB} \lvb\ka a(\Ub_A \ga^a iD_\mu U_B)(\ph\da \ph)				
						+	\half(\uk^{(6)}_{DD\ph,2})^{\ka\mu}_{AB} \lvb\ka a(\Db_A \ga^a iD_\mu D_B)(\ph\da \ph)	+	\hc	$	\\
\hline
\hline
\end{tabular}
\end{table*}

In the fermion sector,
the generations are distinguished by an index $A = 1,2,3$.
The three charged leptons are denoted 
$l_A \equiv (e, \mu, \ta)$, 
the three neutrinos are 
$\nu_A \equiv (\nu_e, \nu_\mu, \nu_\ta)$.
For simplicity,
we consider the minimal SM 
with massless neutrinos.
The general effects of masses 
and combinations of Dirac and Majorana couplings 
for operators of arbitrary dimension affecting neutrino propagation 
are discussed in Ref.\ \cite{km04}.
The six quark flavors are
$u_A \equiv (u,c,t)$, $d_A \equiv (d,s,b)$,
with the color index suppressed.
Left- and right-handed spinor components
are defined as usual by
$\ps_L \equiv \frac 1 2 ( 1 - \ga_5 ) \ps$,
$\ps_R \equiv \frac 1 2 ( 1 + \ga_5 ) \ps$.
The right-handed leptons and quarks are SU(2) singlets,
$R_A = (l_A)_R$,
$U_A = (u_A)_R$,
$D_A = (d_A)_R$,
while the left-handed leptons and quarks form SU(2) doublets,
$L_A = ( (\nu_A)_L, (l_A)_L )^T$,
$Q_A = ( (u_A)_L, (d_A)_L )^T$.

Terms involving operators of mass dimension $d\leq 6$
in the Lagrange density 
$\cL_{\rm fermion}=\cL_{\rm lepton}+\cL_{\rm quark}$
are listed in Table \ref{tab:leptonquark},
including all couplings to the gravity and gauge sectors.
For simplicity,
this table is restricted to terms without background derivatives,
and operators that are total derivatives are omitted.
Except for the latter restriction,
the format of the table follows that of Table \ref{tab:SME-gauge}.
The conventions adopted in Table \ref{tab:leptonquark}
agree with standard usage in the literature.
In particular,
backgrounds with different spin and CPT properties
are represented by different symbols in the usual way,
and the index positions on the backgrounds are the standard ones. 
The index symmetry of a given background is determined
by that of the corresponding dynamical operator.
Each occurrence of the symbol $\hc$ for the hermitian conjugate
applies to all terms in the particular component of the Lagrange density. 
In rows without the symbol $\hc$,
the backgrounds can be assumed hermitian in generation space.

In the Higgs sector,
we write the Higgs doublet $\ph$ in unitary gauge
in the form $\ph = ( 0, r_\ph)^T /\sqrt{2}$.
The conjugate Higgs doublet is denoted $\ph^c$.
Table \ref{tab:higgs} provides the explicit form
of the components $\cL^{(d)}_{\rm Higgs}$ with $d\leq 6$
of the Lagrange density in the Higgs sector,
excluding terms with background derivatives
and ones involving total derivatives of the dynamical operators. 
The structure of the table follows that of Table \ref{tab:leptonquark},
and it includes all couplings to the gravity and gauge sectors.
In the table,
each background $\uk$ has indices with symmetry
matching that of the associated dynamical operator,
and all backgrounds can be taken as real.
Note that a few terms have backgrounds $\uk$ without spacetime indices.
For example,
a term with $d=4$ proportional to $\uk^{(4)}_{\ph\ph} (\ph\da \ph)^2$ 
appears in the table.
These types of expressions incorporate 
both scalar coupling constants and position-dependent effects.
The component terms of this form in 
$\cL^{(2)}_{\rm Higgs}$ and $\cL^{(4)}_{\rm Higgs}$ 
replicate conventional SM terms in the first row of the table,
so they can either be omitted or understood
as renormalizations of the conventional terms having no physical effects.
A similar comment applies to the first term in $\cL^{(4)}_{\rm Higgs}$,
where the trace piece of the constant component
of the background $(\uk)_{\ph\ph})^{\mu\nu}$
can be viewed as a renormalization of the usual Higgs kinetic term.

Finally,
we present in Table \ref{tab:yukawa}
all terms involving operators with $d\leq 6$ 
that couple the fermions to the Higgs boson,
restricting attention for simplicity to terms without background derivatives
and disregarding total derivatives of dynamical operators. 
The terms in this table represent generalizations 
of the Yukawa couplings in the SM,
which are listed in the first row.
Note that the first three entries for $\cL^{(4)}_{\rm Yukawa}$ 
contain pieces that can be understood as physically irrelevant
renormalizations of the usual SM Yukawa couplings,
along with other nontrivial background effects.
For entries already discussed in the literature,
the notation in the table follows existing conventions.
Backgrounds in terms written without an accompanying hermitian conjugate $\hc$
can be taken as hermitian in generation space.
All backgrounds have indices with symmetries
constrained according to the structure of the corresponding operators.

\renewcommand\arraystretch{1.6}
\begin{table*}
\caption{
\label{tab:LI}
Terms with $d\leq 6$ involving scalar coupling constants.}
\setlength{\tabcolsep}{6pt}
\begin{tabular}{ccl}
\hline
\hline
Sector & $d$ & Terms\\
\hline
Gravity	&	4	&	$		\tfrac{1}{2\ka}R				$	\\
	&	6	&	$		\tfrac{1}{2\ka}R_{\al\be\ga\de}R^{\al\be\ga\de}					
					,\	\tfrac{1}{2\ka}R^{\al\be}R_{\al\be}					
					,\	\tfrac{1}{2\ka}R^2					
					,\	\tfrac{1}{2\ka}\ep^{\ka\la\mu\nu}\uR\al\be\ka\la \uR\be\al\mu\nu				$	\\
Gauge	&	4	&	$		\tr(F_{\mu\nu}F^{\mu\nu})					
					,\	\tr(F_{\mu\nu}\Ftil^{\mu\nu})				$	\\
	&	6	&	$		\tr(F_{\mu\nu}D^\al D_\al F^{\mu\nu})					
					,\	\tr(F_{\mu\nu}D^\al D_\al \Ftil^{\mu\nu})					
					,\	\tr(F^{\mu\nu}D_{(\mu}D_{\rh)}F^{\rh}_{\pt{\rh}\nu})					
					,\	\tr(F^{\mu\nu}D_{(\mu}D_{\rh)}\Ftil^{\rh}_{\pt{\rh}\nu})			,	$	\\
	&		&	$		R^{\ka\la\mu\nu} \tr(F_{\ka\la}F_{\mu\nu})					
					,\	R^{\ka\la\mu\nu} \tr(F_{\ka\la}\Ftil_{\mu\nu})					
					,\	R^{\nu\rh}\tr(F_{\mu\nu}F^\mu_{\pt{\mu}\rh})					
					,\	R^{\nu\rh}\tr(F_{\mu\nu}\Ftil^\mu_{\pt{\mu}\rh})					
					,\	R\,\tr(F_{\mu\nu}F^{\mu\nu})					
					,\	R\,\tr(F_{\mu\nu}\Ftil^{\mu\nu})			,	$	\\
	&		&	$		R^{\al\be}\tr(D_{(\al}D_{\ga)}F_{\be}^{\pt{\be}\ga})					
					,\	R^{\al\be}\tr(D_{(\al}D_{\ga)}\Ftil_{\be}^{\pt{\be}\ga})					
					,\	i\,\tr(F^\mu_{\pt{\mu}\nu}F^\nu_{\pt{\nu}\rh}F^\rh_{\pt{\rh}\mu})					
					,\	i\,\tr(F^\mu_{\pt{\mu}\nu}F^\nu_{\pt{\nu}\rh}\Ftil^\rh_{\pt{\rh}\mu})				$	\\
Fermion	&	3	&	$		\psb\ps					
					,\	i\psb\ga_5\ps				$	\\
	&	4	&	$		\half\ivb\mu a \psb\ga^a iD_\mu\ps	+	\hc			
					,\	\half\ivb\mu a \psb\ga_5\ga^a iD_\mu\ps	+	\hc		$	\\
	&	5	&	$		\half \psb iD^\mu iD_\mu\ps	+	\hc			
					,\	\half i\psb \ga_5 iD^\mu iD_\mu\ps	+	\hc			
					,\	R\psb\ps					
					,\	iR\psb\ga_5\ps					
					,\	\ivb\mu a\ivb\nu b \psb\si^{ab}F_{\mu\nu}\ps					
					,\	\ivb\mu a\ivb\nu b \psb\si^{ab}\Ftil_{\mu\nu}\ps			,	$	\\
	&	6	&	$		\half g^{\nu\rh}\ivb\mu a\psb\ga^a iD_{(\mu} iD_\nu iD_{\rh)}\ps	+	\hc			
					,\	\half g^{\nu\rh}\ivb\mu a\psb\ga_5\ga^a iD_{(\mu} iD_\nu iD_{\rh)}\ps	+	\hc	,	$	\\
	&		&	$		\half\lvb\ka a R^{\ka\mu}\psb\ga^a iD_\mu\ps	+	\hc			
					,\	\half\lvb\ka a R^{\ka\mu}\psb\ga_5\ga^a iD_\mu\ps	+	\hc			
					,\	\half\ivb\mu a R \psb\ga^a iD_\mu\ps	+	\hc			
					,\	\half\ivb\mu a R \psb\ga_5\ga^a iD_\mu\ps	+	\hc	,	$	\\
	&		&	$		\lvb\nu a (D_\mu R^{\mu\nu})\psb\ga^a\ps					
					,\	\lvb\nu a(D_\mu R^{\mu\nu}) \psb\ga_5\ga^a\ps					
					,\	\ivb\mu a(D_\mu R) \psb\ga^a\ps					
					,\	\ivb\mu a(D_\mu R)\psb\ga_5\ga^a\ps			,	$	\\
	&		&	$		\lvb\ka a\psb\ga^a (D_\mu F^{\ka\mu})\ps					
					,\	\lvb\ka a\psb\ga_5\ga^a (D_\mu F^{\ka\mu})\ps					
					,\	\half\lvb\ka a\psb\ga^a F^{\ka\mu} iD_\mu\ps	+	\hc			
					,\	\half\lvb\ka a\psb\ga_5\ga^a F^{\ka\mu} iD_\mu\ps	+	\hc	,	$	\\
	&		&	$		\half\lvb\ka a\psb\ga^a \Ftil^{\ka\mu} iD_\mu\ps	+	\hc			
					,\	\half\lvb\ka a\psb\ga_5\ga^a \Ftil^{\ka\mu} iD_\mu\ps	+	\hc	,	$	\\
	&		&	$		(\psb\ps)(\psb\ps)					
					,\	i(\psb\ps)(\psb\ga_5\ps)					
					,\ -	(\psb\ga_5\ps)(\psb\ga_5\ps)					
					,\	(\psb\ga^a \ps) (\psb\ga_a \ps)					
					,\ 	(\psb\ga^a \ps) (\psb\ga_5\ga_a \ps)			,	$	\\
	&		&	$		(\psb\ga_5\ga^a \ps) (\psb\ga_5\ga_a \ps)					
					,\	\quar (\psb\si^{ab}\ps) (\psb\si_{ab}\ps)					
					,\	\quar \ep_{abcd}(\psb\si^{ab}\ps) (\psb\si^{cd}\ps)				$	\\
Scalar	&	2	&	$		\ph\da\ph				$	\\
	&	4	&	$		\half \ph\da iD^\mu iD_\mu \ph	+	\hc			
					,\	R\ph\da\ph					
					,\	(\ph\da\ph)^2				$	\\
	&	5	&	$		(\psb\ps)(\ph\da\ph)					
					,\	i(\psb\ga_5\ps)(\ph\da\ph)				$	\\
	&	6	&	$		\half g^{\mu\nu}g^{\rh\si}\ph\da iD_{(\mu}iD_\nu iD_\rh iD_{\si)}\ph	+	\hc			
					,\	\half R^{\mu\nu} \ph\da iD_{(\mu}iD_{\nu)}\ph	+	\hc			
					,\	\half R \ph\da iD^\mu iD_\mu\ph	+	\hc	,	$	\\
	&		&	$		\half (D_\mu R^{\mu\nu}) \ph\da iD_\nu\ph	+	\hc			
					,\	\half (D^\mu R) \ph\da iD_\mu\ph	+	\hc			
					,\	\half \ph\da(D_\mu F^{\mu\nu})iD_\nu\ph	+	\hc			
					,\	\half(\ph\da\ph)(\ph\da iD^\mu iD_\mu\ph)	+	\hc	,	$	\\
	&		&	$		\half(\ph\da iD^\mu\ph)(\ph\da iD_\mu\ph)	+	\hc			
					,\	\half \big((iD^\mu\ph)\da\ph\big)(\ph\da iD_\mu\ph)	+	\hc			
					,\	\half \ivb\mu a (\psb\ga^a\ps)(\ph\da iD_\mu\ph)	+	\hc	,	$	\\
	&		&	$		\half \ivb\mu a (\psb\ga_5\ga^a\ps)(\ph\da iD_\mu\ph)	+	\hc			
					,\	\half \ivb\mu a (\psb\ga^a iD_\mu\ps)(\ph\da\ph)	+	\hc			
					,\	\half \ivb\mu a (\psb\ga_5\ga^a iD_\mu\ps)(\ph\da\ph)	+	\hc	,	$	\\
	&		&	$		(D_\mu D_\nu R^{\mu\nu})\ph\da\ph					
					,\	(D^\mu D_\mu R)\ph\da\ph					
					,\	R_{\al\be\ga\de}R^{\al\be\ga\de}(\ph\da\ph)					
					,\	R^{\al\be}R_{\al\be}(\ph\da\ph)					
					,\	R^2(\ph\da\ph)					
					,\	\ep^{\ka\la\mu\nu}\uR\al\be\ka\la \uR\be\al\mu\nu(\ph\da\ph)			,	$	\\
	&		&	$		\ph\da F_{\mu\nu}F^{\mu\nu}\ph					
					,\	\ph\da F_{\mu\nu}\Ftil^{\mu\nu}\ph					
					,\	\tr(F_{\mu\nu}F^{\mu\nu})(\ph\da\ph)					
					,\	\tr(F_{\mu\nu}\Ftil^{\mu\nu})(\ph\da\ph)					
					,\	R(\ph\da\ph)^2					
					,\	(\ph\da\ph)^3				$	\\
\hline
\hline
\end{tabular}
\end{table*}

\subsection{Constant scalar backgrounds}
\label{sec:LI}

All the examples discussed above
contain special cases with constant scalar backgrounds,
which can be viewed as scalar coupling constants.
In particular,
the corresponding limits of the effective field theories
built on the Einstein-Maxwell theories
and on GR coupled to the SM
can be extracted from the tables provided
in the previous subsections.
For a given theory,
the explicit terms of this type 
are obtained as appropriate restrictions of the backgrounds $\uk$
to maintain local Lorentz and diffeomorphism invariance.

As discussed in Sec.\ \ref{Backgrounds},
a background $k$ transforms as a spacetime tensor
under observer local Lorentz and general coordinate transformations,
but it remains invariant under all particle transformations
including both local Lorentz transformations and diffeomorphisms. 
This behavior is compatible 
with local Lorentz and diffeomorphism invariance of the Lagrange density 
only if $k$ carries no indices and is constant,
in which case it acts as a conventional coupling constant.
We can therefore identify all contributions 
that produce scalar coupling constants in a given effective Lagrange density 
by keeping only terms involving components of each combination 
$\uk^{\mu\cdots}{}_{\nu\cdots}{}^{a\cdots}$
that are proportional to products 
of the vierbein $e_{\mu}{}^{a}$, metric $g_{\mu\nu}$,
and Levi-Civita tensor $\ep_{\ka\la\mu\nu}$
and then fixing these components to be spacetime constants.
Note that this implies discarding all terms
involving nonzero background derivatives $Dk$.

With this procedure in hand,
it is straightforward to extract
the limits of the various theories discussed above
that have only scalar coupling constants.
As an illustration,
we provide in Table \ref{tab:LI} 
a listing of terms with $d\leq 6$ having only scalar coupling constants 
that is obtained from the generic Lagrange densities presented 
in Tables \ref{tab:curvature},
\ref{tab:gauge},
\ref{tab:spinor2},
and \ref{tab:scalar}.
The first column of Table \ref{tab:LI} 
specifies the sector,
the second column fixes the value of $d$,
and the third column displays the corresponding terms
with scalar coupling constants.
The complete Lagrange density of this type
is obtained by multiplying each operator displayed by a coupling constant
and adding all the resulting terms. 
Note that the term $\ep^{\ka\la\mu\nu}\uR\al\be\ka\la \uR\be\al\mu\nu$
and the combination
\bea
&&\ep^{\al\be\ga\de}\ep^{\ka\la\mu\nu}R_{\al\be\ka\la}R_{\ga\de\mu\nu}\nn\\
&&\hskip60pt
=4(R_{\al\be\ga\de}R^{\al\be\ga\de}-4R_{\al\be}R^{\al\be}+R^2)
\qquad
\eea
can be expressed as total derivatives.
For simplicity,
other total-derivative operators are omitted from the table.

In the limit of scalar coupling constants,
all indices on the dynamical operators must be contracted.
The curvature, the gauge field strength, and the scalar field
all have even numbers of indices
and appear in dynamical combinations that have even mass dimension $d$.
Since the vierbein, metric, and Levi-Civita tensors
also have even numbers of indices,
no terms with odd $d$ can appear in the gravity or gauge sectors.
Since the quadratic combination of fermions has odd mass dimension $d=3$,
terms with odd $d$ can appear in the fermion sector
and in fermion couplings to scalars. 
These features are reflected in the result in Table \ref{tab:LI}.

The gravity sector of Table \ref{tab:LI} 
consists of terms of dimension $d\leq 6$ 
in the effective field theory for GR
that involve only scalar coupling constants.
Various combinations of the sectors in the table
form other effective field theories of this type including,
for example,
ones based 
on Einstein-Yang-Mills and Einstein-Maxwell theory,
on nonabelian gauge theory and Maxwell electrodynamics 
in Minkowski spacetime,
and theories with fermions and scalars.
Using the same procedure,
the effective field theory for GR coupled to the SM
can be obtained directly from
Tables \ref{tab:SME-gauge},
\ref{tab:leptonquark},
\ref{tab:higgs},
and \ref{tab:yukawa}.
In Minkowski spacetime,
this reduces to the SM effective field theory with $d\leq 6$.
The operators for arbitrary $d$ involving only scalar coupling constants
in these various realistic theories
have been presented elsewhere in the literature,
including ones for arbitrary $d$ in 
the graviton sector
\cite{km16},
the neutrino sector
\cite{km04},
the photon sector
\cite{km09},
and the fermion sector
\cite{km13}.

\section{Summary}
\label{Summary}

In this work,
we develop the framework for gravitational effective field theory
in the presence of backgrounds
and provide a methodology for constructing 
operators of arbitrary mass dimension $d$ in the Lagrange density.
Explicit terms with $d\leq 6$ are obtained for several theories,
including the realistic cases of GR coupled to the SM
and some of its limits.
The results presented here are achieved through a combination
of conceptual developments and technical results.

The underpinnings of the framework
are discussed in Sec.\ \ref{Framework}.
Table \ref{tab:transformations} summarizes 
the relevant spacetime transformations 
in both curved and approximately flat spacetimes.
The various types of backgrounds 
and their implications for violations of symmetries in curved spacetimes
are described,
with examples provided in Table \ref{tab:fulltypes}.
Linearizing in approximately flat spacetimes
produces limiting cases of the spacetime transformations,
listed in Table \ref{tab:lineartypes}.
The links between properties of terms 
in the full and linearized Lagrange densities
are schematically displayed in Fig.\ \ref{fig:linear},
and the symmetry properties of the various cases
are illustrated with examples in Table \ref{tab:relation}.
We also revisit the no-go constraints
arising from the compatibility 
of the variational procedure with the Bianchi identities,
showing that a large class of potential perturbative terms
in the effective Lagrange density 
cannot arise from a Riemann geometry or its extensions
but instead must have an alternative geometric or nongeometric origin
in the underlying theory.
These results are illustrated pictorially in Fig.\ \ref{fig:geometry}.

Using this framework,
the methodology for constructing terms 
in a generic effective Lagrange density 
is presented in Sec.\ \ref{Effective gravity}.
The use of compact notation for backgrounds
and technical results for construction of gauge-covariant operators 
permit the enumeration and classification of terms
in the Lagrange density.
Tables \ref{tab:gravity}, \ref{tab:curvature}, and \ref{Dk}
provide explicit results for the pure-gravity and background sector
involving operators with $d\leq6$.
Analogous forms for terms in the matter-gravity sector are obtained.
Tables \ref{fullgauge} and \ref{tab:gauge}
consider operators containing gauge fields,
Tables \ref{tab:fullspinor} and \ref{tab:spinor2}
present results for Dirac fermions,
and Tables \ref{tab:fullscalar} and \ref{tab:scalar}
treat scalars.

Applications of the methodology
to cases of practical importance
are considered in Sec.\ \ref{Applications}.
For Einstein-Maxwell effective field theories,
terms in the matter-gravity Lagrange density with $d\leq 6$
are provided for photons in Table \ref{tab:photon}
and for uncharged scalars in Table \ref{tab:realscalar}.
For the realistic effective field theory consisting of 
GR coupled to the SM,
we present all terms with $d\leq 6$ in the matter-gravity sector
involving gauge, lepton, quark, Higgs, and Yukawa couplings 
in Tables \ref{tab:SME-gauge}, \ref{tab:leptonquark},
\ref{tab:higgs}, and \ref{tab:yukawa}.
Terms for the limiting case with backgrounds 
acting only as scalar coupling constants
are displayed in Table \ref{tab:LI}.

The results obtained in this work establish the foundation 
for further investigations of gravitational effective field theories.
The explicit characterization of terms in realistic scenarios
provided here opens the way 
for future phenomenological and experimental searches,
with promising potential for detecting 
observable signals from the Planck scale.

\section{Acknowledgments}

This work was supported in part
by the U.S.\ Department of Energy under grant {DE}-SC0010120
and by the Indiana University Center for Spacetime Symmetries.

\end{document}